\documentclass[11pt,a4paper]{article}
\usepackage{jheppub}
\usepackage{array}
\usepackage{mathtools}
\usepackage{multirow}
\usepackage{rotating}
\usepackage{amsmath}

\usepackage{color}
\definecolor{darkred}{rgb}{0.7,0,0}
\definecolor{purple}{rgb}{0.58,0,0.84}

\newcommand{\be}{\begin{eqnarray}}
\newcommand{\bse}{\begin{subequations}}
\newcommand{\ese}{\end{subequations}}
\newcommand{\bea}{\begin{eqnarray}}
\newcommand{\eea}{\end{eqnarray}}
\newcommand{\Beq}{\begin{eqnarray}}
\newcommand{\Eeq}{\end{eqnarray}}
\newcommand{\ba}{\begin{array}}
\newcommand{\ea}{\end{array}}
\newcommand{\ee}{\end{eqnarray}}

\newcommand{\ga}{g_{\rm{A}}}
\def\p{\partial}
\def\bz{{\bar z}}

\def\bw{{\bar w}}

\newcommand{\mH}{\mathcal{H}}

\newcommand{\xp}{\xi_{\rm{A}}}
\newcommand{\hs}{h_{\sigma}}
\newcommand{\Ths}{B_{\sigma}}

\newcommand{\an}{\quad \textmd{and} \quad}
\newcommand{\where}{\quad \textmd{where} \quad}
\newcommand{\x}{\tilde{\tau}}

\newcommand{\mpl}{M_{\rm Pl}}
\newcommand{\kl}{\kappa_{\sigma}}
\newcommand{\bk}{\vec{k}}

\newcommand{\vk}{\vec{k}}
\newcommand{\da}{\delta_{_{\rm B}}}

\newcommand{\xz}{\xi_{Z_0}}
\newcommand{\xL}{\tilde{\tau}_{\Lambda}}
\newcommand{\Bs}{B_{\sigma}}
\newcommand{\tD}{\tilde{D}}
\newcommand{\C}{\mathcal{C}}
\newcommand{\mC}{\tilde{\mathcal{C}}}
\definecolor{darkred}{rgb}{0.7,0,0}
\definecolor{purple}{rgb}{0.58,0,0.84}

\title{Production and Backreaction of Spin-2 Particles of $SU(2)$ Gauge Field during Inflation}
\author[a]{A. Maleknejad,}
\author[a,b]{and E. Komatsu}

\affiliation[a]{Max-Planck-Institute for Astrophysics, Karl-Schwarzschild-Str. 1, 85741 Garching, Germany}
\affiliation[b]{Kavli Institute for the Physics and Mathematics of the Universe (Kavli IPMU,
WPI), UTIAS, The University of Tokyo, Chiba, 277-8583, Japan}

\emailAdd{amalek@MPA-Garching.MPG.DE}
\emailAdd{komatsu@MPA-Garching.MPG.DE}

\abstract{Primordial SU(2) gauge fields with an isotropic background lead to the production of spin-2 particles during inflation. We provide a unified formalism to compute this effect in all of the inflation models with isotropic SU(2) gauge fields such as Gauge-flation and Chromo-Natural inflation with and without spectator axion fields or the mass of the gauge field from the Higgs mechanism. First, we calculate the number and energy densities of the spin-2 particles. We then obtain exact analytical formulae for their backreaction on the background equations of motion of SU(2) and axion fields in (quasi) de Sitter expansion, which were calculated only numerically for one particular model in the literature. We show that the backreaction is directly related to the number density of the spin-2 field. Second, we relate the number density of the spin-2 particles to the power spectrum and the energy density of the gravitational waves sourced by them. Finally, we use the size of the backreaction to constrain the parameter space of the models. We find that the tensor-to-scalar ratio of the sourced gravitational waves can at most be on the order of that of the vacuum contribution to avoid a large backreaction on slow-roll dynamics of the gauge and axion fields in quasi-de Sitter expansion.  }
\begin{document}
\maketitle

\date{\today}

%%%%%%%%%%%%%%%%%%%%%%%%%%%%%%%%%%%%%%%%%%%%%%%%%%%%%%%%%%%

\section{Introduction}\label{sec:Introduction}
Inflation \cite{Guth:1980zm, Sato:1980yn, Linde:1981mu, Albrecht:1982wi} with SU(2) gauge fields \cite{Maleknejad:2011sq, Maleknejad:2011jw, Adshead:2012kp, Adshead:2013nka} has a rich phenomenology that is not shared by canonical single-scalar-field inflation models (see \cite{Maleknejad:2012fw} for a review). As was first discovered by one of the authors (A.M.), when the conformal symmetry of Yang-Mills theory is broken by an effective $(F\tilde F)^2$ term in the Lagrangian, non-Abelian gauge fields acquire an isotropic and homogeneous background (vacuum expectation value; VEV) solution during inflation  \cite{Maleknejad:2011sq, Maleknejad:2011jw}. This VEV produces a copious amount of spin-2 particles which, in turn, linearly mix with tensor perturbations in the metric, i.e., gravitational waves. The same phenomenology is obtained when the conformal symmetry is broken by a Chern-Simons interaction with an axion field $\varphi F\tilde F$ \cite{Adshead:2012kp, Adshead:2013nka}.

In addition to the original models\footnote{The original models of
gauge-flation and chromo-natural inflation have been ruled out by the
Planck data \cite{Namba:2013kia, Adshead:2013nka}.}, there are several more inflationary
models with the $SU(2)$ VEV which share the above features
\cite{Maleknejad:2016qjz, Nieto:2016gnp, Caldwell:2017chz,
Adshead:2016omu, Dimastrogiovanni:2016fuu, Adshead:2017hnc}. Despite
differences in details of the models, their tensor sector can be
presented in a unified manner. As the
sourced tensor power spectrum is proportional to the density parameter of the
gauge field during inflation, these models violate the Lyth
bound \cite{Dimastrogiovanni:2012ew,Adshead:2013qp}.
Depending on the details of the slow-roll
dynamics of the gauge field VEV, e.g., a form of the axion potential,
the tensor spectral index $n_T$ can be negative or positive, and thus
violates the conventional consistency relation of single-field slow-roll
inflation, $n_T =-r/8$.
Moreover, parity-violating interactions in linear perturbations make this
spin-2 field chiral and hence generate observable circularly
polarized gravitational waves as well as parity-odd correlations of
cosmic microwave background (CMB) anisotropies, i.e.,
non-zero $TB$ and $EB$ \cite{Thorne:2017jft}.
Due to self-interactions of gauge fields, the gravitational waves can be
highly non-Gaussian, yielding a large tensor bispectrum with
approximately an equilateral shape \cite{Agrawal:2017awz,
Agrawal:2018mrg,Dimastrogiovanni:2018xnn}.
Finally, chiral gravitational waves can generate baryon
asymmetry via a gravitational anomaly
\cite{Maleknejad:2014wsa, Maleknejad:2016dci, Caldwell:2017chz,
Adshead:2017znw}, and can serve as a natural
leptogenesis mechanism during inflation to explain the observed baryon
asymmetry in the Universe.

All of these signatures are robust consequences of having gauge fields
during inflation and carry important information about the matter
content of the early universe. The stochastic background of
gravitational waves can be within reach of future CMB experiments
\cite{Matsumura:2013aja,Ade:2018sbj,Abazajian:2016yjj}, and that of
$n_T>0$ can be within reach of future gravitational wave interferometers
\cite{Thorne:2017jft, Caldwell:2017chz, Domcke:2018rvv}.
 As none of these features exists in canonical single-scalar-field inflation models, we can use them to distinguish the particle physics models of inflation. For example, these models could be embedded in supergravity \cite{DallAgata:2018ybl} and string theory  \cite{McDonough:2018xzh}.

In this paper, we take a closer look at the phenomenology of the spin-2
particles generated from SU(2) gauge fields. Particularly significant is
the backreaction of spin-2 particles on the background equations of
motion of the gauge and axion fields, as it could spoil significant
properties of inflation with SU(2) gauge fields.  We also gain better
insights into the power spectrum and energy density of primordial
gravitational waves by relating them to the number density of the spin-2
particles.

%
% EK: This is too much for introduction.
%
%{\tcdr{We realize that the strong backreaction excludes $r_{source}/r_{vac}\gg1$ region in the massless (un-Higgsed) models in our class of inflationary scenarios. In the Higgsed models, e.g., Higgsed gauge-flation and Higgsed chromo-natural, both scalar and tensor perturbations are amplified by the Higgs VEV, however, the scalar modes are more strongly enhanced. As a result, Higgs VEV generally reduces the tensor-to-scalar ratio \cite{Adshead:2017hnc, Adshead:2016omu}. In the current work, we find out the size of backreaction is much stronger in the Higgsed models. Therefore, the Higgsed models cannot evade $r_{source}/r_{vac}\sim 1$ bound as well.}}

This paper is organized as follows. In section \ref{review}, we briefly review the existing models of inflation with an $SU(2)$ gauge field in a unified approach. We study the spin-2 particle production in section \ref{P-P}. In section \ref{backreact}, we compute the backreaction of this spin-2 field on the background field equations. We then relate the power spectrum and energy density of the sourced gravitational waves to the number density of the spin-2 particles in section \ref{gw}. In section \ref{PA}, we use the size of the backreaction to constrain the parameter space of the models. Finally, we conclude in \ref{conclusion}. In appendix \ref{BG-VEV}, we discuss the symmetry structure of the background $SU(2)$ gauge field. In appendix \ref{app-spin-2}, first, we discuss the action of the transverse-traceless perturbed $SU(2)$ field around its VEV in a unified approach. In \ref{spin-2-SU(2)}, we prove that a perturbed $SU(2)$ field around its VEV has a spin-2 field. The details of our analytical study as well as some necessary mathematical tools are presented in appendices \ref{Math}-\ref{GW-app}.

\section{Review of theory}\label{review}
Consider an inflationary model with a Friedmann-Lema\^{i}tre-Robertson-Walker (FRLW) background
\be\label{FRW}
ds^2= - dt^2 +a^2(t)\delta_{ij} dx^i dx^j,
\ee
 that can support slow-roll inflation with a slowly varying  SU(2) gauge field VEV given by \cite{Maleknejad:2011jw, Maleknejad:2011sq}
\be\label{BG-A}
\bar{A}_{\mu}(t) \equiv \bar{A}^a_{\mu}(t) T_a= \begin{cases}
 0 & \mu=0 \\
 a \psi(t) \delta^a_{i}~T_{a}  &  \mu=i \,,
\end{cases}
\ee
where $\{T_a\}$ are the generators of the $su(2)$ algebra with $a=1,2,3$
\be
T_aT_b = \frac14 \delta_{ab}{\rm{I}}_{n} + \frac12 i\epsilon^{abc}T_c,
\ee
where ${\rm{I}}_n\delta_{ab}$ is the identity matrix and $\epsilon^{abc}$ is the totally antisymmetric matrix.
In appendix \ref{BG-VEV}, we show that ansatz \eqref{BG-A} is the general background solution for a gauge field with an isotropic and homogeneous energy-momentum tensor. 
We unify all the inflation models with an $SU(2)$ field in the literature in the following Lagrangian
\be\label{g-action}
S = \int d^4x \sqrt{-g} \bigg[  \mathcal{L}_{\rm{A}}(A_{\mu},\varphi) +\alpha_s \mathcal{L}_0(\chi) + \alpha_H \mathcal{L}_H (A_{\mu},H) \bigg],
\ee 
where $\mathcal{L}_{\rm{A}}$ is the gauge field theory sector (with possibly an axion field $\varphi$), $\mathcal{L}_{0}$ is a (possible) scalar theory and $\mathcal{L}_H$ is a (possible) Higgs sector which makes the gauge field massive. The parameters $\alpha_s$ and $\alpha_H$ classify models as (see table \ref{T0})
\be
\alpha_s = \begin{cases}
 0 & \textmd{(axion)-SU(2) gauge field inflaton}, \\
 1  &  \textmd{spectator (axion)-SU(2) gauge field} \,,
\end{cases} \an \alpha_H = \begin{cases}
 0 & \textmd{massless models}, \\
 1  &  \textmd{Higgsed models} \,.
\end{cases} \nonumber
\ee       
 In these models, the conformal symmetry of Yang-Mills 
theory is broken by either adding a $(F\tilde F)^2$ effective term to the gauge theory, e.g. Gauge-flation \cite{Maleknejad:2011jw, Maleknejad:2011sq}, or by coupling the gauge field sector to an axion field $\varphi$ with slow-roll dynamics, e.g. chromo-natural \cite{Adshead:2012kp, Adshead:2013nka}.
The former models are given by \cite{Maleknejad:2011jw,Maleknejad:2011sq}
\be\label{Gf}
\mathcal{L}_{\rm{A}} \rightarrow \mathcal{L}_{\rm{Gf}} \equiv -\frac14 F_{\mu\nu}F^{\mu\nu} + \frac{\kappa}{96} (F_{\mu\nu}\tilde F^{\mu\nu})^2,
\ee
\\
while the latter are given by \footnote{A more general action including two dimension six operators, $\textmd{tr}(FFF)$ and the (PT violating) Weinberg operator $\textmd{tr}(FF\tilde{F})$ \cite{Weinberg:1989dx}, has been considered in \cite{Maleknejad:2014wsa}. }
\cite{Adshead:2012kp, Adshead:2013nka}
\be\label{Cn}
\mathcal{L}_{\rm{A}} \rightarrow \mathcal{L}_{\rm{Cn}} \equiv -\frac14 F_{\mu\nu}F^{\mu\nu} - \frac{\lambda\varphi}{4f} F_{\mu\nu}\tilde F^{\mu\nu}-\frac{1}{2}\p_{\mu}\varphi \p^{\mu}\varphi -V(\varphi),
\ee
where $F_{\mu\nu}= T_a F^a_{\mu\nu}$ is the field strength tensor $$F^a_{\mu\nu}=\p_{\mu}A^a_{\nu}-\p_{\nu}A^a_{\mu} + \ga \epsilon^{abc} A^b_{\mu} A^c_{\nu},$$  while $\tilde F^{\mu\nu} \equiv \frac12 \epsilon^{\mu\nu\lambda\sigma} F_{\lambda\sigma}$, $\varphi$ is the axion and $V(\varphi)$ is the axion potential. \footnote{In the original chromo-natural model, the potential is the standard cosine potential, $V(\varphi)=\mu^4(1+\cos(\frac{\varphi}{f}))$, with $f\ll \mpl$ and $\lambda \gtrsim 10^3 \frac{f}{\mpl}$.}  Moreover, in the Higgsed version of the models, the gauge field becomes massive by a Higgs field and we have an extra term for the dynamics of the Goldstone boson which in the Stueckelberg form is \cite{Ruegg:2003ps, Kunimasa:1967zza} \footnote{Here, the full Higgs field theory is
\be
\mathcal{L}_{Z}=-\frac12 D_{\mu} ZD^{\mu}Z^{\dag}-V(z), 
\ee
with a VEV given as $\bar{Z}_A=Z_0(t)\delta_A^a ~T_a$ in which $A=1,2,3$ is the field's internal index, and $D_{\mu}=\p_{\mu}-i\ga A_{\mu}$ is the covariant derivative. However, we are interested in the limit that the Higgs mass is much greater than the Hubble scale. Therefore, the only relevant sector is the Goldstone boson part given in \eqref{Higgs}. Notice that $\mathcal{L}_H$ is gauge invariant and it can be written as $\mathcal{L}_H= -\ga^2 Z_0^2 {\rm{tr}}(UD_{\mu}U^{-1}/(-i\ga))^2$. }
\be\label{Higgs}
\mathcal{L}_{H} = -\ga^2Z_0^2 {\rm{tr}}\bigg(A_{\mu} -\frac{i}{\ga} U^{-1} \p_{\mu} U\bigg)^2,
\ee
where $U=\exp(i\ga \pi)$ and $\pi$ is the Goldstone mode corresponding to the Higgs fluctuations around its VEV, $\pi=\pi^a T_a$.
We relate these models to the literature in table \ref{T0}.

\begin{table}
\begin{center}
\begin{tabular}{ | m{3cm} | m{0.5cm} | m{0.5cm} | m{0.6cm}| m{3.4cm}|} 
\hline
~~~~~~~ Model & $\alpha_{s}$  & $\alpha_{H}$  & $\mathcal{L}_{A}$ & Original references   \\ 
\hline
\multirow{2}{*}{} & \multirow{2}{*}{0} & \multirow{2}{*}{0} &  $\mathcal{L}_{Gf}$ & ~~~~~~~ \cite{Maleknejad:2011jw,Maleknejad:2011sq}  \\
& & & $\mathcal{L}_{Cn}$ & ~~~~~~~ \cite{Adshead:2012kp, Adshead:2013nka} \\
 \hline
\multirow{2}{*}{~~~~Spectator} & \multirow{2}{*}{1} & \multirow{2}{*}{0} & $\mathcal{L}_{Gf}$ & ~~~~~~~ $ \large{-}$  \\
& & & $\mathcal{L}_{Cn}$ & ~~~~~~~ \cite{Dimastrogiovanni:2016fuu} \\ 
\hline
\multirow{2}{*}{~~~~Higgsed} & \multirow{2}{*}{0} & \multirow{2}{*}{1} & $\mathcal{L}_{Gf}$ & ~~~~~~~ \cite{Nieto:2016gnp,Adshead:2017hnc} \\
& & & $\mathcal{L}_{Cn}$ & ~~~~~~~ \cite{Adshead:2016omu} \\
  \hline
\multirow{2}{*}{Spectator-Higgsed} & \multirow{2}{*}{1} &  \multirow{2}{*}{1} & $\mathcal{L}_{Gf}$ & ~~~~~~~ $-$ \\
& & & $\mathcal{L}_{Cn}$ & ~~~~~~~ $-$ \\
  \hline
\end{tabular}
\end{center}
\caption{Inflationary models involving an $SU(2)$ gauge field in the literature and their relation to $\mathcal{L}_{A}$, $\alpha_s$ and $\alpha_H$.
}\label{T0}
\end{table}

 The $(F\tilde F)^2$ term in \eqref{Gf} comes as an effective theory of \eqref{Cn} by integrating out the massive axion on energy scales below the mass of the axion, $M \simeq \mu^2/f$. In that case, the parameter $\kappa$ is given as $\kappa = \frac{3\lambda^2}{\mu^4}$ \cite{SheikhJabbari:2012qf, Adshead:2012qe, Maleknejad:2012fw}. Therefore, gauge-flation models are effectively equivalent to chromo-natural models in the limit that the axion is very massive, and they have the same tensor and vector perturbations \cite{Maleknejad:2012fw}. (See also appendix \ref{app-spin-2}.)

These models can be specified in terms of three dimensionless parameters, $\xp$, $\xi$ and $\xi_{Z_0}$, which are defined in the following.  First, the almost constant gauge field configuration of the form \eqref{BG-A} leads to a slowly-varying dimensionless parameter
\be
\xp \equiv  \frac{\ga \psi}{H}.
\ee
Validity of perturbation theory in the scalar sector of the $SU(2)$ gauge field requires \cite{Namba:2013kia,Dimastrogiovanni:2012ew,Adshead:2013nka}
\be\label{xp-condition}
\xp>\sqrt{2}.
\ee
A scalar mode in these models would have a negative frequency at $\frac{k}{a} \propto (2-\xp^2)$ which is unstable for $\xp<\sqrt{2}$. 
In refs.~\cite{Dimastrogiovanni:2016fuu, Agrawal:2018mrg, Agrawal:2017awz}, $\xp$ has been called $m_Q$.
The second dimensionless parameter is 
\be 
\xi\equiv \frac{\lambda\dot\varphi}{2f H}.
\ee
In the Higgsed version of the models ($\alpha_H=1$), we also have
\be
\xi_{Z_0} \equiv \frac{\ga Z_0}{H}.
\ee
Another important quantity in this setup is
\bea\label{epsilon-A}
\epsilon_A \equiv 2 ~ \frac{\bar{\rho}_{_{\rm YM}} }{\bar{\rho}}  = (1+\xp^2) \bigg(\frac{\psi}{\mpl}\bigg)^2,
\eea
which is the contribution of the gauge field to the total slow-roll parameter and equals twice the ratio of the energy density of the gauge field background $\bar\rho_{\rm YM}$ to the total energy density $\bar\rho$. In refs.~\cite{Dimastrogiovanni:2016fuu, Agrawal:2018mrg, Agrawal:2017awz}, $\epsilon_B= \epsilon_A\xp^2/(1+\xp^2)$ is used instead of $\epsilon_A$.

The background field equation of the gauge field is given by the $\mu=i$ component of the following equation \footnote{The explicit form of the background field equation of $\bar{A}_{\mu}$ is 
\be\label{eq--psi}
\delta^a_i\bigg(\frac{(a\psi\ddot{)}}{a}+\frac{H(a\psi\dot{)}}{a}+2\ga^2\psi^3+\alpha_H \ga^2Z_0^2\psi - 2\dot{\bar{\alpha}}_A \ga\psi^2\bigg)=0.
\ee
} 
\be\label{field-eq-A}
\bar{D_{\nu}} \bigg( \bar{F}^{\mu\nu} + \bar{\alpha}_{A} \epsilon^{\mu\nu\lambda\sigma} \bar{F}_{\lambda\sigma} \bigg) + \alpha_H \ga^2Z_0^2 \bar{A}^{\mu} = 0,
\ee
where a bar denotes a background quantity and $D_{\nu}$ is the covariant derivative 
$$D_{\mu} \equiv \nabla_{\mu} - i\ga A_{\mu} \quad (\bar{D}_{\mu}=\bar{\nabla}_{\mu} - i\ga \bar{A}_{\mu}),$$
and $\bar{\alpha}_A$ is a function of the background fields which, depending on the form of $\mathcal{L}_A$, is given as \footnote{In \eqref{alpha-bar}, we have $\overline{F\tilde{F}}=12\ga(H\psi+\dot{\psi})\psi^2$.}
\be\label{alpha-bar}
\bar{\alpha}_{A} = \begin{cases}
& -\frac{2\kappa}{96} \overline{F\tilde{F}} \quad \textmd{for} \quad \mathcal{L}_A=\mathcal{L}_{Gf},\\
&  ~~\frac{\lambda}{2f}\bar{\varphi} \qquad \textmd{for} \quad  \mathcal{L}_A=\mathcal{L}_{Cn}.
\end{cases}
\ee
Note that the zeroth component of \eqref{field-eq-A} is a constraint equation which is equivalent to zero for our background ansatz.

Assuming slow-roll dynamics in the background, equation \eqref{field-eq-A} relates $\xp$, $\xz$, and $\dot{\bar{\alpha}}_A/H$ as
\be\label{xi--sl}
\frac{\dot{\bar{\alpha}}_A}{H}\simeq \frac{(1+\xp^2+\frac{\alpha_{H}}{2} \xi^2_{Z_0})}{\xp}.
\ee

In the $\mathcal{L}_{Cn}$ models, the background field equation of the axion is 
\be
\ddot\varphi +3H\dot\varphi + V_{\varphi} +\frac{3\lambda \ga}{f}\psi^2(\dot\psi+H\psi)= 0.
\ee

\subsection{Tensor perturbations}

 The existence of a spin-2 degree of freedom in the gauge field is a unique feature of the $SU(2)$ inflation models. This is the primary focus of our work. Once we have a slow-roll background dynamics, the tensor perturbations in this family of models are entirely determined by the background quantities $\xi$, $\xp$, $\xz$, and by the perturbed gauge field sector of the model, $\mathcal{L}_A + \alpha_H \mathcal{L}_H$. The vector and tensor perturbations in $\mathcal{L}_{Gf}$ and $\mathcal{L}_{Cn}$ are the same. 
Let us first briefly review the spin-2 part of the perturbed $SU(2)$ gauge field. More details are presented in appendix \ref{app-spin-2}. See \cite{Maleknejad:2011jw, Maleknejad:2011sq} for the full decomposition of the field into the scalar, vector, and tensor perturbations. 

Once we perturb the metric and the $SU(2)$ gauge field around their homogeneous and isotropic solutions \eqref{BG-A}, we have the following spin-2 fluctuations 
\bea
\delta\!_{_{T}} g_{ij}(t,\vec{x})&=& a^2 \gamma_{ij}(t,\vec{x})  ,\\   \label{pert-A}  
\delta\!_{_{T}} A^a_{i}(t,\vec{x})&=&  \mpl ~ \delta^{aj} B_{ij}(t,\vec{x}),
\eea
where $\delta\!_{_{T}}$ denotes the spin-2 subsector of the perturbed field. In appendix \ref{spin-2-SU(2)}, we prove that $\tilde{\gamma}_{ij} = B_{ij}/a$ is a (pseudo) spin-2 field. Nonetheless, throughout this paper, we shall call $B_{ij}$ a spin-2 field. In Fourier space, the vacuum (free) $\gamma_{ij}$ and $B_{ij}$ can be expanded as
\be\label{h}
a\mpl \gamma_{ij}(t,\vec{x}) &=& \sqrt{2} \sum_{\sigma=\pm 2}\int d^3k  e^{i\vec{k}.\vec{x}}e_{ij}(\sigma,\hat{k})\left[\hat{a}_{\sigma}(\vec{k})h_{\sigma}(\vec{k})+\hat{a}^{\dag}_{\sigma}(-\vec{k})h_{\sigma}^{*}(-\vec{k})\right],\\
\label{htild}
\mpl B_{ij}(t,\vec{x}) &=& \frac{1}{\sqrt{2}} \sum_{\sigma=\pm 2}\int d^3k e^{i\vec{k}.\vec{x}}e_{ij}(\sigma,\hat{k})\left[\hat{b}_{\sigma}(\vec{k})B_{\sigma}(\vec{k})+\hat{b}^{\dag}_{\sigma}(-\vec{k})B_{\sigma}^{*}(-\vec{k})\right],
\ee
where $h_{\sigma}$ and $B_{\sigma}$ are the canonically normalized fields, $e_{ij}(\pm,\hat{k})$ are the polarization tensors associated with the $\pm 2$ helicity states, \footnote{The polarization tensor of the spin-2 field in the direction $\hat{k}=-\hat{r}$ is given as
\be\label{polariz-tensor}
e_{ij}(\pm 2,\hat{k})=\sqrt{2} e_{i}(\pm 1,\hat{k})e_{j}(\pm 1,\hat{k}) \quad \textmd{where} \quad \vec{e}(\pm 1,-\hat{r})=\frac{1}{\sqrt{2}}(\hat{\theta}\mp i\hat{\phi}),
\ee
where $\hat{r}$, $\hat{\theta}$ and $\hat{\phi}$ are the local orthogonal unit vectors in the directions of increasing $r$, $\theta$, and $\phi$. Note that $e_{ij}(\sigma,\vec{k})=e^{*}_{ij}(\sigma,-\vec{k})$, and $\vec{k}\times \vec{e}(\pm 1,\vec{k})=\mp i k \vec{e}(\sigma,\vec{k})$.}
which are normalized as $e_{ij}(\sigma,\vec{k})e_{ij}^{~*}(\sigma',\vec{k})=2\delta_{\sigma \sigma'}$, and $\hat{a}_{\sigma}(\vec{k})$ and $\hat{b}_{\sigma}(\vec{k})$ are the annihilation operators of the spin-2 modes of the metric and gauge field, respectively, satisfying
$$[\hat{a}_{\sigma}(\vec{k}),\hat{a}^{\dag}_{\sigma'}(\vec{k}')] = [\hat{b}_{\sigma}(\vec{k}),\hat{b}^{\dag}_{\sigma'}(\vec{k}')] = \delta^3(\vec{k}-\vec{k}') \delta_{\sigma \sigma'}.$$

\begin{table}
\begin{center}
\begin{tabular}{ | m{0.5cm} | m{1.5cm} | m{1.6cm}| m{2.1cm} |  m{0.4cm} |  m{1.3cm} |   m{1.9cm} |  m{2.5cm} |} 
\cline{2-8}
\multicolumn{1}{c|}{} & $\mathcal{L}_A$ & references  & $\frac12\frac{m^2}{H^2} (\frac{\psi}{\mpl})^{-2}$  & $\beta_{\rm{c}}$  & $\theta_{\rm{c}}$  & $\delta_{\rm{c}}$  & $\frac{\tilde{m}^2}{H^2}$ \\ 
\hline
\multirow{3}{*}{\begin{sideways} $\alpha_H=0$~ \end{sideways}} &
$\mathcal{L}_{Gf}$  & \cite{Maleknejad:2011jw,Maleknejad:2011sq} & $(\xp^2-1)$ & $\xp$ & $\xp^2$ & $\frac{2(1+2\xp^2)}{\xp}$ & $2(2+\xp^2)$ \\
& & & & & & &\\
& $\mathcal{L}_{Cn}$ & \cite{Adshead:2012kp, Adshead:2013nka, Maleknejad:2016qjz,Caldwell:2017chz,Dimastrogiovanni:2016fuu} &  $(\xp^2-1)$ & $\xp$ & $\xp^2$ & $2(\xp+\xi)$ & $2(1+\xi\xp)$ \\ 
\hline
\multirow{2}{*}{\begin{sideways} $\alpha_H=1$~ \end{sideways}} &
$\mathcal{L}_{Gf}$ & \cite{Nieto:2016gnp,Adshead:2017hnc} & $(\xp^2-1+\xi_{Z_0}^2)$ & $\xp$ & $\xp^2+\xi_{Z_0}^2$ & $\frac{2(1+2\xp^2)+\xi_{Z_0}^2}{\xp}$ & $2(2+\xp^2+\xi_{Z_0}^2)$\\
& & & & & & & \\
& $\mathcal{L}_{Cn}$  & \cite{Adshead:2016omu} & $(\xp^2-1+\xi_{Z_0}^2)$ & $\xp$ & $\xp^2+\xi_{Z_0}^2$ & $2(\xp+\xi)$ & $2(1+\xi\xp)+\xi_{Z_0}^2$ \\
\hline \\ \hline
\multicolumn{3}{|c|}{ Unified} & $(\xp^2-1+\alpha_{H} \xz^2)$ & $\xp$ & $\xp^2+\alpha_H \xz^2$ & $2(\xp+\frac{\dot\alpha_A}{H})$ & $2(1+\xp \frac{\dot\alpha_A}{H}) +\alpha_H \xz^2$ \\
\hline
\end{tabular}
\end{center}
\caption{Definition of parameters in the equations of tensor perturbations in \eqref{hs-eq}-\eqref{eq-h} in terms of $\xp$, $\xi$ and $\xi_{Z_0}$ for each type of models. The last row shows the parameters in the unified form for the generic action given in \eqref{g-action}.}\label{T}
\end{table}

The tensor perturbations obey the following equations of motion 
\bea\label{hs-eq}
&&\hs'' + \left[k^2 - \frac{a''}{a}  + \frac{m^2}{H^2}\mH^2\right] \hs = \frac{2\psi}{\mpl} \mH \left[ ( -\lambda_{\sigma} \beta_{\rm c} k+ \theta_{\rm{c}} \mH)  \Ths - \Ths'\right],\\
\label{eq-h} 
&&\Ths'' + \left[k^2 - \lambda_{\sigma} \delta_{\rm c} k\mH - \frac{a''}{a} + \frac{\tilde{m}^2}{H^2}\mH^2\right]\Ths = \mathcal{O}(\frac{\psi}{\mpl}h_{\sigma}),
\eea
where $\lambda_{\pm} = \pm 1.$ See \cite{Adshead:2013nka,  Maleknejad:2011sq, Adshead:2016omu} for the expression on the right hand side of \eqref{eq-h} which we ignore here. \footnote{The neglected term in RHS of \eqref{eq-h} is proportional to $\frac{\psi}{\mpl}\ll 1$ and therefore is subleading inside the horizon. However, after the horizon crossing when the homogeneous solution of $B_{\sigma}$ decays due to its mass, this term acts like a small source term for $B_{\sigma}$. See for instance \cite{Adshead:2013qp}. However, this effect makes a negligible correction to the sourced gravitational waves and the backreaction.} 
The primes denote a derivative with respect to conformal time, $\tau$, while $\frac{m^2}{H^2}$, $\frac{\tilde{m}^2}{H^2}$, $\beta_{\rm c}$, $\theta_{\rm c}$ and $\delta_{\rm c}$ are dimensionless slowly varying parameters defined in table \ref{T} for each model. The field equation \eqref{eq-h} can be written as a Whittaker equation as
\be\label{Whittaker}
\p_z^2 B_{\sigma} + \bigg[ -\frac14 + \frac{\kl}{z} + \frac{1}{z^2}\bigg(\frac14-\mu^2\bigg) \bigg] B_{\sigma} = 0,
\ee
where $z=2ik\tau$ and we used the slow-roll relation $aH \simeq -\frac{1}{\tau}$. The parameters $\kappa_{\sigma}$ and $\mu$ are given as
\be\label{kl-mu}
\kl= - \frac{i\lambda_{\sigma}\delta_{\rm c}}{2}   \an    \mu^2= \frac94 - \frac{\tilde{m}^2}{H^2}.
\ee
Since $\lvert \kappa_{+} \rvert = \lvert \kappa_{-} \rvert$, we write
$$\lvert \kappa \rvert \equiv \lvert \kappa_{\sigma} \rvert = \frac12 \delta_c.$$
General solutions are given by linear combinations of the Whittaker functions $W_{\kappa_{\lambda},\mu}(z)$, $M_{\kappa_{\lambda},\mu}(z)$.  Imposing the Bunch-Davies vacuum condition in the asymptotic past, we have   (see \eqref{WM-asymp})
\Beq
\label{eq:usExactSolution}
B_{\sigma}(\tau,\vec{k}) =\frac{e^{i\kappa_{\sigma}\pi/2}}{(2\pi)^{\frac32}\sqrt{2k}}W_{\kappa_{\sigma},\mu}(2i k\tau)\,.
\Eeq
Using the above in the field equation of $h_{\sigma}$ \eqref{hs-eq}, we find the sourced part of the gravitational waves. 

 Here, we summarize the main features of the spin-2 field with the field equation of \eqref{eq-h} and the quadratic action of \eqref{Action-quad}.

\begin{itemize}
 \item{$B_{\pm}$ evolves as a massive field in de Sitter space with a parity breaking linear derivative interaction term, $\mp \delta_c k\mH B_{\pm}$, with $\delta_c$ given by table \ref{T}. In terms of  $\bar{\alpha}_A$ \eqref{alpha-bar}, we can write it in a unified form, $\delta_c = 2(\xp + \frac{\dot{\bar{\alpha}}_A}{H})$.}
 \item{The first term in $\delta_c$ ($2\xp$) comes from the interaction of $B_{\sigma}$ with the VEV of the gauge field through the covariant derivative $D_{\mu}$, and is due to the self-interactions of gauge field in Yang-Mills theory.}
 \item{The second contribution in $\delta_c$ ($2\frac{\dot{\bar{\alpha}}_A}{H}$) is a time derivative of $\bar{\alpha}_A$. As shown in \eqref{alpha-bar}, for $\mathcal{L}_{A}=\mathcal{L}_{Gf}$, this parameter is due to the VEV of $F\tilde F$ while for $\mathcal{L}_{A}=\mathcal{L}_{Cn}$, it is due to the derivative interaction with the VEV of the axion field.}
\item{The sound speeds of $B_{\pm}$ field and GWs are unity in all of the models in this family. \footnote{This is also valid in the presence of dimension six operators, $\textmd{tr}(FFF)$ and the (PT violating) Weinberg operator $\textmd{tr}(FF\tilde{F})$ \cite{Maleknejad:2014wsa}. }} 
\item{Due to the self-interactions of the gauge field, the $B_{\sigma}$ is massive with a mass term $\frac{\tilde{m}^2}{H^2}$, given in table \ref{T}. The mass of the spin-2 field can be written in the unified form $\frac{\tilde{m}^2}{H^2} = 2(1+\frac{\dot{\bar{\alpha}}_A}{H}\xp)+ \alpha_{H}\xz^2$.}
\item{In a similar Abelian field case (see \eqref{ActionU1}), the first derivative interaction as well as the mass term are missing. Thus, the non-Abelian nature of the gauge field makes i) a more efficient particle production, while making ii) the transverse field massive and therefore decaying after horizon crossing.}
 \item{As we will see in section \ref{P-P}, these derivative interactions are responsible for production of the spin-2 particle by the background fields.}
 \item{In $SU(2)$ gauge field setups, the right hand side of the field equation of the gravitational waves in \eqref{hs-eq} is non-zero and is given by an anisotropic inertia proportional to $\frac{\psi}{\mpl}$. Therefore, the efficiency of the mixing between the spin-2 field and the gravitational waves is specified by the VEV of the $SU(2)$ gauge field.}
\item{This anisotropic inertia is parametrized in terms of $\theta_c$, $\beta_c$, and a small mass term for the graviton, $\frac{m^2}{H^2}$, given in table \ref{T}.}
 \item{ The term $\beta_c$ is the coefficient of a linear derivative interaction which is equal to $\xp$ regardless of the model. The other parameter can be written as $\theta_c= \xp^2+\alpha_H \xz^2$. The mass term is $\frac{m^2}{H^2}=2(\psi/\mpl)^2(\xp^2-1+\alpha_H \xz^2)$.}
 \item{All the interaction and parameters in the tensor perturbation sector are specified only by $\mathcal{L}_A$ and $\mathcal{L}_H$, and therefore independent of whether the gauge field sector is a spectator or not. }
 \end{itemize}
 
In this work, we assume (quasi) de Sitter expansion and keep terms up to first order in slow-roll. The slow-roll time evolution of $\xp$ and $\xz$, which is model dependent, contributes to the spectral tilt of the sourced gravitational waves. Depending on the details of the evolution of the gauge field VEV, the spectral tilt of the sourced gravitational waves can be positive or negative \cite{Fujita:2018ndp}. Since we are interested in the number density and the backreaction of the $B_{\sigma}$ particle as well as in the amplitude of the sourced gravitational waves which are model independent, we neglect this effect in this paper.

\section{Spin-2 Schwinger-type particle production}\label{P-P}
In this section, we study the spin-2 particle production due to their interactions with the VEV of the background fields. The background fields act as a classical source for the quantum fluctuations analogous to the well-known Schwinger effect \cite{Schwinger:1951nm}. However, unlike the standard Schwinger process in which the quantum field is sourced only by a background gauge field, here the spin-2 quantum field is sourced by both the backgrounds of axion and gauge fields. The derivation given in this section follows closely section 3.2 of \cite{Lozanov:2018kpk}.

\begin{figure}[h!]
\begin{center}
\includegraphics[height=4.7cm,width=0.48\textwidth]{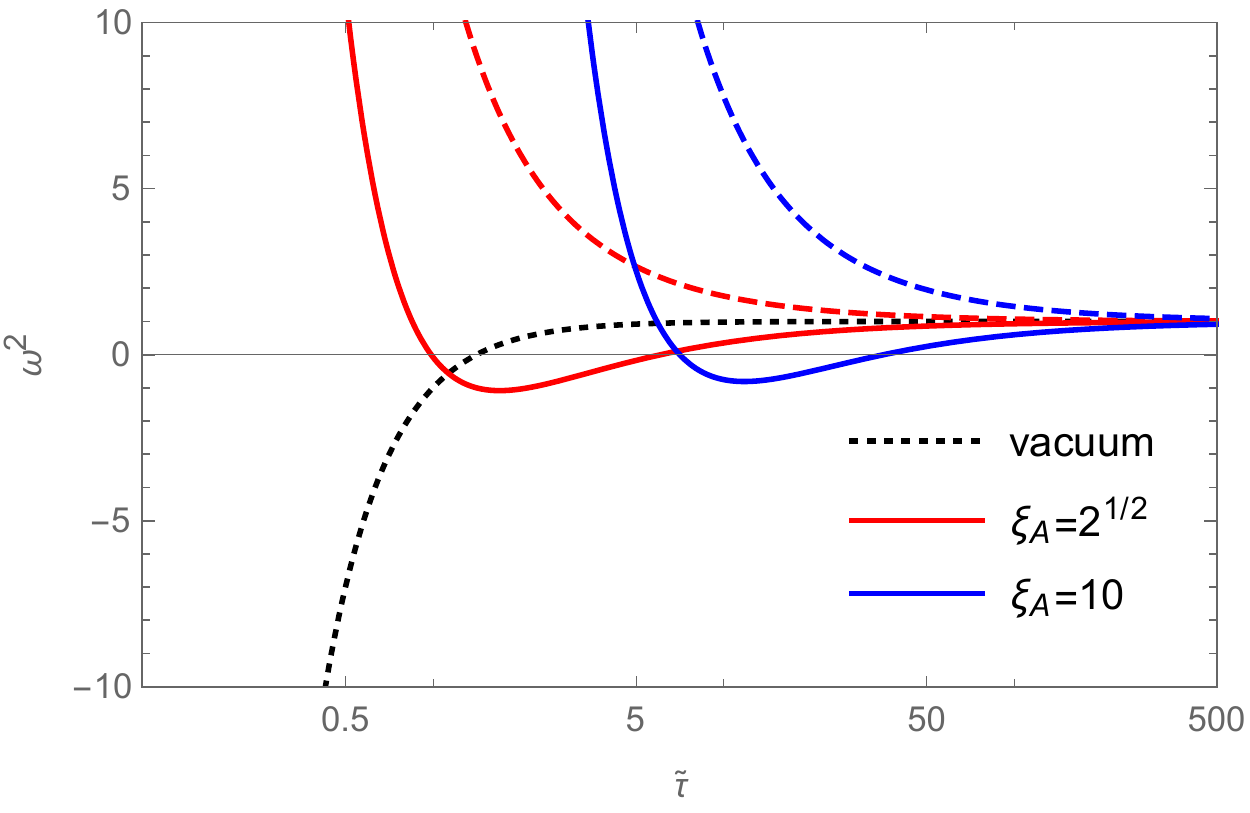} \includegraphics[width=0.48\textwidth]{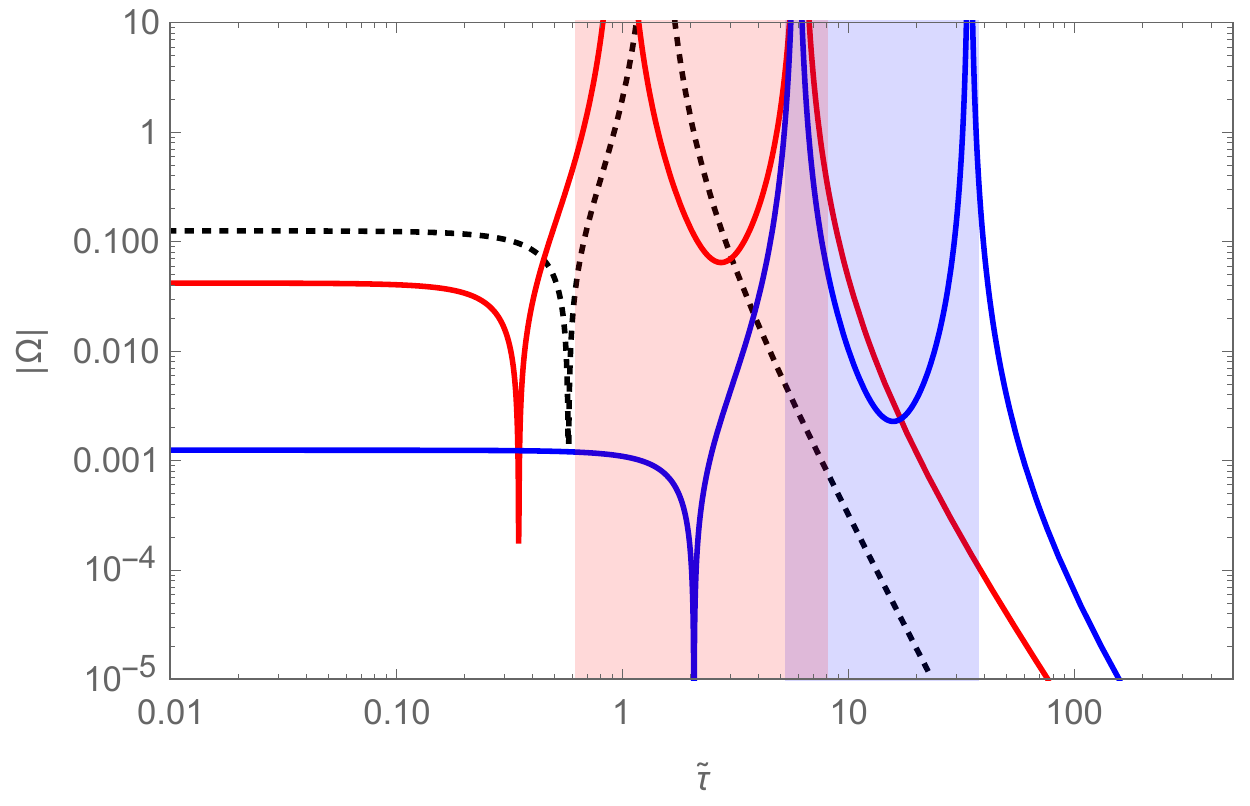}\\
\caption{Effective frequency squared ($\omega^2$) and deviation from adiabaticity ($\lvert \Omega \rvert$). The left panel shows $\omega^2$ \eqref{omega} as a function of $\x$ while the right panel shows $\lvert \Omega_{\pm}(\x) \rvert$ \eqref{Omega} for massless systems ($\alpha_H=0$) with $\xp=\sqrt{2}$ (red line) and $10$ (blue line). For comparison, the dotted black line shows the vacuum gravitational waves. Here, the solid red and blue lines show the plus polarization modes while the dashed blue and red lines show the minus polarization. The shaded areas in the right panel show the particle production regime.}\label{adiab}
\end{center}
\end{figure}

\begin{figure}[h!]
\begin{center}
\includegraphics[height=4.5cm,width=0.51\textwidth]{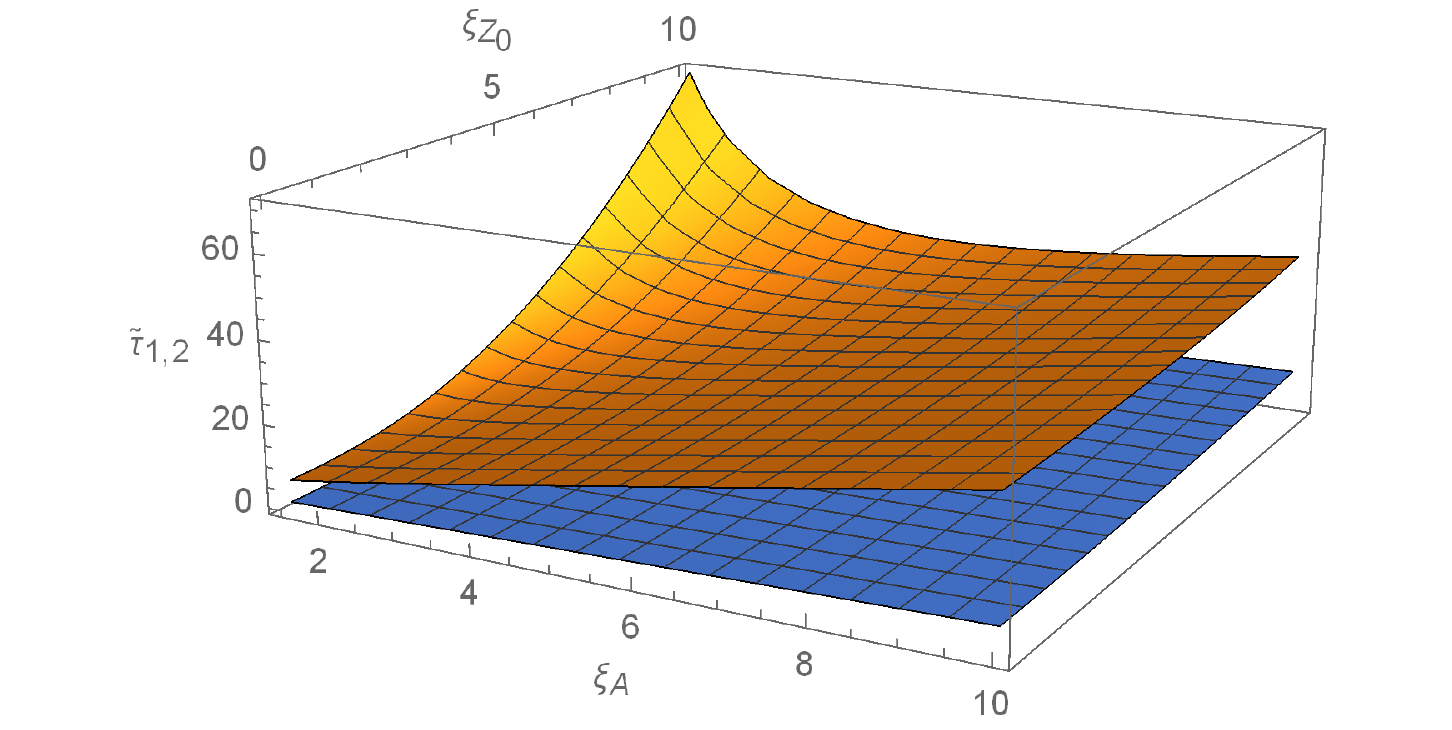} \includegraphics[width=0.48\textwidth]{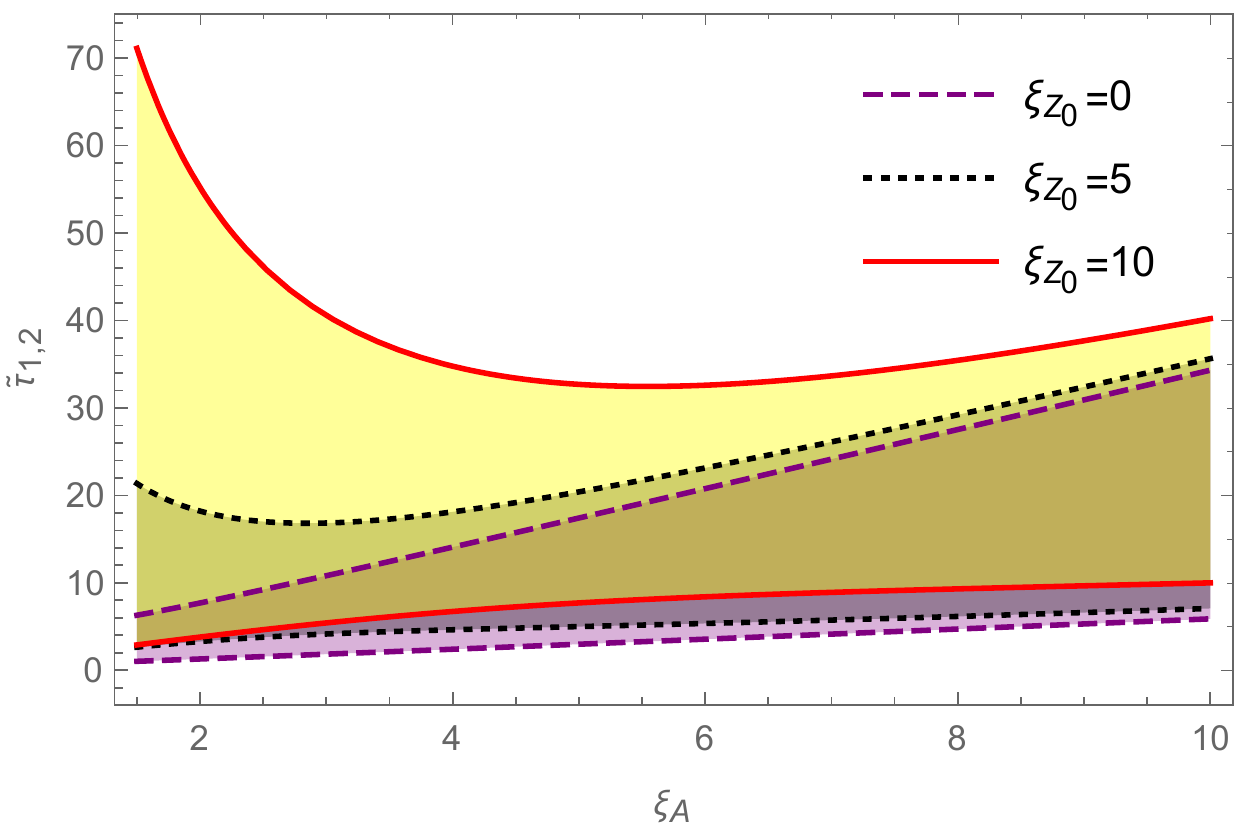}
\caption{(left) $\x_{1,2}$ interval for which $B_{\sigma}$ has instability as a function of $\xp$ and $\xi_{Z_0}$. (Right) three slices of the left panel. We show $\x_{1,2}$ as a function of $\xp$ for $\xi_{Z_0}=0$ in (dashed) purple, $\xi_{Z_0}=5$ in (dotted) black, and $\xi_{Z_0}=10$ in (solid) red.}\label{roots}
\end{center}
\end{figure}

To have a better qualitative understanding of the particle production process, let us write the field equation of Fourier modes using the (normalized) physical momentum
$$\x=\frac{k}{aH},$$
as
\be
\p_{\x}^2 B_{\sigma}(\tau,\vk) + \omega^2_{\sigma}(\x) B_{\sigma}(\tau,\vk) = 0,
\ee
where $\omega_{\sigma}(\x)$ is the (time-dependent) effective frequency of the modes 
\be\label{omega}
\omega^2_{\pm}(\x) =  1 \mp \frac{\delta_{\rm c}}{\x} + (-2+\frac{\tilde{m}^2}{H^2})\frac{1}{\x^2}.
\ee
In the limits that the effective frequency is slowly varying and 
\be
\label{eq:AdiabCond}
\left(\frac{\p_{\x}\omega_{\sigma}(\x)}{\omega^{2}_{\sigma}(\x)}\right)^2\ll1\, \an \left|\frac{\p_{\x}^2\omega_{\sigma}(\x)}{\omega^{3}_{\sigma}(\x)}\right|\ll1\,,
\ee
the particle production is zero and the modes are in an adiabatic vacuum state. Then the solution can be well-approximated by the WKB form,
\be\label{WKB-app}
B^{\rm WKB}_{\sigma1,2}(\tau,\vk) =  \frac{1}{(2\pi)^{\frac32}\sqrt{2k\omega_{\sigma}(\x)}} \exp\bigg( \pm i \int \omega_{\sigma}(\x) d\x\bigg),
\ee
in which $B^{\rm WKB}_{\sigma1}$ and $B^{\rm WKB}_{\sigma2}$ are the positive and negative frequency modes respectively.
The WKB approximation is the exact solution of 
\be
\p_{\x}^2 B_{\sigma} + \omega^2_{\sigma}(\x) \bigg( 1 - \Omega_{\sigma}(\x)\bigg) B_{\sigma} = 0,
\ee
where $\Omega(\x)$ is defined as
\be\label{Omega}
\Omega_{\sigma}(\x) \equiv  \frac34\left(\frac{\p_{\x}\omega_{\sigma}(\x)}{\omega^2_{\sigma}(\x)}\right)^2 - \frac12  \frac{\p_{\x}^2\omega_{\sigma}(\x)}{\omega^3_{\sigma}(\x)},
\ee
which quantifies the deviation of our mode function from the exact adiabatic solution (see figure \ref{adiab}). We find that $\Omega_{-}$ is always very small and thus $B_{-}(\tau,\vk)$ remains adiabatic. However, $\Omega_{+}$ becomes large around the roots of $\omega_{+}$ 
\be\label{q12}
\x_{1,2} = \frac12 \left(\delta_{\rm c} \pm \sqrt{8+\delta^2_{\rm c}-4\frac{\tilde{m}^2}{H^2}} \right) \simeq \big( \lvert \kappa \rvert \pm \sqrt{\lvert \kappa \rvert^2 - \lvert \mu \rvert^2}\big),
\ee
 and the system experiences a large deviation from adiabaticity. 
The above roots are presented in figure \ref{roots}.

When the adiabatic conditions hold, i.e., $\lvert \Omega \rvert\ll1$, we have a well-defined adiabatic vacuum, and the field excitation about it describes particles. Deviations from adiabaticity in the asymptotic past and future are
\Beq
\label{eq:OmegaPMdS}
\Omega_{\pm}(\x) \rightarrow \begin{cases}
    \frac{1}{2\x^3}\bigg( \mp \delta_{\rm c} +\frac{3}{\x}(\frac78 \delta^2_{\rm c} +2 -\frac{\tilde{m}^2}{H^2})\bigg) \simeq 0 ,& \text{if } \x\rightarrow \infty\\\
    \frac{1}{8(1-\frac{\tilde{m}^2}{H^2})},              & \text{if } \x\rightarrow0\,.
\end{cases}
\Eeq
In the asymptotic past, $\x\rightarrow \infty$, the adiabaticity conditions are satisfied and the WKB solution \eqref{WKB-app} is the Bunch-Davies vacuum \eqref{eq:usExactSolution}. At later times the adiabaticity conditions are violated and particles are produced. In the asymptotic future, recalling that $\xp> \sqrt{2}$ and using Table \ref{T}, we have $\lim_{\x\rightarrow 0} \lvert\Omega_{+}(\x)\rvert\lesssim 10^{-2}$. Thus, the positive frequency modes (vacuum mode functions) in the asymptotic future, $v_{\sigma}(\tau,\vk)$, are given by the WKB solution in \eqref{WKB-app}, as
\be
v_{\sigma}(\tau,\vk) = \lim_{-k\tau\rightarrow 0} B^{\rm WKB}_{\sigma1}(\tau,\vk)  \simeq \frac{\sqrt{\x}}{(2\pi)^{\frac32}\sqrt{2k\lvert \mu \rvert}} e^{i \lvert \mu \rvert \ln\x},
\ee
in which we used $\mu=i\lvert \mu\rvert$ and $\lim_{\x\rightarrow 0} \omega(\x) \simeq \lvert \mu \rvert/\x $.
Using \eqref{WM-asymp}, the asymptotic future vacuum mode functions can be well approximated by the $M$-Whittaker functions as 
\Beq
\label{eq:AsympFutVacModeFunc}
v_{\sigma}(\tau,\vk)=\frac{e^{i\mu\pi/2}}{2(2\pi)^{\frac32}\sqrt{k|\mu|}}M_{\kl,\mu}(-2i\x)\,.
\Eeq
Using \eqref{MvsW}, the asymptotic past and future vacuum modes are related as 
\be\label{f-p}
v_{\sigma}(\tau,\vk) = e^{i(\kappa_{\sigma}-\mu)\frac{\pi}{2}}\frac{\sqrt{2|\mu|}\Gamma(2\mu)}{\Gamma(\frac12+\mu+\kappa_{\sigma})}  B_{\sigma}(\tau,\vk) + ie^{i(\kappa_{\sigma}+\mu)\frac{\pi}{2}}\frac{\sqrt{2|\mu|}\Gamma(2\mu)}{\Gamma(\frac12+\mu-\kappa_{\sigma})} B^{*}_{\sigma}(\tau,-\vk).
\ee
Therefore, the spin-2 field can be either expanded in terms of the positive frequency modes in the asymptotic past \eqref{eq:usExactSolution} as in \eqref{htild}, or in terms of the positive frequency modes in the asymptotic future \eqref{eq:AsympFutVacModeFunc}, as
\be\label{u-1}
\mpl B_{ij}(\tau,\vec{x}) &=& \frac{1}{\sqrt{2}}\sum_{\sigma=\pm 2}\int d^3k e^{i\vec{k}.\vec{x}}e_{ij}(\sigma,\hat{k})\left[\tilde{b}_{\sigma}(\vec{k})v_{\sigma}(\vec{k},\tau)+\tilde{b}^{\dag}_{\sigma}(-\vec{k})v_{\sigma}^{*}(-\vec{k},\tau)\right],
\ee
where $\tilde{b}_{\sigma}(\vk)$ and $\tilde{b}^{\dag}_{\sigma}(\vk)$ are the annihilation and creation operations of a particle with respect to the asymptotic future vacuum respectively.
By definition, we have
\Beq
b_{\sigma}(\vk) \lvert 0_{\textmd{in}}\rangle =0 &\an& \tilde{b}_{\sigma}(\vk)\lvert 0_{\textmd{out}}\rangle=0,
\Eeq
where $\lvert 0_{\textmd{in}}\rangle$ and $\lvert 0_{\textmd{out}}\rangle$ are the vacuum states in the asymptotic past and future of the (quasi) de Sitter spacetime, respectively.

\begin{figure}[h!]
\begin{center}
\includegraphics[width=0.7\textwidth]{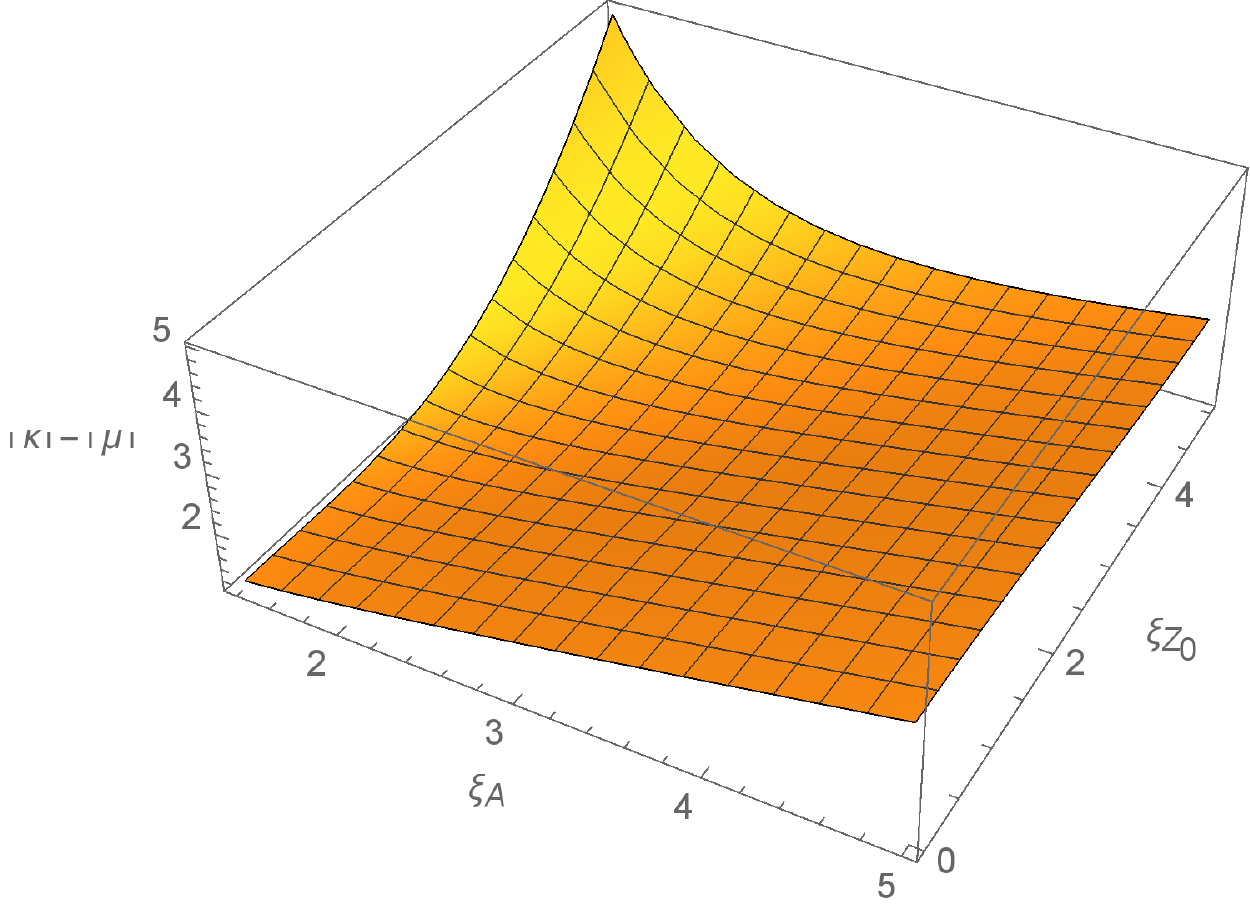} 
\caption{Efficiency of the particle production, $\lvert \kappa \rvert - \lvert \mu \rvert$, as a function of $\xp$ and $\xz$. }\label{kappa-mu}
\end{center}
\end{figure}

Using Bogoliubov transformation, we can write $\tilde{b}_{\sigma}(\vk)$ in terms of $b_{\sigma}(\vk)$ and $b^{\dag}_{\sigma}(\vk)$ as 
\Beq
\tilde{b}_{\sigma}(\vk)&=& \alpha_{\sigma,\vk} b_{\sigma}(\vk) + \beta^{*}_{\sigma,\vk} b^{\dag}_{\sigma}(-\vk),\Eeq
where $\alpha_{\sigma,\vk}$ and $\beta_{\sigma,\vk}$ are Bogoliubov coefficients which satisfy the normalization condition
\Beq
|\alpha_{\sigma,\vk}|^2-|\beta_{\sigma,\vk}|^2=1\,.
\Eeq
From the combination of Eqs. \eqref{htild}, \eqref{f-p} and \eqref{u-1}, we find
\Beq\label{alpha-beta}
\alpha_{\sigma,\vk} &=& \sqrt{2\lvert \mu\rvert } e^{(\lambda_{\sigma} \lvert\kappa\rvert+\lvert \mu\rvert)\pi/2} \frac{\Gamma(-2\mu)}{\Gamma(\frac12-\mu-\kl)}  , \\
\beta_{\sigma,\vk} &=& -i \sqrt{2\lvert \mu\rvert } e^{(\lambda_{\sigma} \lvert\kappa\rvert-\lvert \mu\rvert)\pi/2} \frac{\Gamma(2\mu)}{\Gamma(\frac12+\mu-\kl)}  .
\Eeq
Having the $\beta_{\sigma,\vk}$ coefficients, we are ready to determine the particle number density as well as the vacuum-vacuum transition amplitude. The efficiency of the particle production for $\sigma=+2$ is given by the exponent $\lvert \kappa \rvert - \lvert \mu \rvert$, which, recalling \eqref{xp-condition} and using \eqref{kl-mu}, can be approximated as 
\be\label{kappa-mu--}
\lvert \kappa \rvert - \lvert \mu \rvert \approx
\frac{\xp}{2} \bigg[ 1 + \left(\sqrt{1+\alpha_{H}(\frac{\xz}{\xp})^2} -\sqrt{2}\right)^2 \bigg] > \frac{\xp}{2}.
\ee 
This quantity is presented in figure \ref{kappa-mu} as a function of $\xp$ and $\xz$ and has the following asymptotic forms
\be\label{Kappa-Mu-}
\lvert \kappa \rvert - \lvert \mu \rvert \approx \begin{cases}
 ~ (2-\sqrt{2}) \xp  &  \text{if } \frac{\xz}{\xp}\ll 1 \\\
  \frac12 \alpha_{H} \frac{\xz^2}{\xp}   & \text{if } \frac{\xz}{\xp}\gg 1\,.
\end{cases}
\ee

The number density of the created particles with a given comoving momentum, $\vk$, in the asymptotic future is
\Beq\label{nd}
n_{\sigma}(\vk) = \langle 0_{\textmd{in}}\rvert \tilde{b}^{\dag}_{\sigma,\vk}\tilde{b}_{\sigma,\vk} \lvert 0_{\textmd{in}}\rangle = \lvert \beta_{\sigma} \rvert^2= \frac{e^{2\lambda_{\sigma} \lvert \kappa\rvert \pi}+e^{-2\lvert\mu\rvert\pi}}{2\sinh(2\lvert\mu\rvert\pi)},
\Eeq
which has a $k$-independent spectrum for each polarization state. As we see, there is a large pair production in the plus polarization while it is almost zero for the minus state 
\be
n_{+}(\vk) \gtrsim e^{\xp \pi} \an n_{-}(\vk) \lesssim e^{-6\xp \pi}.
\ee

\begin{figure}[h!]
\begin{center}
\includegraphics[width=0.7\textwidth]{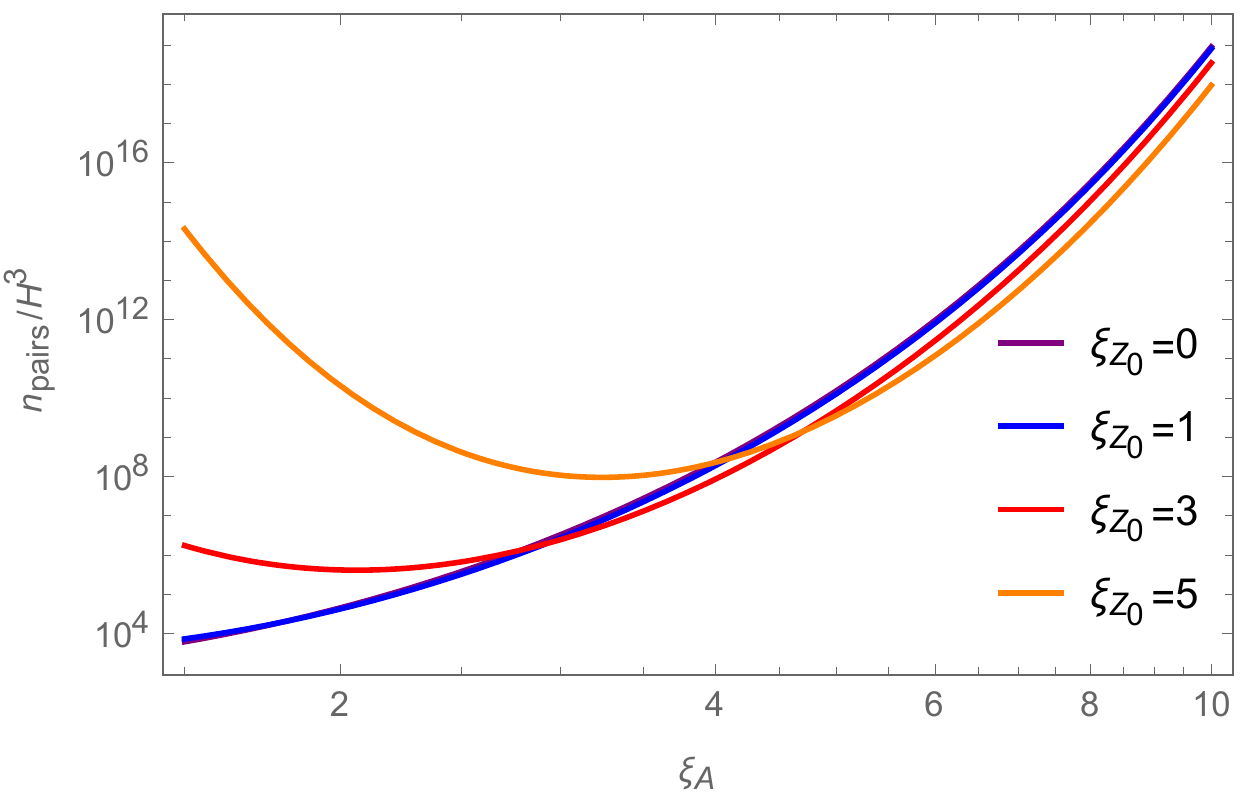}
\caption{The number density of the created pairs, $n_{\rm pairs}/H^3$, as a function of $\xp$ and $\xz$. }\label{n-B}
\end{center}
\end{figure}

The total particle creation from the asymptotic past to the asymptotic future, therefore, is
\be
N_{\sigma} = \frac{1}{(2\pi)^3}  \lvert \beta_{\sigma} \rvert^2 \int d^3k = \frac{1}{(2\pi)^2}  \frac{e^{2\lambda_{\sigma} \lvert\kappa\rvert \pi}+e^{-2\lvert\mu\rvert\pi}}{\sinh(2\lvert\mu\rvert\pi)} \int^{\infty}_{0} k^2dk,
\ee 
which is divergent since it expresses the number of pairs created for all times. The physically meaningful quantity, however, is the pair production rate, i.e., the number of pairs produced per unit time per unit physical volume
\Beq
\label{eq:DecayDef}
\Gamma^{\sigma}_{\rm{pairs}}=\frac{1}{a(\tau)^{4}}\frac{dN_{\sigma}}{d\tau}\,.
\Eeq
To calculate the derivative we need to convert the wavenumber integral into a time integral. It has been shown in the right panel of figure \ref{adiab} that the system has two sharp deviations from adiabaticity around the roots of $\omega_{+}$. Therefore, pairs of particles with $+2$ helicity state of a given comoving momentum, $k$, are produced mostly around \eqref{q12}
\be\label{tau-k}
\tau_{1,2}(k)\simeq -(\lvert \kappa \rvert \pm \sqrt{\lvert \kappa \rvert^2 - \lvert \mu \rvert^2})/k \,.
\ee
Note that $\x=k/(aH)\simeq -k\tau$.
As a result, the total particle creation at $\tau_1(k)$ and $\tau_2(k)$ are given respectively by
\be
N_{1}\approx (\lvert \kappa \rvert + \sqrt{\lvert \kappa \rvert^2 - \lvert \mu \rvert^2})^{3} \frac{e^{2 \lvert\kappa\rvert \pi}+e^{-2|\mu|\pi}}{(2\pi)^2\sinh(2|\mu|\pi)}\int_{-\infty}^0 d\tau \,(a(\tau)H)^4\,,\\
N_{2}\approx (\lvert \kappa \rvert - \sqrt{\lvert \kappa \rvert^2 - \lvert \mu \rvert^2})^3\frac{e^{ 2 \lvert\kappa\rvert \pi}+e^{-2|\mu|\pi}}{(2\pi)^2\sinh(2|\mu|\pi)}\int_{-\infty}^0 d\tau \,(a(\tau)H)^4\,.
\ee
The corresponding production rates are
\be
\Gamma^{1}_{\rm{pairs}} &\approx& (\lvert \kappa \rvert + \sqrt{\lvert \kappa \rvert^2 - \lvert \mu \rvert^2})^3 \frac{e^{ 2 \lvert\kappa\rvert\pi}+e^{-2|\mu|\pi}}{(2\pi)^2\sinh(2|\mu|\pi)}H^4\,,\\
\Gamma^{2}_{\rm{pairs}} &\approx& (\lvert \kappa \rvert - \sqrt{\lvert \kappa \rvert^2 - \lvert \mu \rvert^2})^3 \frac{e^{ 2 \lvert\kappa\rvert\pi}+e^{-2|\mu|\pi}}{(2\pi)^2\sinh(2|\mu|\pi)}H^4.
\ee
We find that the particle production during $\tau_1$ is much more efficient than during $\tau_2$, i.e. 
\be
\frac{\Gamma_{\rm{pairs}}^{2}}{\Gamma_{\rm{pairs}}^{1}} \simeq 10^{-2}. 
\ee
Therefore, we can neglect the burst of particles created at $\tau_2$.
After integrating Eq. \eqref{eq:DecayDef}, we find that the physical number densities of pairs created up to time $\tau$ are also time independent
\Beq
\label{eq:nPairs}
n_{\rm{pairs}}=\frac{1}{a(\tau)^3}\int_{-\infty}^{\tau'} d\tau\, a(\eta)^4 \Gamma_{\rm{pairs}}^{1}\approx\frac{\Gamma_{\rm{pairs}}^{1}}{3H}\,,
\Eeq
i.e., gravitational and Schwinger-type particle production are exactly balanced by the the gravitational redshifting. 
We can approximate the above as 
\Beq
\label{eq:nPairs-III}
n_{\rm{pairs}}\approx (\lvert \kappa \rvert + \sqrt{\lvert \kappa \rvert^2 - \lvert \mu \rvert^2})^3\frac{H^3}{6\pi^2}  e^{2(|\kappa|-|\mu|) \pi}\,.
\Eeq
The particle production increases exponentially with $(\lvert \kappa \rvert - \lvert \mu \rvert)$ and has the following asymptotic forms
\be\label{n-pairs-cases}
\frac{n_{\rm{pairs}}}{H^3} \rightarrow \begin{cases}
 ~  \frac{(2+\sqrt{2})^3}{6\pi^2} \xp^3 e^{2(2-\sqrt{2}) \xp\pi}  &  \text{if } \frac{\xz}{\xp}\ll 1 \\\
  \frac{1}{6\pi^2}\big(\frac{\xi_{Z_0}}{\xp}\big)^3 e^{\frac{\xz^2}{\xp}\pi}  & \text{if } \frac{\xz}{\xp}\gg 1 ~~(\alpha_{H}=1)\,,
\end{cases}
\ee
which may cause a large backreaction on the background VEV fields. This is the next subject of our study, given in section \ref{backreact}. The number density of the created particles is presented in figure \ref{n-B}.

Another interesting quantity to compute is the vacuum-vacuum transition amplitude defined by
\be
\lvert \langle 0_{\textmd{out}}\vert 0_{\textmd{in}} \rangle \rvert ^2 \equiv e^{(-\iint d^3xd\tau a^4 \Upsilon_{\rm vac})} = \exp \bigg[ - \frac{1}{(2\pi)^3}\int d^3x  \int dk^3 \ln(1+ \lvert \beta_{+,\vk}\rvert^2) \bigg],
\ee 
where $\Upsilon_{\rm vac}$ is the vacuum decay rate. Using Eq. \eqref{nd} and \eqref{tau-k}, we obtain
\be
 \Upsilon_{\rm vac} =  - 2  \frac{H^4}{(2\pi)^2} (\lvert \kappa \rvert + \sqrt{\lvert \kappa \rvert^2 - \lvert \mu \rvert^2})^3 \ln \bigg[ 1+  \bigg( \frac{e^{2\lvert \kappa \rvert\pi}+e^{-2\lvert \mu \rvert\pi} }{2\sinh(2\lvert \mu \rvert \pi)} \bigg)\bigg]\,,
\ee
which is well approximated by
\be\label{vacuum-decay}
 \Upsilon_{\rm vac} \approx  -  \frac{H^4}{\pi} (\lvert \kappa \rvert + \sqrt{\lvert \kappa \rvert^2 - \lvert \mu \rvert^2})^3 \big( \lvert \kappa \rvert -\lvert \mu \rvert \big) \,,
\ee
implying a sizable vacuum decay rate. Finally, in the Minkowski limit with $H\rightarrow 0$, this setup with an isotropic $SU(2)$ gauge field does not experience Schwinger-type particle production \cite{Lozanov:2018kpk}.

\section{Backreaction}\label{backreact}

In this section, we compute the induced current and backreaction of the spin-2 field on the inflationary background.

\subsection{Induced current and Backreaction}

The continuous global $SU(2)$ symmetry \eqref{U} leads to the conserved Noether current as
\be\label{J_A}
J^{\mu}_{A} = - T^a \frac{\delta(\da \mathcal{L})}{\delta \bar{A}^a_\mu}  =  \da \bigg[D_{\nu} \bigg(\bar{g}^{\mu\lambda}\bar{g}^{\nu\sigma} + \bar{\alpha}_A \epsilon^{\mu\nu\lambda\sigma}\bigg)F_{\lambda\sigma}\bigg],
\ee
where $\da$ denotes quadratic order action with respect to $B_{ij}$ field, a bar denotes background quantities, and $D_{\mu} \equiv \p_{\mu} -i\ga A_{\mu}$ is the covariant derivative. The current affects the background equations of motion as (see \eqref{field-eq-A})
\be\label{field-eq-A-}
(\nabla_{\nu} - i\ga \bar{A}_{\nu}) \bigg( \bar{F}^{\mu\nu} + \bar{\alpha}_{A} \epsilon^{\mu\nu\lambda\sigma} \bar{F}_{\lambda\sigma} \bigg) + \alpha_H \ga^2Z_0^2 \bar{A}^{\mu} = - \langle J_A^{\mu} \rangle,
\ee
where $\bar{\alpha}_A$ is a function of background fields given in \eqref{alpha-bar}. The expectation value of the zero-component of $J^{\mu}_{A}$ vanishes
\be
\langle J^{0}_{A}\rangle =0.
\ee
Thus, the background field equation of $\psi$ in \eqref{eq--psi} is sourced by $\mathcal{J}_A$ defined by
\be
 \mathcal{J}_{A}   &\equiv &  \frac{a}{3}\delta_i^a J^{i}_{Aa} = \frac{\ga\mpl^2}{3a^2}\bigg(\frac1a \epsilon^i_{qp}B_{jq}\p_i B_{jp} + \dot{\bar{\alpha}}_A B_{ij}^2\bigg) \nonumber\\
 &=& \frac{\ga}{3a^3} \sum_{\sigma} \int d^3k \bigg( - \lambda_{\sigma}k + \mH \frac{\dot{\bar{\alpha}}_A}{H} \bigg) \lvert \Ths(\vk)\rvert^2.
\ee

\begin{figure}[h!]
\begin{center}
\includegraphics[width=0.49\textwidth]{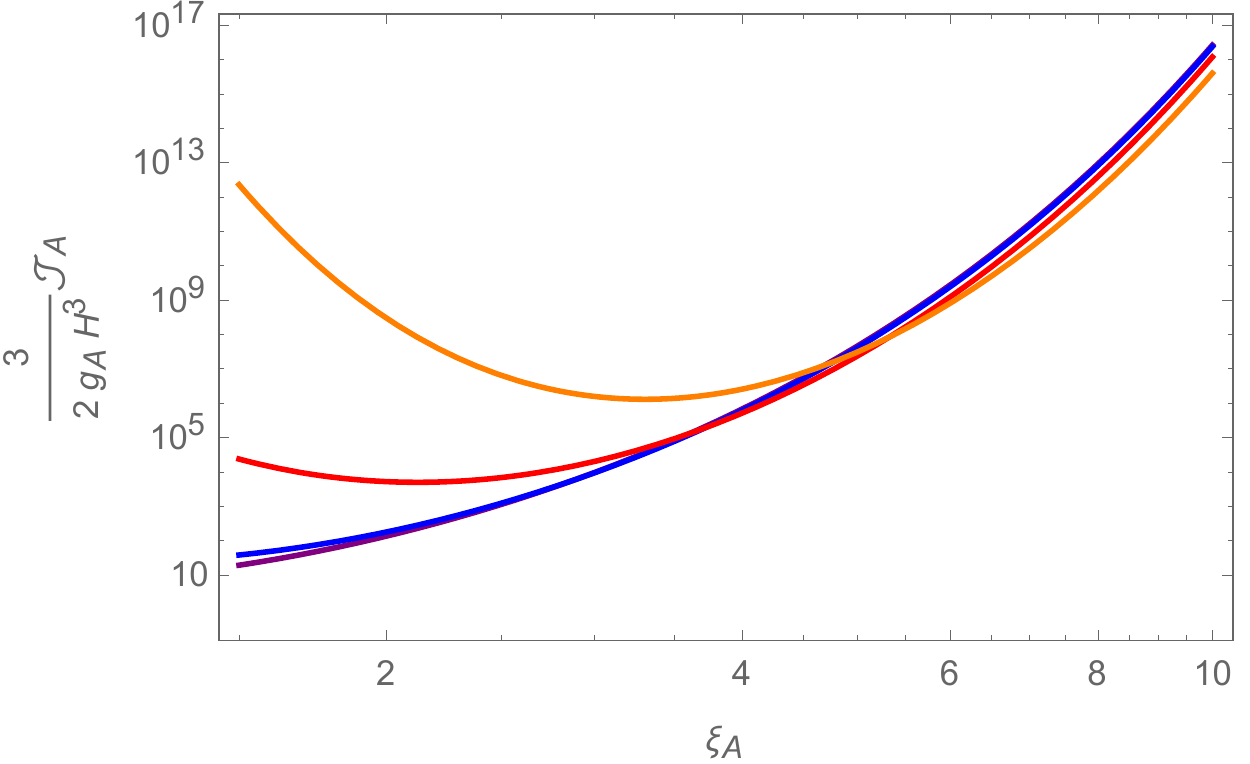} \includegraphics[width=0.49\textwidth]{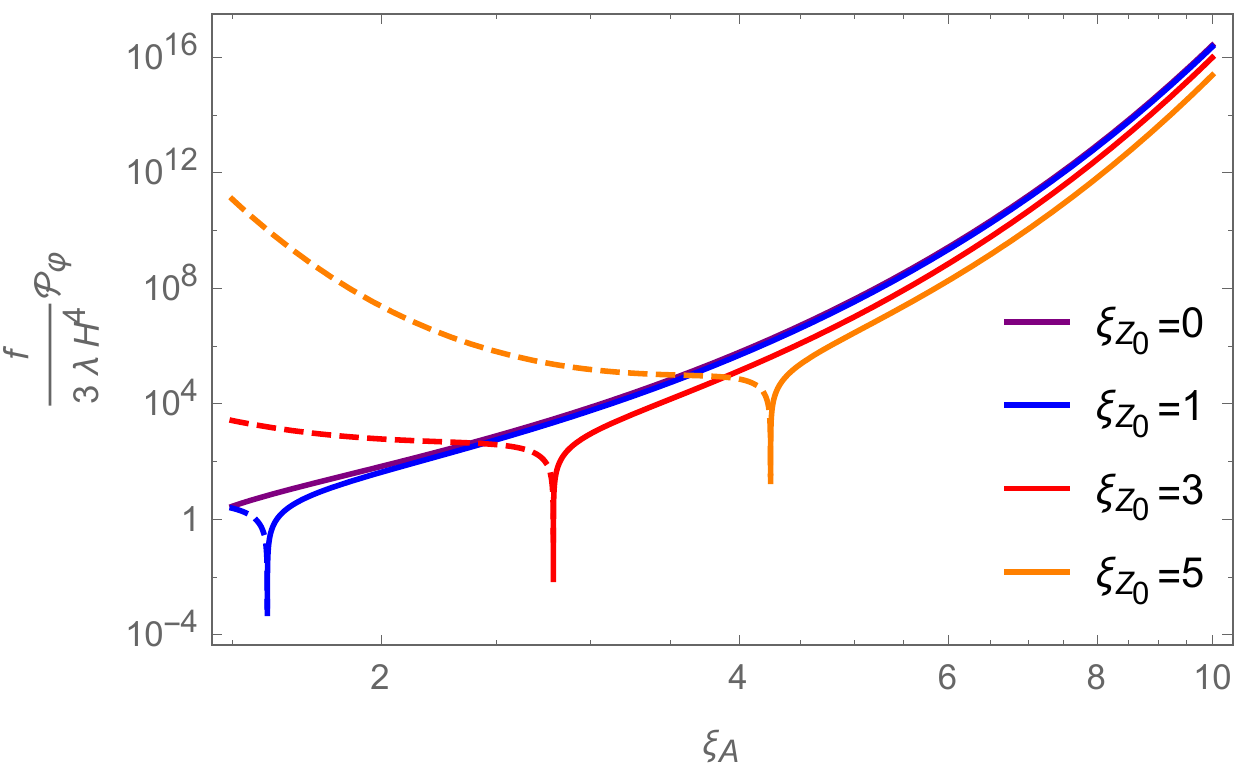}
\caption{Backreaction terms on the gauge ($\mathcal{J}_A$) and axion ($\mathcal{P}_{\varphi}$) fields as a function of $\xp$ and $\xz$. The dashed lines show negative values. We show $\frac{3}{2\ga H^3}\mathcal{J}_A$ and $\frac{f}{3\lambda H^4}\mathcal{P}_{\varphi}$.}\label{currents-}
\end{center}
\end{figure}

In the $\mathcal{L}_A=\mathcal{L}_{Cn}$ models with axion-gauge field coupling, $B_{ij}$ induces another backreaction term of the form $\nabla_{\mu} P^{\mu}_{\varphi}$ where 
\be\label{J_phi}
P^{\mu}_{\varphi} =  \frac{\lambda \mpl^2}{2a^{3}f} \delta^{\mu}_0 \big( \epsilon^{i}_{~qp} B_{qj} \p_i B_{pj} + \ga a\psi  B_{ij} B^{ij}\big),
\ee
with $P^{i}_{\varphi}=0$. The divergence of the zero-competent then backreacts on the background axion field as
\be
&&\ddot\varphi +3H\dot\varphi +V_{\varphi} +\frac{3\lambda \ga}{f}\psi^2(\dot\psi+H\psi)= \langle \mathcal{P}_{\varphi} \rangle,
\ee
where 
\be\label{cB}
\langle \mathcal{P}_{\varphi} \rangle \equiv \langle \nabla_\mu P^\mu_{\varphi} \rangle = \frac{\lambda}{2a^3f} 
\sum_{\sigma}~ \frac{d}{dt}\bigg[\int d^3k ~ \bigg( - \lambda_{\sigma} k + \mH \xp\bigg) \lvert \Ths(\vec{k}) \rvert^2\bigg].
\ee
Note that $P_{\varphi}^{\mu}$ is \textit{not} a Noether current.  In figure \ref{currents-}, we plotted $\mathcal{J}_{A}$ and $\mathcal{P}_{\varphi}$ with respect to $\xp$.

\begin{figure}[h!]
\begin{center}
\includegraphics[width=0.6\textwidth]{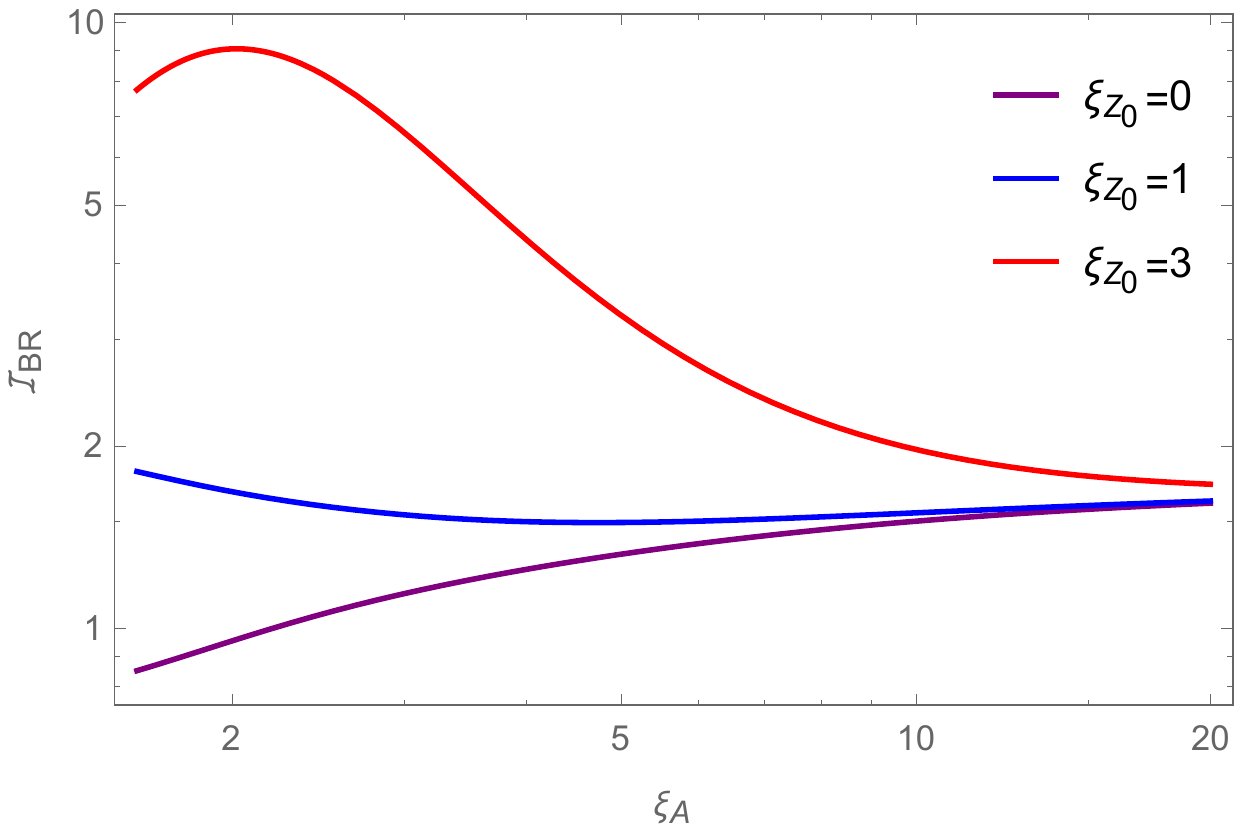} %
\caption{ The prefactor $\mathcal{I}_{BR}[(1+\xp^2+\frac{\alpha_{H}}{2} \xi^2_{Z_0})/\xp]$ as a function of $\xp$ and for different values of $\xz$. Note that for massive cases with $\xz\neq 0$, we set $\alpha_{H}=1$. }\label{I-BR-Plot}
\end{center}
\end{figure}

Both $\langle \mathcal{J}_A\rangle$ and $\langle \mathcal{P}_{\varphi}\rangle$ can be written in terms of the following momentum integral
\be\label{cJ-}
\mathcal{K}[X] \equiv   \sum_{\sigma} (2\pi)^2 \int \frac{d^3k}{2\mH^3} ~ \big( - \lambda_{\sigma} k + \mH X \big) \lvert \Ths(\vec{k}) \rvert^2.
\ee
We find 
\be\label{cJA}
\langle \mathcal{J}_A\rangle &=& \frac{2\ga H^3}{3(2\pi)^2}  \mathcal{K}[\dot{\bar{\alpha}}_A/H],\\\label{cJP}
\langle \mathcal{P}_{\varphi}\rangle &=&  \frac{\lambda}{(2\pi)^2a^3f} \frac{d}{dt}\bigg( a^3H^3 \mathcal{K}[\xp] \bigg) \simeq \frac{3\lambda H^4}{(2\pi)^2f}  \mathcal{K}[\xp],
\ee
where we have used $\frac{\dot{\bar{\alpha}}_A}{H}\simeq \frac{(1+\xp^2+\frac{\alpha_{H}}{2} \xi^2_{Z_0})}{\xp}$ from \eqref{xi--sl}.
We work out the integral of $\mathcal{K}[X]$ and its renormalization (using the adiabatic subtraction technique) in Appendix \ref{Current-appendix}. Here we present its final regularized form 
\be\label{G-reg}
\mathcal{K}_{reg}[X] &=&
\frac{1}{6} e^{2(\lvert \kappa \rvert - \lvert \mu \rvert)\pi} \bigg( \lvert \mu \rvert (-4\lvert\mu\rvert^2 + 15\lvert\kappa\rvert^2-4 - 9 \lvert \kappa \rvert X)  +  \frac{\C(X)}{4} {\rm{Re}}\bigg[ \psi^{(0)}(\frac12 + i \lvert \kappa \rvert - i \lvert \mu \rvert ) \nonumber\\ 
&& - \psi^{(0)}(\frac12 + i \lvert \kappa \rvert + i \lvert \mu \rvert ) \bigg] \bigg) , 
\ee
where $\psi^{(0)}(z)\equiv\frac{d}{dz}\ln\Gamma(z)$ is the digamma function, and $\C(X)$ is 
\be
\C(X) \equiv  3X(\frac12+6\kappa^2-2\mu^2)-\frac32 (7+20\kappa^2-12\mu^2) \vert \kappa \vert.
\ee 
We find that $\mathcal{K}_{reg}[X]$ is  proportional to $e^{2(\lvert \kappa \rvert - \lvert \mu \rvert)\pi} \gg 1$ (see figure \ref{kappa-mu}).
In the limit that $\lvert \kappa \rvert - \lvert \mu \rvert \gg1$, we can further simply $\mathcal{K}_{reg}[X]$ to
\be\label{G-reg-simple}
\mathcal{K}_{reg}[X] =
\frac{1}{6} e^{2(\lvert \kappa \rvert - \lvert \mu \rvert)\pi} \bigg[ \lvert \mu \rvert (-4\lvert\mu\rvert^2 + 15\lvert\kappa\rvert^2-4 - 9 \lvert \kappa \rvert X)  +  \frac{\C(X)}{4}  \ln \bigg(\frac{\lvert \kappa \rvert -  \lvert \mu \rvert}{\lvert \kappa \rvert + \lvert \mu \rvert}\bigg) \bigg],~~
\ee 
in which we used \eqref{exp-dgamma} to expand $\psi^{(0)}(z)$. This completes our derivation of the analytical formulae for the backreaction terms. \footnote{As we see in figure \ref{kappa-mu}, the generated pair particle number is very large and thus we are in the classical regime. Therefore, we can estimate the induced gauge field current, $J^{\mu}_A$, by semi-classical approximations. Specifically, we can approximate it as $\mathcal{J}_{A}\sim 2 \ga n_{\rm pairs} v$. Assuming that the particles
travel with the speed of light $v\sim 1$, we can approximate the induced gauge field current as
\be
\mathcal{J}_{A} \sim \frac{2\ga H^3}{6\pi^2} \big( \lvert \kappa \rvert + \sqrt{\lvert \kappa \rvert^2- \lvert \mu \rvert^2}\big)^3  e^{2(\lvert \kappa \rvert-\lvert \mu \rvert)\pi},
\ee
which is in agreement with the result of our exact solution in \eqref{cJA}.}  Using \eqref{eq:nPairs-III}, we can relate $\mathcal{K}_{reg}[X]$ to the number density of the spin-2 field as
\bea
\mathcal{K}_{reg}[X] \simeq \frac{\pi^2}{10} \mathcal{I}_{BR}[X] \bigg( \frac{n_{\rm{pairs}}}{H^3}\bigg),
\eea
where $\mathcal{I}_{BR}[X]$ is of order unity and given as 
\bea
\mathcal{I}_{BR}[X] \equiv 10\left[\frac{\lvert \mu \rvert (-4\lvert\mu\rvert^2 + 15\lvert\kappa\rvert^2-4 - 9 \lvert \kappa \rvert X)  +  \frac{\C(X)}{4}  \ln \big(\frac{\lvert \kappa \rvert -  \lvert \mu \rvert}{\lvert \kappa \rvert + \lvert \mu \rvert}\big)}{(\lvert \kappa \rvert + \sqrt{\lvert \kappa \rvert^2 - \lvert \mu \rvert^2})^3} \right].
\eea
The backreaction is also directly related to $n_{\rm{pairs}}$. In particular, the backreaction to the field equation of the gauge field is
\bea\label{JA-npair}
\mathcal{J}_A \simeq \frac{\ga \mathcal{I}_{BR}[(1+\xp^2+\frac{\alpha_{H}}{2} \xi^2_{Z_0})/\xp]}{60} ~ n_{\rm{pairs}},
\eea
where $\mathcal{I}_{BR}[(1+\xp^2+\frac{\alpha_{H}}{2} \xi^2_{Z_0})/\xp]$ is of order unity as presented   in figure \ref{I-BR-Plot}. Similar to $\mathcal{J}_A$, the backreaction to the axion field equation is also directly given by the number density of the spin-2 field, i.e. $\mathcal{P}_\varphi \sim \frac{\lambda H}{f} n_{\rm{pairs}}$. This relation is valid for all the models considered in this paper.

 Equations \eqref{cJA}-\eqref{G-reg} and \eqref{JA-npair} are the first main results of this paper. Equations \eqref{cJA}-\eqref{G-reg} have been estimated only numerically for one model in this family previously in \cite{Fujita:2017jwq} which are in agreement with our formulae over the region where the comparison is possible. The relation between the backreaction and the number density of the spin-2 field in \eqref{JA-npair} is derived here for the first time.

\subsection{Energy density of spin-2 fields}

The extra spin-2 field has a sizable energy-momentum density. The expectation value of its energy density adds to the total energy density in the background, $\bar{\rho}$, as
\be
\bar{\rho}(t) = \sum_{I} \bar{\rho}_{I}(t) + \langle \da\rho(t)\rangle_{\vk=0}.
\ee
Perturbing the energy momentum tensor and considering only terms quadratic in $B_{ij}$, we have 
\be
\da T_{00} = - \bar{g}_{00} ~ \da \rho  \an \da T_{ij} =  \bar{g}_{ij} ~\da P .
\ee
The contribution of $B_{ij}$ to the energy density, $\da\rho$, is
\be\label{drho-2}
\langle \da\rho \rangle = \mpl^2 \langle \frac{1}{2a^2}(\p_0 B_{ij})^2 +\frac{1}{2a^4} (\p_k B_{ij})^2  + \frac{\ga  \psi}{a^3} \epsilon^{qpi} B_{jq}\p_{i} B_{jp}+ \frac12\alpha_H \ga^2 Z_0^2 B_{ij}^2 \rangle.
\ee
The contribution to the isotropic pressure is 
\be
\langle \da P \rangle =  \langle \frac13 \da\rho\big \rangle\rvert_{\alpha_H=0} - \frac16 \alpha_H \ga^2 Z_0^2\mpl^2 \langle B_{ij}^2 \rangle.
\ee
In the absence of interaction with the Higgs field, $\alpha_{H}=0$, $B_{ij}$ field has the equation of state of radiation. On the other hand, for a massive gauge field with $\alpha_{H}=1$, the $B_{ij}$ field gets closer to dust.

\begin{figure}[h!]
\begin{center}
\includegraphics[width=0.6\textwidth]{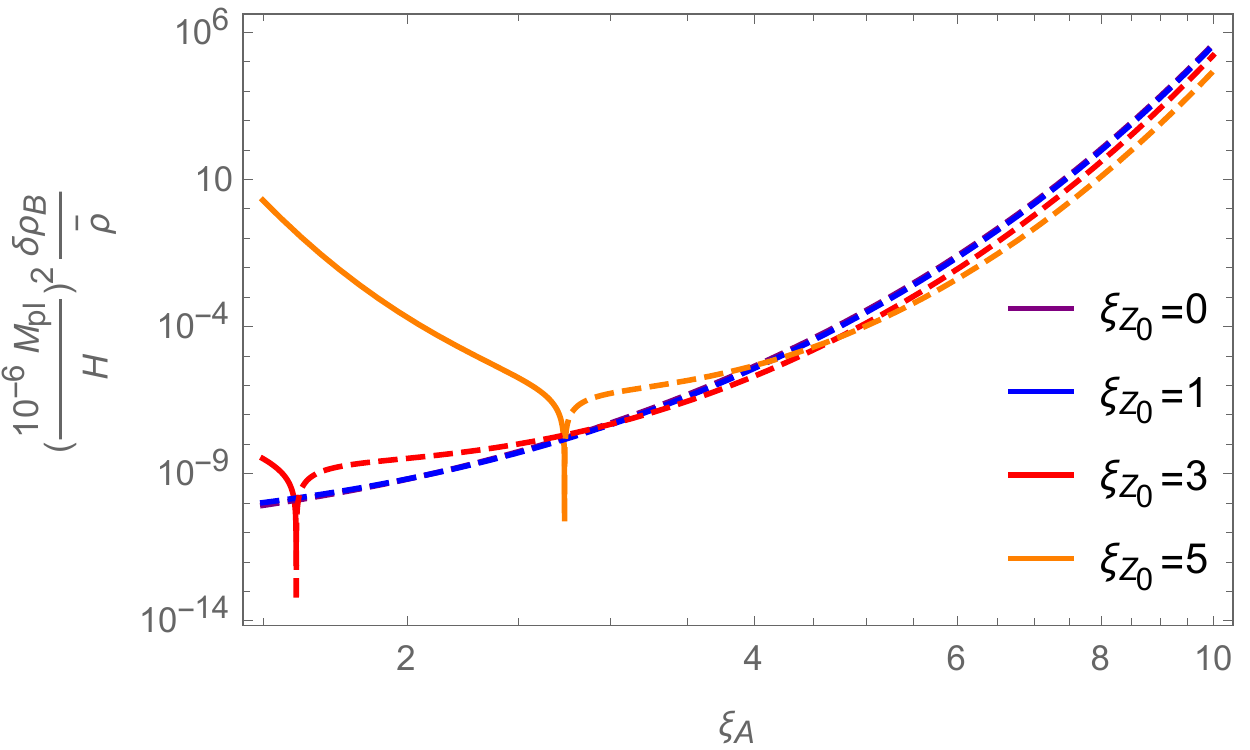} %
\caption{The energy density of the spin-2 field, $\da \rho$, as a function of $\xp$ and for different values of $\xz$. The dashed lines show negative values.  }\label{energy}
\end{center}
\end{figure}

We compute the normalized energy density in Appendix \ref{energy-app}, and only show the result here. Using \eqref{rho-reg}, we find the energy density fraction in the spin-2 field $\Bs$ as
\be\label{delta-rho-B}
\frac{\langle \da\rho \rangle_{reg}}{\bar{\rho}} \approx  \bigg(\frac{H}{\mpl}\bigg)^2  \frac{(\delta_c + 2\xp)}{3(2\pi)^2} \mathcal{K}_{reg}\big[\frac{ 3\xp (\delta_c - 3 \xp) -1 }{\delta_c + 2\xp} \big],
\ee
where $\frac{ 3\xp (\delta_c - 3 \xp) -1 }{\delta_c + 2\xp} \gtrsim 1$.
We show the energy density as a function of $\xp$ and $\xz$ in figure \ref{energy}.
Let us summarize the main features of the energy density:
\begin{itemize}
\item{The energy density of $B_{-}$ is always positive while the energy density of $B_{+}$ can be negative. }
\item{The total energy density in $B_{ij}$ is negative for $\xz\lesssim \frac95 \xp$.}
\item{The energy density fraction is of order $\frac{\langle \da\rho \rangle}{\bar{\rho}} \sim  \big(\frac{H}{\mpl}\big)^2 \delta_c \lvert \mu \rvert^3  e^{2(\lvert \kappa \rvert - \lvert \mu \rvert)\pi}$. Therefore, reducing the energy scale of inflation decreases the energy fraction in $B_{ij}$ as $\big(\frac{H}{\mpl}\big)^2$.}
\item{Validity of perturbation theory requires $\frac{\langle \da\rho \rangle}{\bar{\rho}}\lesssim 10^{-5}$, which constrains the parameter space of the models as a function of the energy scale of inflation.}
\end{itemize}

Although $\langle \da\rho \rangle_{reg}$ is negative in most of the parameter space, it is always a small part of the total energy density of the setup. The total energy-momentum tensor satisfies null and weak energy conditions, while the $B_{+}$ field violates both. The reason underlying the negative energy of the plus polarization is the existence of a short phase of instability for each $\vec{k}$-mode of $B_{+}$ around horizon crossing. However, this phase ends as soon as the mode exists the horizon; thus, the existence of the cosmic horizon evades (dangerous) infinite energy extraction of negative energy systems, unlike in flat space.

\section{Gravitational Waves}\label{gw}
\begin{figure}[h!]
\begin{center}
\includegraphics[width=0.49\textwidth]{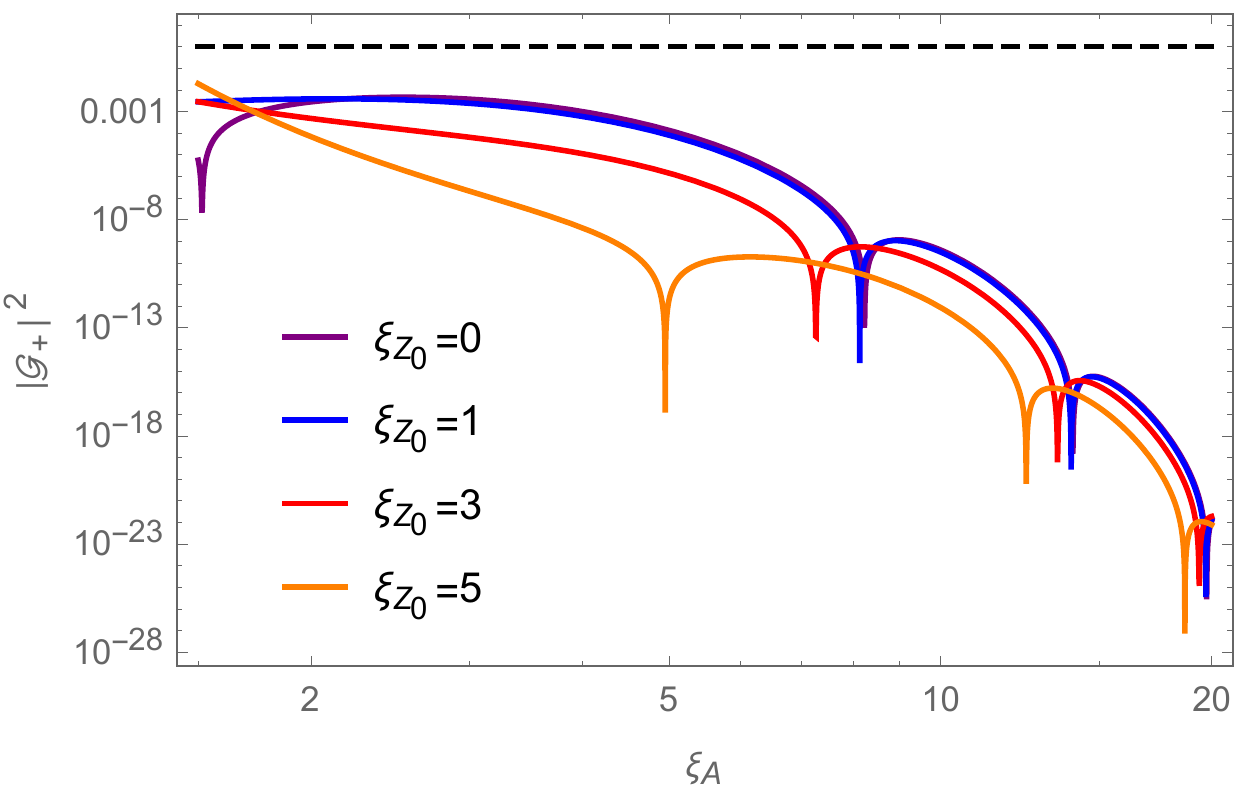}
\includegraphics[width=0.5\textwidth]{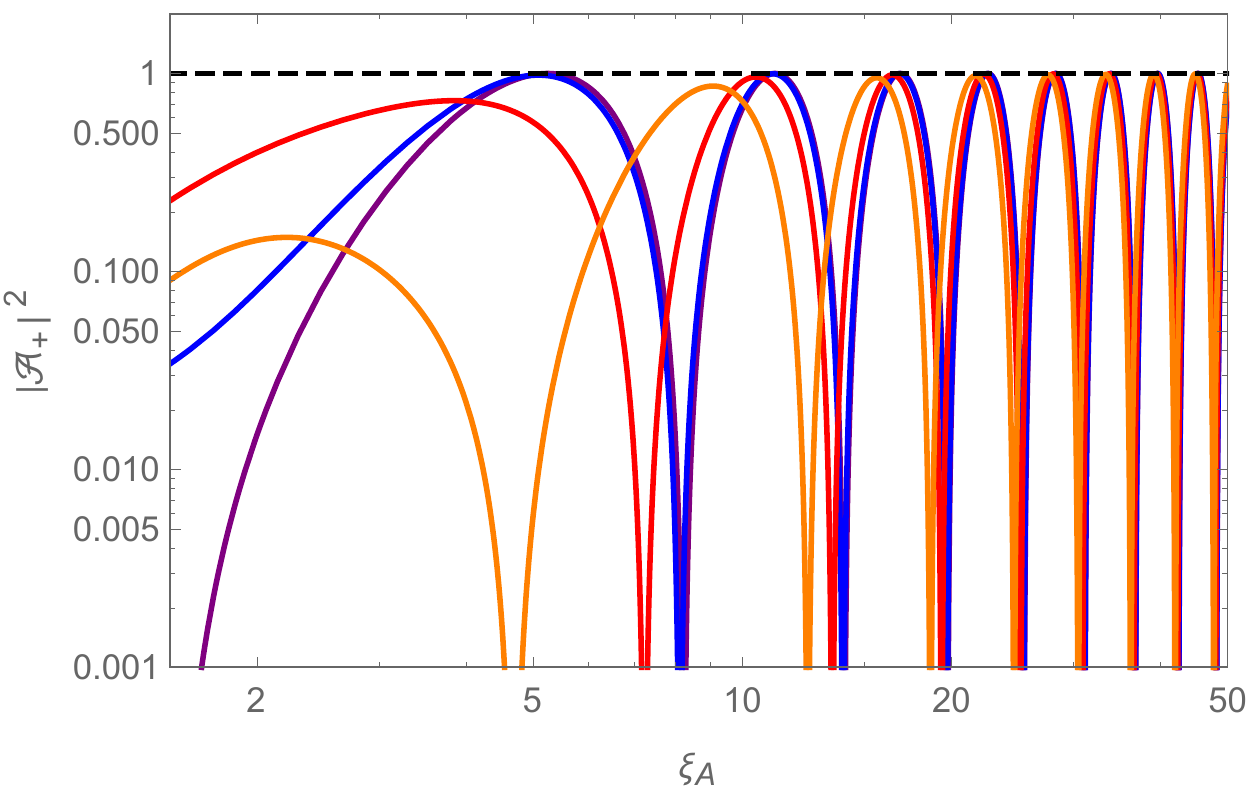} 
\caption{ The prefactors $\lvert \mathcal{G}_{+} \rvert^2$ (left) and $\lvert \mathcal{A}_{+} \rvert^2$ (right) as functions of $\xp$ and $\xz$. The (black) dashed line shows unity.}\label{A22}
\end{center}
\end{figure}

Each polarization state of the spin-2 field $B_{ij}$ mixes with the corresponding polarization of the gravitational waves, $\gamma_{ij}$ (see \eqref{hs-eq}). In particular, in the presence of the gauge field, we have
\be
\gamma_{\sigma}(t,\vec{x}) = \gamma^{\rm vac}_{\sigma}(t,\vec{x}) + \gamma^{\rm s}_{\sigma}(t,\vec{x}),
\ee
where $\gamma^{\rm vac}_{\sigma}(t,\vec{x})$ is the vacuum gravitational
waves (i.e. by quantum fluctuations of the spacetime
\cite{Starobinsky:1979ty, Grishchuk:1995ia}) with helicity $\sigma$
while $\gamma^{\rm s}_{\sigma}(t,\vec{x})$ is the part sourced by the
spin-2 field. Here, we compute the power spectrum and the energy density
of $\gamma^{\rm s}_{\sigma}$ in terms of $\xp$ and $\xz$ and relate them
with the number density of $B_{\sigma}$, $n_{\rm pairs}$. We work out the exact form of the sourced gravitational waves in (quasi) de Sitter in Appendix \ref{GW-app}. The result in the super horizon limit is 
\be
\hs^{\rm s}(\tau,\vk) = \frac{e^{i\kappa_{\sigma}\!\pi/2}}{(2\pi)^{\frac32}} \bigg(\frac{\psi}{\mpl}\bigg) \bigg( \frac{aH}{\sqrt{2}k^{\frac32}}\bigg)
\mathcal{G}_{\sigma}(\xp,\xi_{Z_0}),
\ee
where the explicit form of $\mathcal{G}_{\sigma}(\xp,\xi_{Z_0})$ is given in \eqref{Int-IV} and shown in figure \ref{A22}. It can be well approximated as
\bea\label{G-simple}
&&  \mathcal{G}_{\sigma}(\xp,\xi_{Z_0}) = \frac{\pi}{\cos(\pi\mu)\Gamma(-\kappa_{\sigma})} \bigg[\! \frac{(i+\lambda_{\sigma}\beta_c)}{\kappa_{\sigma}} +  \frac{( i - \lambda_{\sigma}\beta_c )\Gamma^2(-\kappa_{\sigma})}{\Gamma(\frac12-\kappa_{\sigma}-\mu)\Gamma(\frac12-\kappa_{\sigma}+\mu) }  \bigg].\nonumber\\
\eea
Taking the (classical) limit, $\lvert \mu \rvert \gg1$, and using \eqref{Int-VI}, we have 
\be
\gamma_{+}^{\rm s}(\tau,\vk) \simeq \frac{  \mathcal{A}_{+}}{(2\pi)^{\frac32}} e^{(\lvert \kappa\rvert - \lvert \mu \rvert)\pi }  \bigg(\frac{\psi}{\mpl}\bigg) \bigg( \frac{H}{ k^{\frac32}}\bigg) \sqrt{\pi\lvert \kappa \rvert},
\ee
where $\mathcal{A}_{+}$ is (see \eqref{cal-A})
\be\label{cal-A--}
\mathcal{A}_{+} &\simeq & \bigg[ \frac{( i -\xp )\Gamma^2\big(i(2\xp +\frac{\xz^2+2}{2\xp} )\big)}{\Gamma\big(\frac12+i[2\xp +\frac{\xz^2+2}{2\xp}  - \sqrt{2}(1+\xp^2+\xz^2)^{\frac12}]\big)\Gamma\big(\frac12+i[2\xp +\frac{\xz^2+2}{2\xp} +\sqrt{2}(1+\xp^2+\xz^2)^{\frac12}]\big) } \nonumber\\
& + & \frac{i\xp-1}{2\xp +\frac{\xz^2+2}{2\xp}} \bigg].
\ee 
Figure \ref{A22} shows $\lvert \mathcal{A}_{+} \rvert^2$ as a function
of $\xp$ and $\xz$. We find that $\lvert \mathcal{A}_{+} \rvert^2$
oscillates between zero and unity as a function of $\xp$.

\begin{figure}
\begin{center}
\includegraphics[width=0.45\textwidth]{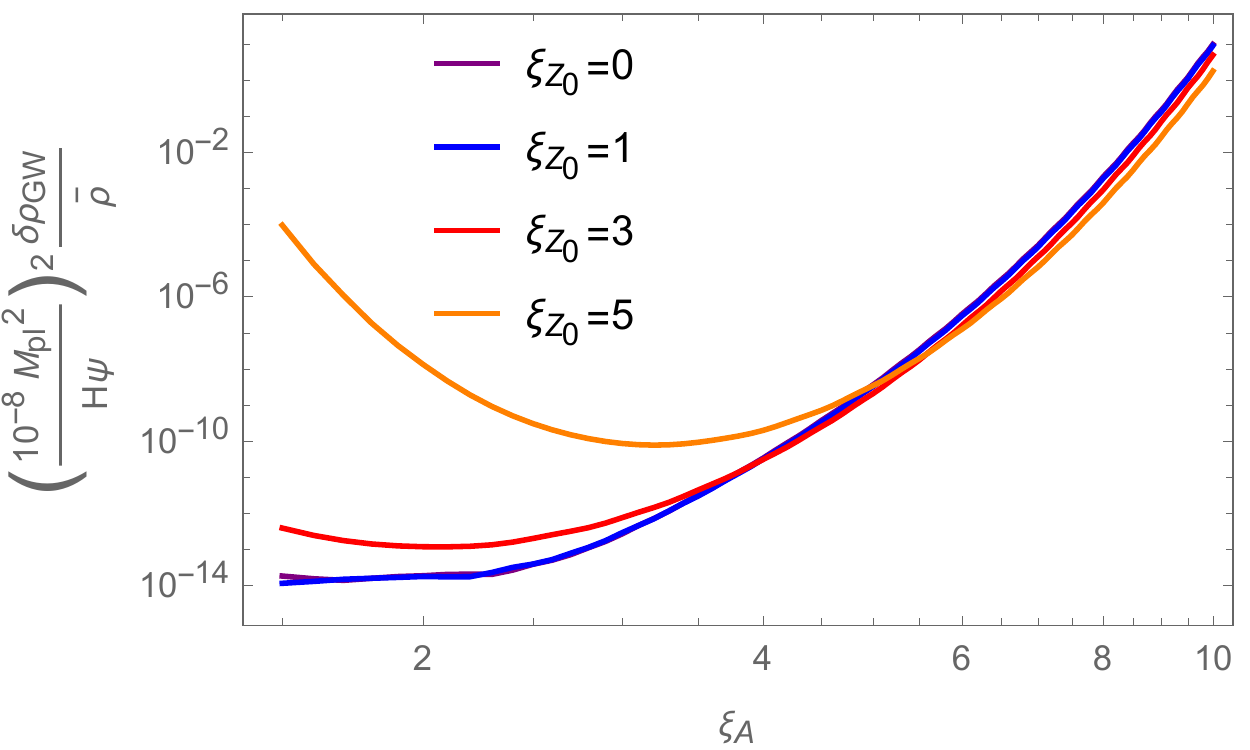} \includegraphics[width=0.45\textwidth]{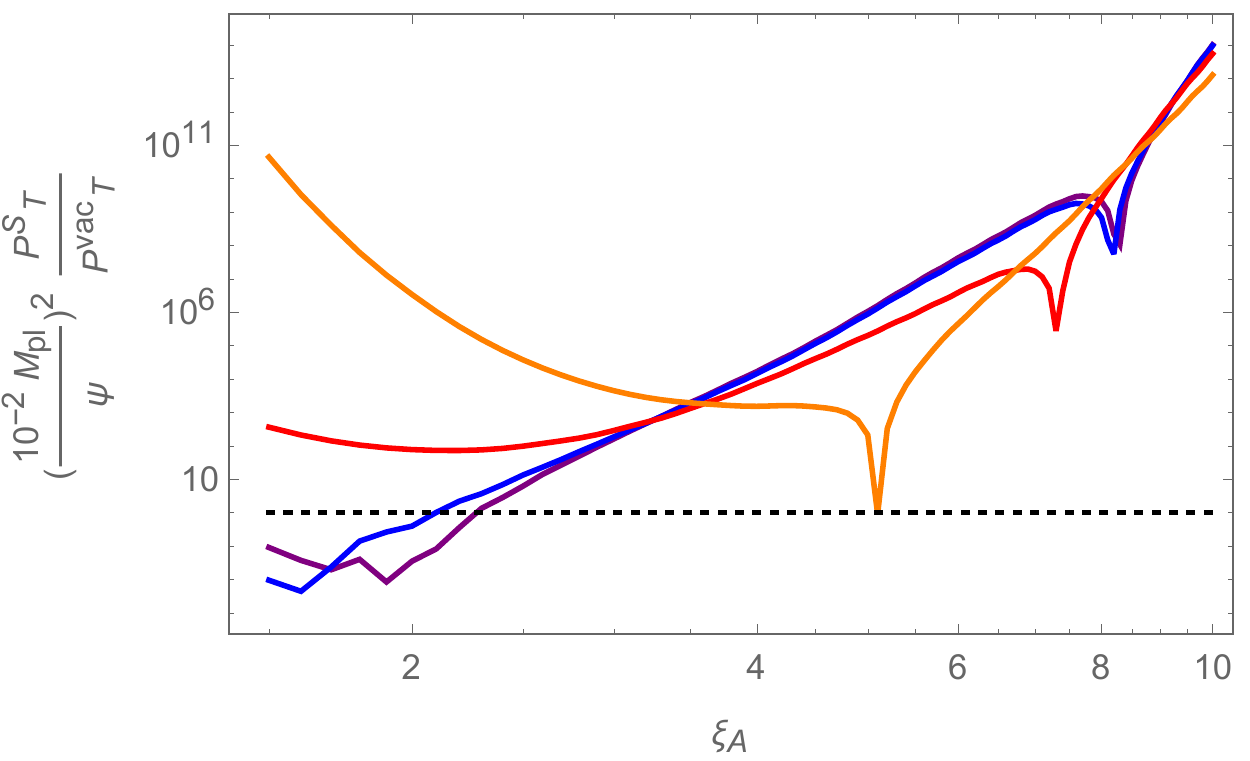}
\caption{The energy density (left) and power spectrum (right) of gravitational waves sourced by the gauge field as a function of $\xp$ and $\xz$. We show $\big(10^{-8}\mpl^2/ H\psi\big)^2 \delta\rho^{\rm GW}_{\rm s}/\bar{\rho}$ (left) and $\big(10^{-2}\mpl/\psi\big)^2 P^{\rm s}_T/P^{\rm vac}_T $ (right). The dashed line in the right panel shows unity.}\label{energy-GW}
\end{center}
\end{figure}

 Finally, we can write the ratio of the power spectra of 
sourced and vacuum gravitational waves in term of the spin-2 fields number density as
\be\label{power-gw}
\frac{P_{T}^{\rm s}}{P_T^{\rm vac}} = \frac{\langle \gamma^{\rm s}_{+}\gamma^{\rm s}_{+}\rangle }{\langle \gamma^{\rm vac}_{+}\gamma^{\rm vac}_{+} \rangle }\bigg\vert_{-k\tau\ll 1} \simeq  \bigg(\frac{\psi}{\mpl}\bigg)^2 
  \bigg(\frac{n_{\rm pairs}}{H^3}\bigg) \frac{6 \pi^3\lvert\mathcal{A}_{+}\rvert^2 \lvert \kappa_+ \rvert }{(\lvert \kappa_+ \rvert+ \sqrt{\lvert \kappa_+ \rvert^2-\lvert \mu \rvert^2})^3},
\ee
where $P^{{\rm{x}}}_T(k) \delta^{(3)}(\vec{k}+\vec{k}') \equiv 8 \pi k^3
\langle
\gamma^{{\rm{x}}}_{+}(\vec{k})\gamma^{{\rm{x}}}_{+}(\vec{k'})\rangle$ is
the power spectrum of $\gamma^{{\rm{x}}}_{+}$ where ${\rm{x}}=({\rm s,vac})$. Moreover, $\langle \gamma^{\rm vac}_{+}(\vec{k})\gamma^{\rm vac}_{+}(\vec{k'})\rangle=\frac{H^2}{(2\pi)^3k^3}$ where $a\gamma_{+}(\vec{k'})$ is a canonically normalized field. In the right panel of figure \ref{energy-GW}, we show $\big(10^{-2}\mpl/\psi\big)^2 P^{\rm s}_T/P^{\rm vac}_T $. This is the second  main result of this paper. 

The next interesting quantity is the energy density of the gravitational waves. The energy-momentum tensor of the gravitational wave is 
\be
t^{\rm GW}_{\mu\nu} = \frac{\mpl^2}{4} \langle \p_{\mu} \gamma_{ij} \p_{\nu} \gamma_{ij}\rangle,
\ee
which gives the energy density in the sourced part of the gravitational waves, $\delta \rho^{\rm GW}_{\rm s}=t^{\rm GW}_{00}$, as 
\be
\delta\rho^{\rm GW}_{\rm s} = \frac{\mpl^2}{4a^2} \langle \gamma^{s'}_{ij} \gamma^{s'}_{ij}\rangle.
\ee
In terms of the physical momentum, it can be written as
\be
\delta\rho^{\rm GW}_{\rm s} = \frac{2 H^4}{(2\pi)^2} \sum_{\sigma=\pm}\int \x^3d\x k \langle \big(\p_{\x} h^{s*}_{\sigma} + \frac{\mH}{k} h^{s*}_{\sigma}\big) \big(\p_{\x} h^{\rm s}_{\sigma} + \frac{\mH}{k} h^{\rm s}_{\sigma}\big)\rangle.
\ee
In the left panel of figure \ref{energy-GW}, we show $\big(10^{-8}\mpl^2/ H\psi\big)^2 \delta\rho^{\rm GW}_{\rm s}/\bar{\rho}$. The energy density can be written as
\bea
\delta\rho_{\rm s}^{\rm GW} =  \frac{4 H^4 }{3\pi^2} \big(\frac{\psi}{\mpl}\big)^2 \mathcal{I}_{\rm{GW}}(\xp,\xz) e^{2(\lvert \kappa\rvert - \lvert \mu\rvert)\pi},
\eea
where $\mathcal{I}_{\rm{GW}}(\xp,\xz)$ is roughly  
\be\label{rho-gw}
\mathcal{I}_{\rm{GW}}(\xp,\xz) \sim (\lvert \kappa \rvert + \sqrt{\lvert \kappa \rvert^2 - \lvert \mu \rvert^2})^2.
\ee
Comparing to the number density of the (gauge field's) spin-2 particles in \eqref{eq:nPairs-III}, we find 
\be\label{delta-rho-GW}
\delta\rho^{\rm GW}_{\rm s} \sim \frac{8\pi}{(\lvert \kappa_+ \rvert + \sqrt{\lvert \kappa_+ \rvert^2 - \lvert \mu \rvert^2})}\bigg(\frac{\psi}{\mpl}\bigg)^2   H n_{\rm pairs}.
\ee
The above relation is correct up to order unity coefficients.
The energy density in the sourced gravitational wave is proportional to $\big(\frac{\psi}{\mpl}\big)^2$, the number density of $B_{+}$, $H n_{\rm pairs}$, and the inverse of $\x_1$ in \eqref{q12}.

\section{Constraints on the parameter space}\label{PA}

\begin{figure}[h!]
\begin{center}
\includegraphics[width=0.49\textwidth]{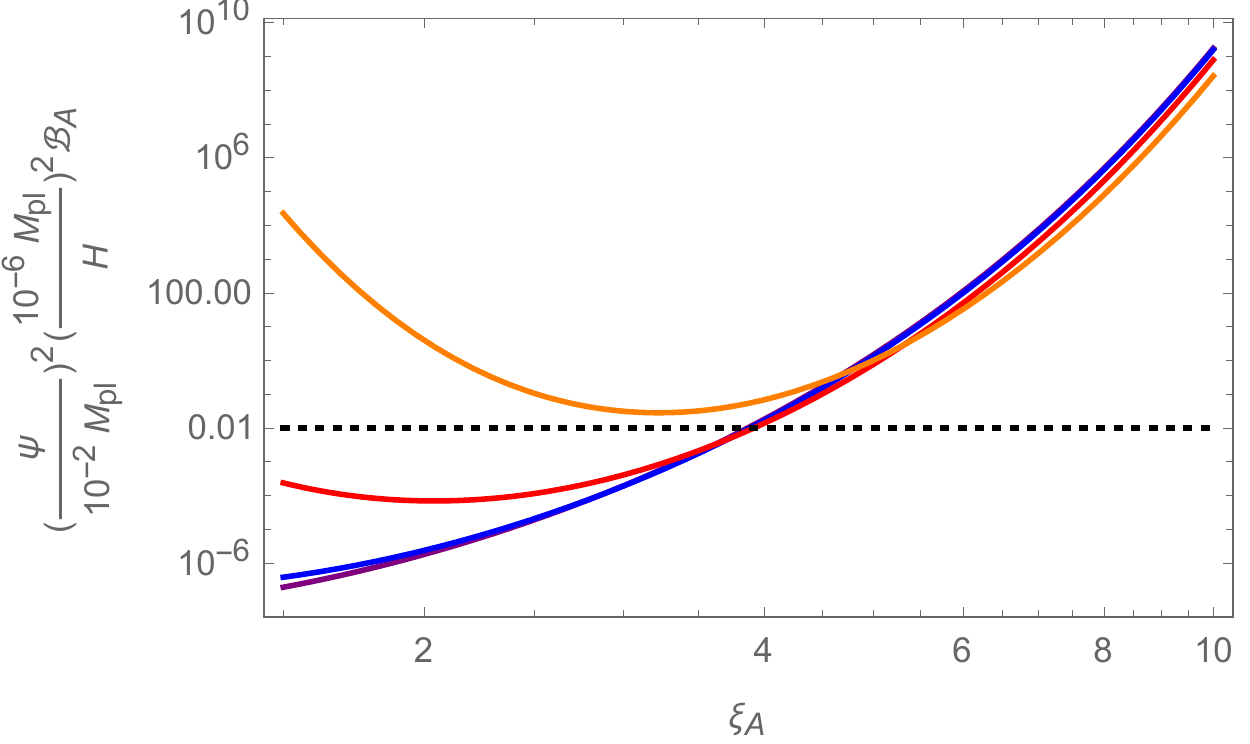} \includegraphics[width=0.49\textwidth]{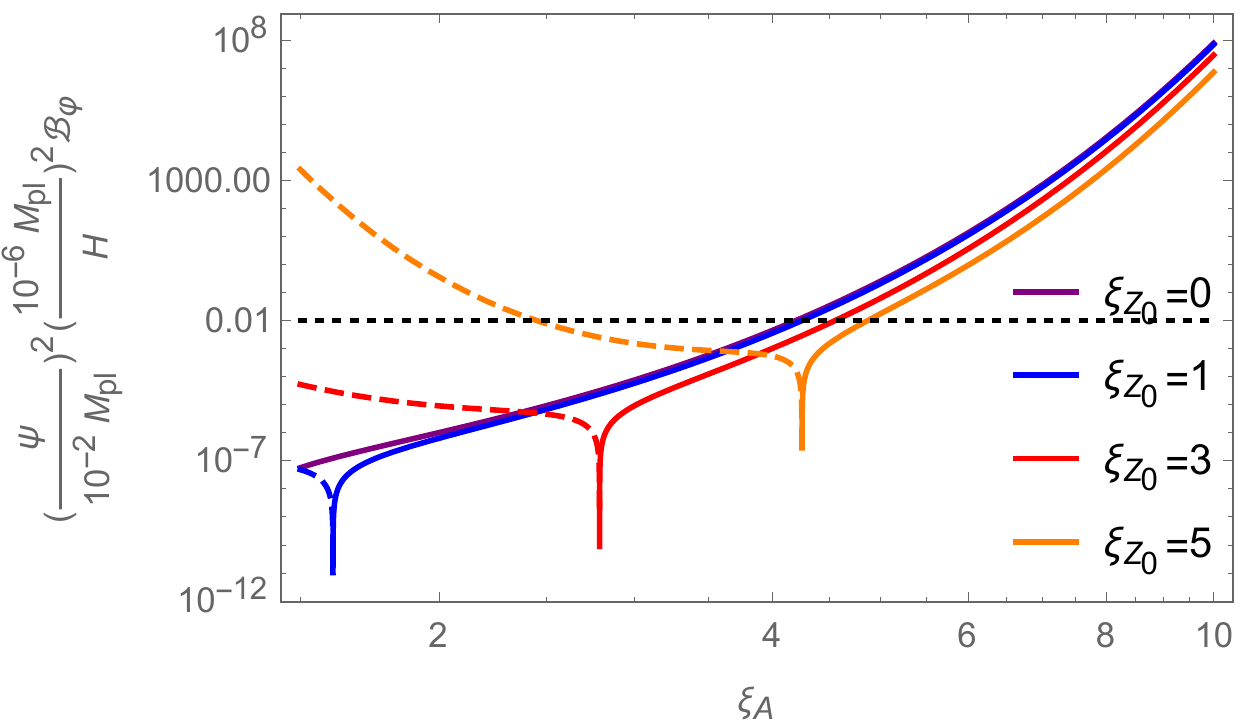}
\caption{The dimensionless backreaction terms to the gauge and axion field equations as a function of $\xp$. The dotted black line shows $0.01$. We show $\big(10^{2}\psi/\mpl\big)^2 \big(10^{-6}\mpl/H\big)^2\mathcal{B}_A$ and $\big(10^{2}\psi/\mpl\big)^2 \big(10^{-6}\mpl/H\big)^2\mathcal{B}_{\varphi}$. The dashed lines show negative values.}\label{B-B}
\end{center}
\end{figure}

\begin{figure}[h!]
\begin{center}
\includegraphics[width=0.6\textwidth]{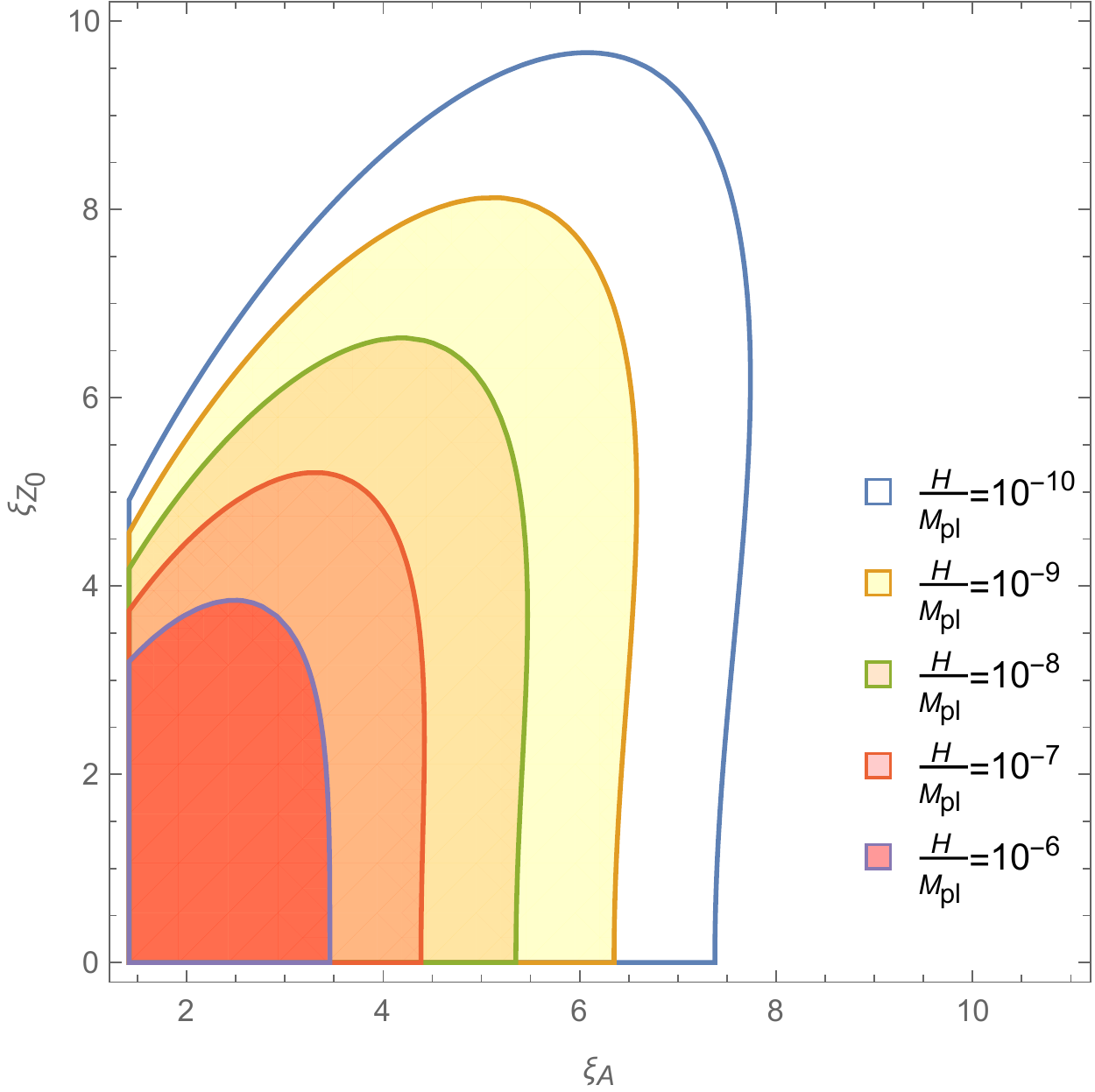} 
\caption{The parameter space $(\xp,\xz, \frac{H}{\mpl}, \epsilon_A=2\times 10^{-4})$ with dimensionless backreactions less than $10^{-2}$.}\label{parameter}
\end{center}
\end{figure}

In this section, we use the size of the backreaction to constrain the parameter space of the models. Validity of perturbation theory requires that the backreaction terms be much smaller than the other terms in the field equations. To do this, we normalize $\mathcal{J}_A$ and  $\mathcal{P}_{\phi}$ by $H^2\psi$ and $\lambda H \ga \psi^3/f$ respectively to construct two dimensionless backreaction terms
\be
\mathcal{B}_{A} &\equiv& \frac{\mathcal{J}_A}{H^3} \frac{H}{\mpl} \frac{\mpl}{\psi} = \bigg(\frac{10^{-2}\mpl}{\psi}\bigg)^2 \bigg(\frac{H}{10^{-6}\mpl}\bigg)^2 \frac{2\times 10^{-8}}{3(2\pi)^2} \xp   \mathcal{K}[\frac{\dot{\bar{\alpha}}_A}{H}],\\
\mathcal{B}_{\varphi} &\equiv&  \frac{\mathcal{P}_{\varphi}}{\lambda H\ga\psi^3/f}  =   \bigg(\frac{10^{-2} \mpl}{\psi}\bigg)^2  \bigg(\frac{H}{10^{-6}\mpl}\bigg)^2  \frac{3\times 10^{-8}}{(2\pi)^2} \frac{1}{\xp}\mathcal{K}[\xp],
\ee
where the explicit forms of $\mathcal{K}[X]$ for $X=\dot{\bar{\alpha}}_A/H\simeq (1+\xp^2+\frac{\alpha_{H}}{2} \xi^2_{Z_0})/\xp$ and $X=\xp$ are given in \eqref{G-reg} and a good approximation is given in \eqref{G-reg-simple}. Another backreaction is the contribution of the spin-2 field, $B_{\sigma}$ particle, to the total energy density which is presented in figure \ref{energy} and is subleading comparing to the above quantities. As we see, both $\mathcal{B}_A$ and  $\mathcal{B}_{\varphi}$ are proportional to the scale of inflation as $(H/\mpl)^2$ and hence decrease by reducing the scale of inflation. These dimensionless backreaction terms are shown in figure \ref{B-B}. We require both $\mathcal{B}_A$ and $\mathcal{B}_{\varphi}$ to be lower than $0.01$ so that they are smaller than the slow-roll suppressed terms in the background field equation. Figure \ref{parameter} shows the available parameter space corresponding to each scale of inflation.

Among the three backreactions, $\mathcal{B}_A$, $\mathcal{B}_{\varphi}$,
and $\frac{\delta_B\rho}{\bar{\rho}}$, the first one is the largest.
$\mathcal{B}_A $ is an exponential function of $\xp$ and $\xz$ while it
depends on the scale of inflation as $H^2$. For a given $\psi$ and a
bound on $\mathcal{B}_A$, lowering $H$ by a factor of $10$ enlarges the
acceptable domain of $\xp$ and $\xz$ by one unit (see figure \ref{parameter}). 
For instance, for a $\frac{\psi}{\mpl}\sim 10^{-2}$ and a GUT scale
inflation with $H\sim 10^{-6} \mpl$, we find that $0\leq\xz<4$ and
$\sqrt{2}<\xp \lesssim 3.5$ (lower bound comes from stability of the
scalar sector) are allowed. However, for a lower scale of inflation with
$H\sim 10^{-7} \mpl$, we find that $0\leq\xz<5$ and $\sqrt{2}<\xp
\lesssim 4.5$ are allowed. In this region, the energy density fraction
of the spin-2 field is $\frac{\delta_B\rho}{\bar{\rho}}\lesssim 10^{-7}$
and decreases linearly with the decrease of $H$.

\subsection{Parameter space of massless models}

Up to now, our formulae have been presented in the unified form and are
valid for any models assuming slow-roll dynamics of the VEV of the
gauge field and quasi-de Sitter expansion. In this section, we use our
backreaction formulae to further constrain the massless models, i.e.,
$\xz=0$ \cite{Maleknejad:2016qjz, Dimastrogiovanni:2016fuu, Fujita:2017jwq}.

The total slow-roll parameter, $\epsilon$, is given by
\be\label{epsi}
\epsilon = \epsilon_{\varphi} + \alpha_{s} \epsilon_{\chi} + \epsilon_A,
\ee
where $\epsilon_{\varphi}$ and $\epsilon_A$ are respectively the
contributions of the axion and the gauge fields to the slow-roll
parameter, whereas $\epsilon_{\chi}$ is the contribution of the inflaton
field in the spectator case (with $\alpha_{s}=1$).

 In massless models with $\xz=0$ and in the regime $\xp \gtrsim 2.5$,
 the extra scalar fields have little effect on the scalar power spectrum
 and scalar tilt, and thus we have the standard result given by
$ P_{\zeta}\simeq \frac{1}{2\epsilon} \big( \frac{H}{2\pi\mpl}\big)^2$ and $n_s-1\simeq -2(3\epsilon-\eta)$. 
On the other hand, adding the sourced tensor power spectrum given in
 \eqref{power-gw} to the standard vacuum part, we have the total tensor
 power spectrum given by
\be\label{r-sourced}
P_T \simeq 2 \bigg[ 1 + \bigg(\frac{\psi}{\mpl}\bigg)^2 \frac{e^{\lvert \kappa_+ \rvert \pi}}{2} \lvert \mathcal{G}_+ (\xp)\rvert^2
 \bigg] \bigg(\frac{~H}{\pi\mpl}\bigg)^2.
\ee
The tensor-to-scalar ratio, $r\equiv P_T/P_\zeta$, is
\bea
r=r_{\rm vac}+r_{\rm source}\simeq 16\epsilon \bigg[  1 + \bigg(\frac{\psi}{\mpl}\bigg)^2 \frac{e^{\lvert \kappa_+ \rvert \pi}}{2} \lvert \mathcal{G}_+(\xp)\rvert^2 \bigg],
\eea
where $r_{\rm vac}=16\epsilon$ is the standard vacuum value, whereas
 $r_{\rm source}$ is the contribution from the gauge field.
Using $ P_{\zeta} = 2.2 \times 10^{-9}$ and $\epsilon_A$ in
 \eqref{epsilon-A}, we write $r$ and  $\mathcal{B}_{A}$ in terms of
 $r_{\rm vac}$, $\epsilon_A/\epsilon$ and $\xp$ as
\begin{eqnarray}
%\nonumber
 r
 %&\simeq& \frac{10}{1.1\pi^2} \bigg( \frac{10^4H}{\mpl}\bigg)^2  \bigg[  1 +  \frac{1}{3.52 \pi^2} \bigg( \frac{10^4H}{\mpl}\bigg)^2 ~ \frac{\epsilon_A}{\epsilon} \frac{e^{\frac{(1+2\xp^2)}{\xp} \pi}}{(1+\xp^2)} \lvert \mathcal{G}_+(\xp)\rvert^2 \bigg]\\
 %&=&
 \simeq
r_{\rm vac}  \bigg[  1 + \frac{\epsilon_A}{2}~
\frac{e^{\frac{(1+2\xp^2)}{\xp}
\pi}}{(1+\xp^2)} \lvert \mathcal{G}_+(\xp)\rvert^2 \bigg]
=
r_{\rm vac}  \bigg[  1 + \frac{r_{\rm vac}}{32}~
\frac{\epsilon_A}{\epsilon} \frac{e^{\frac{(1+2\xp^2)}{\xp}
\pi}}{(1+\xp^2)} \lvert \mathcal{G}_+(\xp)\rvert^2 \bigg],
\label{r-rvac-ratio}
\end{eqnarray}
and
\be
\mathcal{B}_{A} \simeq  \frac{8.8}{3\times 10^{9}} \frac{\epsilon}{\epsilon_A} \xp(1+\xp^2)   ~  \mathcal{K}[\frac{1+\xp^2}{\xp}].
\ee
The exact form of $\mathcal{K}[\frac{1+\xp^2}{\xp}]$ is given in
 \eqref{G-reg}. It is approximately given by
 $\mathcal{K}[\frac{1+\xp^2}{\xp}]\sim \xp^3
 e^{2\big((2-\sqrt{2})\xp+\frac{2}{\xp}\big)\pi}$.

\begin{figure}
\begin{center}
\includegraphics[width=0.8\textwidth]{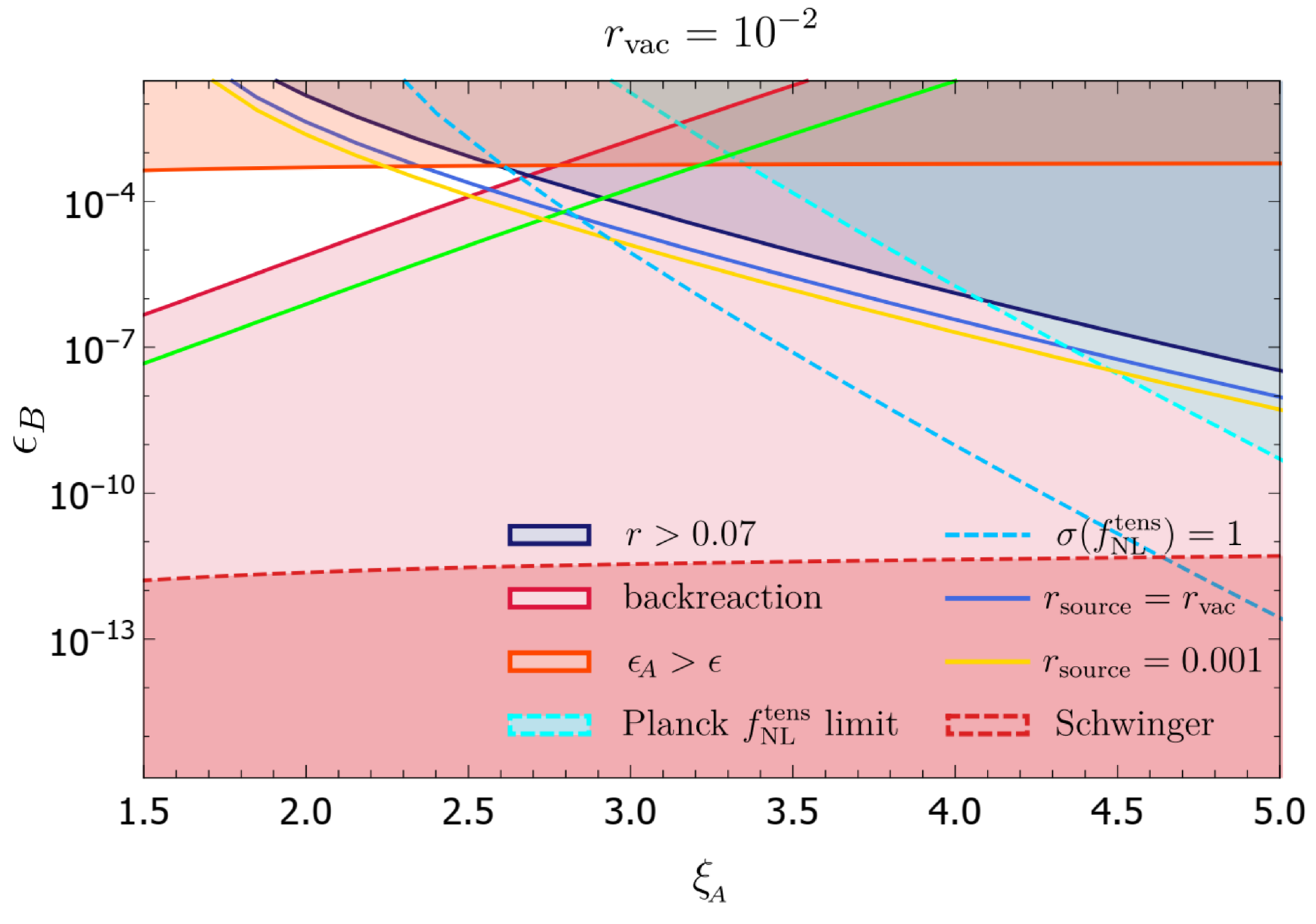}
 \caption{
Excluded parameter space of the massless models with
 $r_{\rm vac}=10^{-2}$. The blue shaded area is excluded by the
 tensor-to-scalar ratio, the light red area by the large backreaction
 ($\mathcal{B}_A>10^{-2}$), the orange area by the inconsistent
 slow-roll condition ($\epsilon_A>\epsilon$), the cyan area by the
 tensor non-Gaussianity, and the dark red area by the Schwinger
 pair-creation of scalar fields. The blue and yellow lines show
 $r_{\rm source}=r_{\rm vac}$ and $r_{\rm source}=10^{-3}$, respectively, while the
 dashed cyan line shows $f_{\rm  NL}^{\rm tens}=1$. The green line corresponds to the the amount of backreaction as large as  $\mathcal{B}_A=0.1$.}\label{BR-r}
\end{center}
\end{figure} 

\begin{figure}
\begin{center}
 \includegraphics[width=0.8\textwidth]{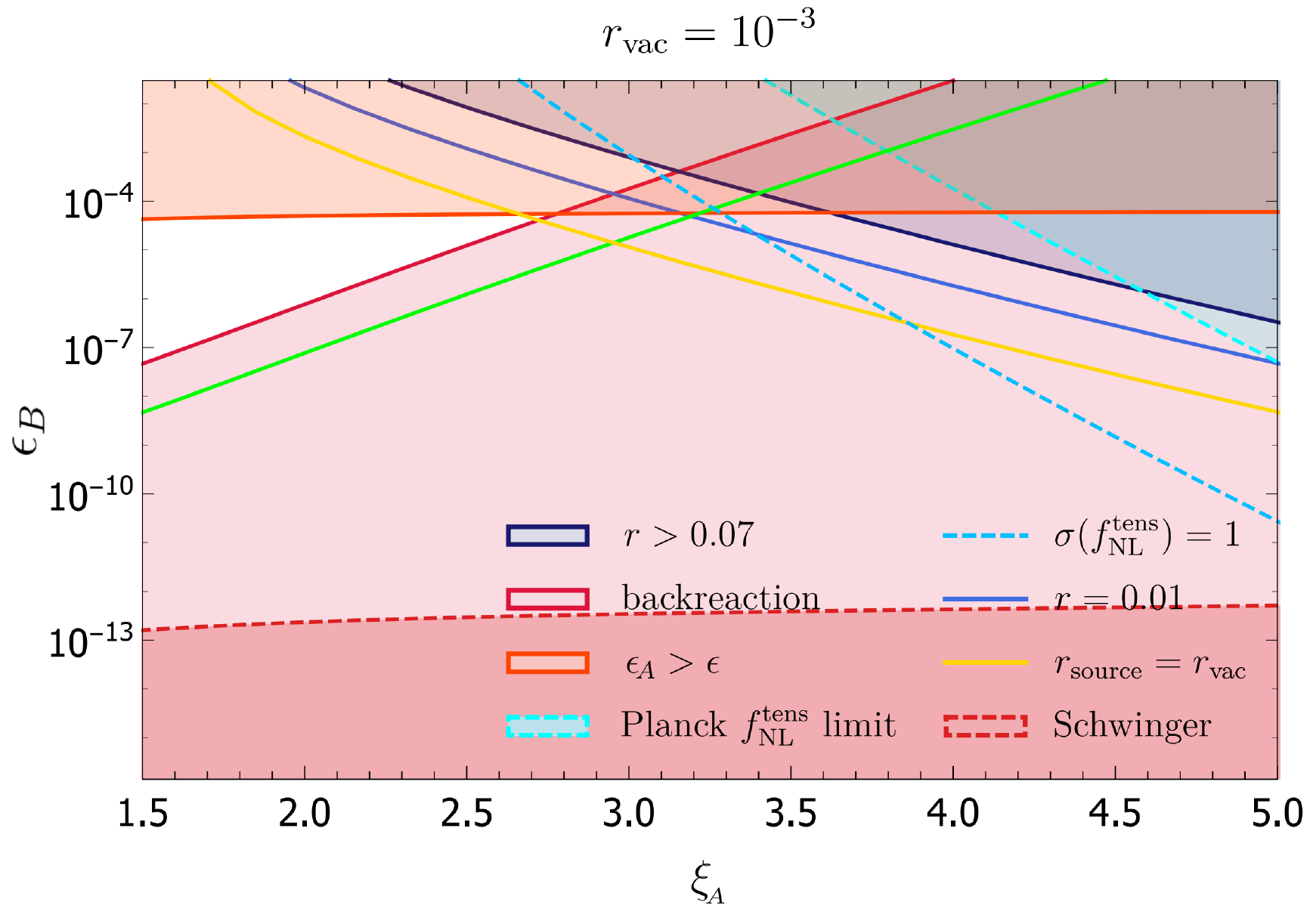}
 \includegraphics[width=0.78\textwidth]{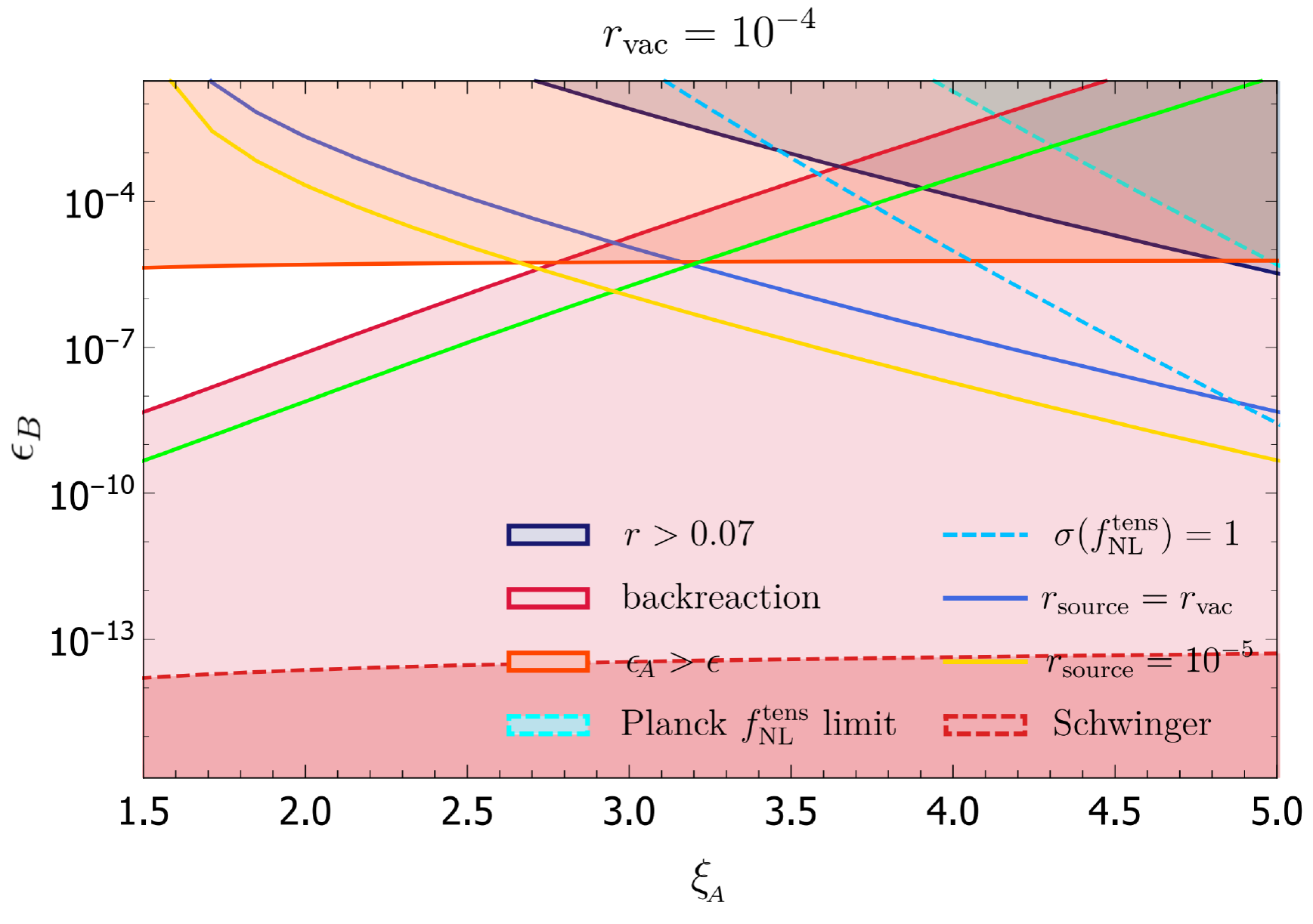}
 \caption{
 Same as figure~\ref{BR-r} but for $r_{\rm vac}=10^{-3}$ (top) and
 $10^{-4}$ (bottom). In the top panel the blue and yellow lines show
 $r=10^{-2}$ and $r_{\rm source}=r_{\rm vac}$, respectively, whereas in the
 bottom panel they show $r_{\rm source}=r_{\rm vac}$ and $r_{\rm source}=10^{-5}$,
 respectively.
 }\label{BR-r-II}
\end{center}
\end{figure}

In this section we constrain the parameter space of models in
 $\epsilon_B$-$\xi_A$ plane, following refs.~\cite{Agrawal:2018mrg,Lozanov:2018kpk}. Here, $\epsilon_B\equiv
 g^2\psi^4/(H^2\mpl^2)$ is related to $\epsilon_A$ as
$\epsilon_B =  \epsilon_A\xp^2/(1+\xp^2)$. We can relate this to
 the backreaction term as
%As we see, the backreaction is independent of $H$ and is inversely proportional to $\frac{\epsilon_A}{\epsilon}$. 
%From that we can write $\frac{\epsilon_B}{\epsilon}$ as
%\bea\label{r-rvac-ratio}
%\frac{\epsilon_B}{\epsilon} \simeq 32 ~ \frac{\xp^2 e^{-\frac{(1+2\xp^2)}{\xp} \pi}}{\rvert \mathcal{G}_+(\xp)\rvert^2} \bigg(\frac{r-r_{\rm vac}}{r_{\rm vac}^2}\bigg),
%\eea
%which by the current upper bound on $r$ constraints $\frac{\epsilon_B}{\epsilon}$ from above as a function of $\xp$ and $r_{\rm vac}$. The other constrain comes from the size of the backreation as
\bea
\frac{\epsilon_B}{\epsilon} \simeq \frac{8.8}{3\times 10^{9}} \frac{
\xp^3  }{\mathcal{B}_A} \mathcal{K}[\frac{1+\xp^2}{\xp}],
\eea
which puts a lower bound on $\frac{\epsilon_B}{\epsilon}$ from
backreaction.

We restrict the parameter space by imposing the following constraints:
\begin{itemize}
 \item The BICEP2/Keck and Planck (BKP) upper bound on the
       tensor-to-scalar ratio, $r<0.07$ (95\%
       C.L)  \cite{Array:2015xqh};
 \item The Planck upper bound on the tensor non-Gaussianity parameter, $f_{\rm
       NL}^{\rm tens}$ \cite{Ade:2015ava,Agrawal:2018mrg};
 \item Small backreaction given by $\mathcal{B}_{A}< 10^{-2}$ or
       $10^{-1}$;
 \item Small Schwinger pair-creation of scalar fields \cite{Lozanov:2018kpk};
 \item Consistency of the slow-roll parameter, $\epsilon_A<\epsilon$;
\end{itemize}
We show the constraints in figure~\ref{BR-r} and \ref{BR-r-II} for
$r_{\rm vac}=10^{-2}$, $10^{-3}$, and $10^{-4}$. We find that large
parameter space is excluded already. In particular, we find that $r_{\rm
source}$ cannot be much greater than $r_{\rm vac}$.

\begin{figure}
\begin{center}
\includegraphics[width=0.49\textwidth]{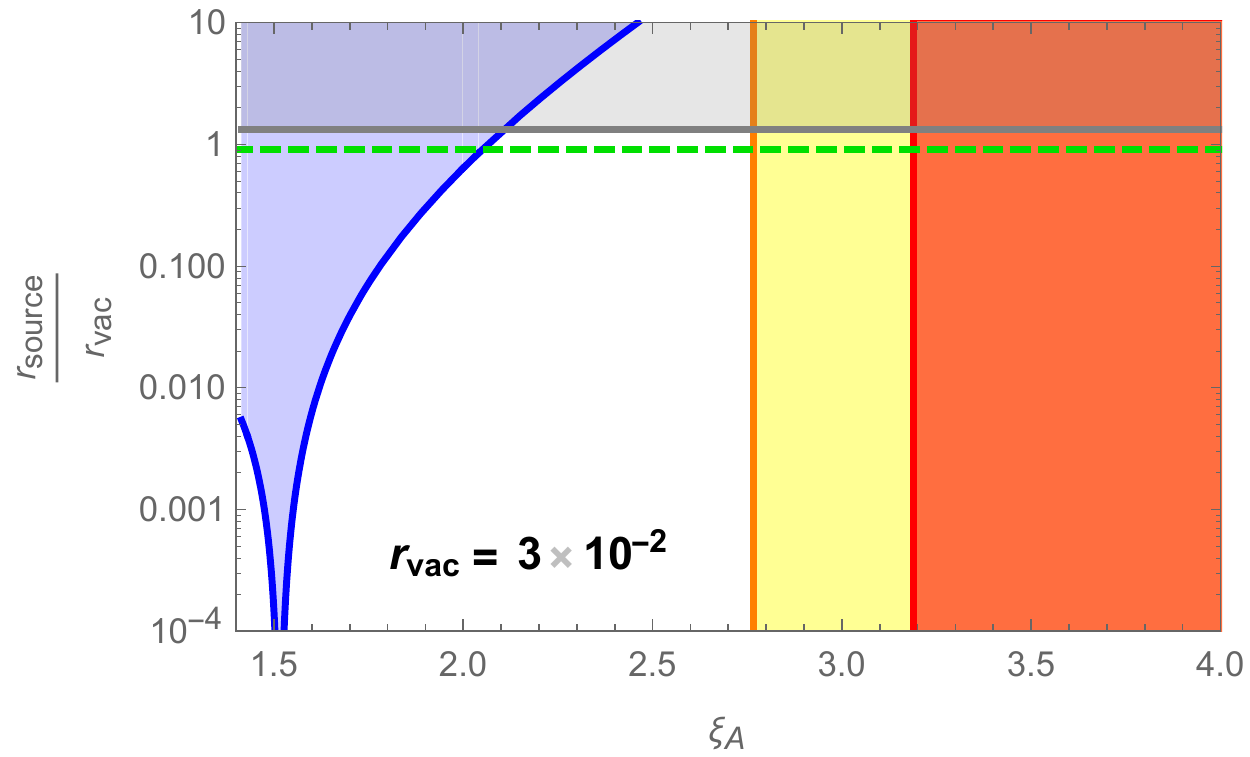} 
\includegraphics[width=0.49\textwidth]{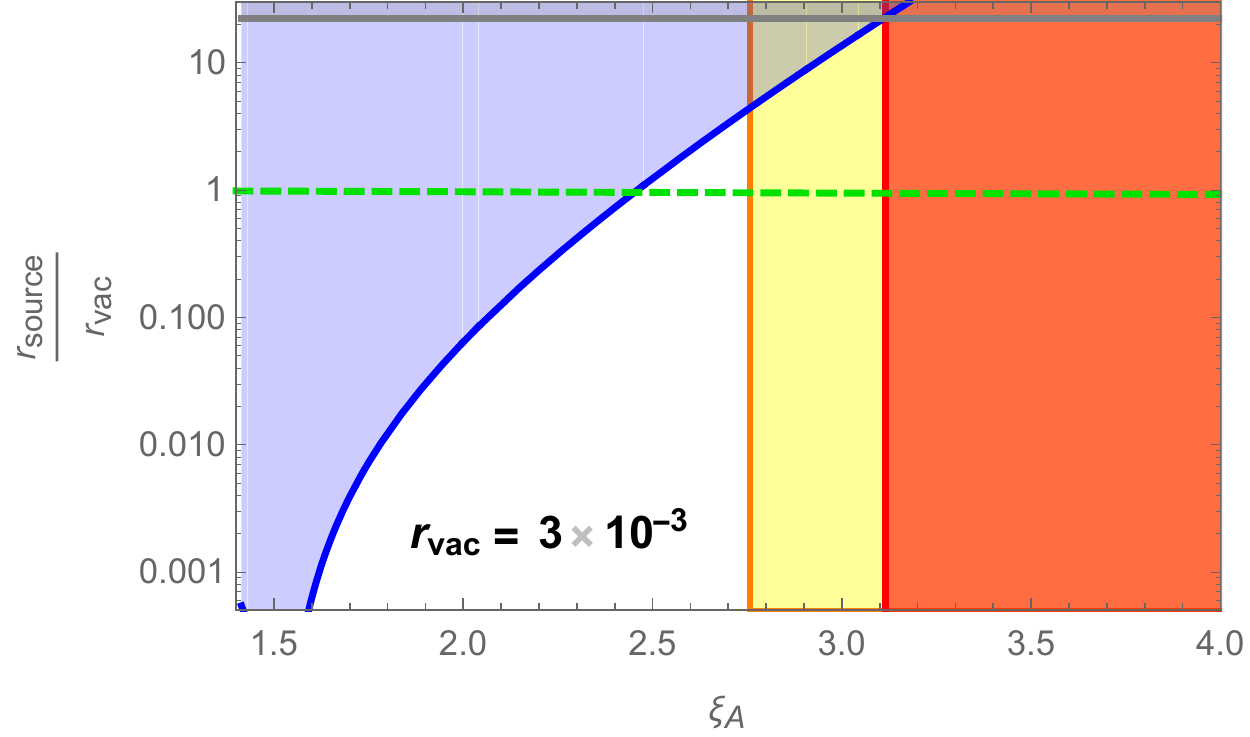} 
\includegraphics[width=0.49\textwidth]{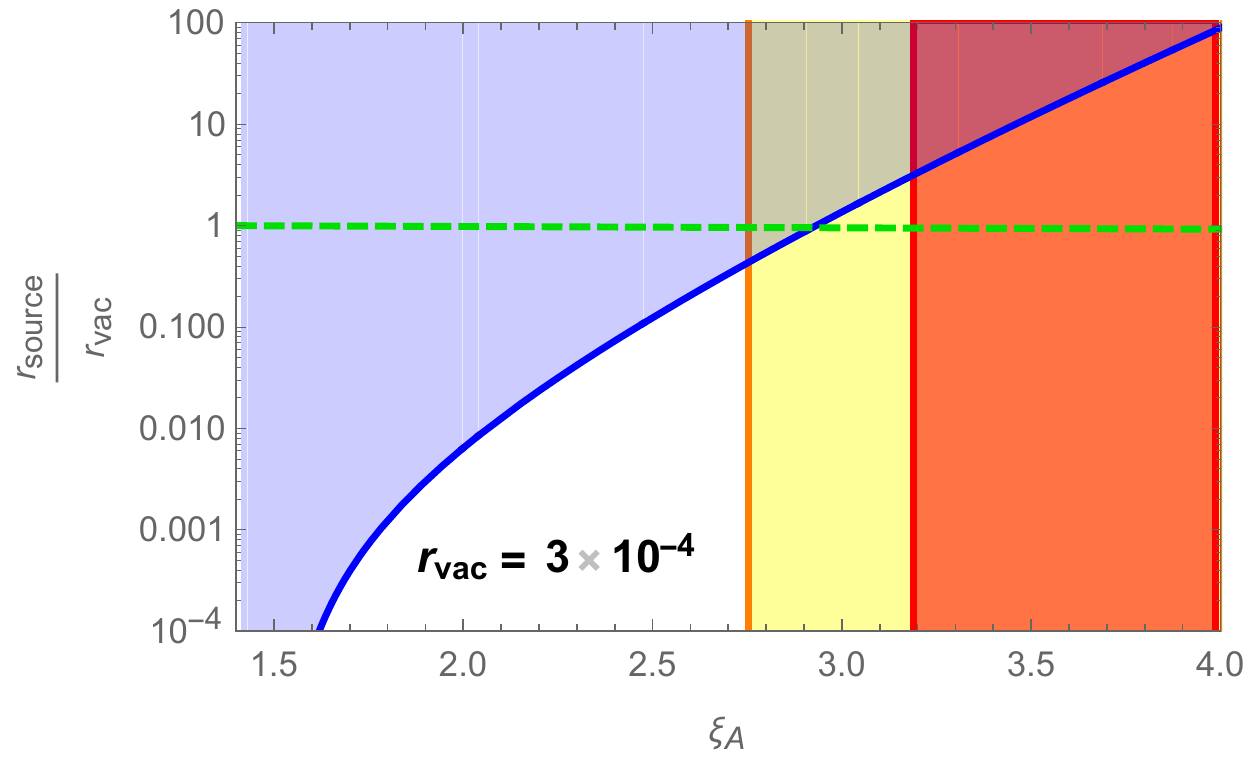}
 \caption{
 Constraints on the fractional contribution of the sourced
 gravitational waves relative to the vacuum one, $r_{\rm source}/r_{\rm
 vac}=(r - r_{\rm vac})/r_{\rm vac}$, as a function of $\xp$ for three
 different values of $r_{\rm vac}=3\times 10^{-2}$ (top left), $3\times
 10^{-3}$ (top right), and $3\times 10^{-4}$ (bottom). The blue shaded
 area is excluded by the inconsistency of the slow-roll condition
 ($\epsilon_A/\epsilon>0.9$), the yellow and orange areas by the large
 backreaction $\mathcal{B}_A>10^{-2}$ and $10^{-1}$, respectively, and
 the gray area by the tensor-to-scalar ratio $r>0.07$. The dashed green line marks $r_{source}/r_{vac}=1$.
}\label{r-rvac}
\end{center}
\end{figure}

To see this in more detail, in figure \ref{r-rvac} we show the ratio
$r_{\rm source}/r_{\rm vac} = (r-r_{\rm vac})/r_{\rm vac}$ as a function
of $\xp$ for given values of $\epsilon_A/\epsilon$ and $r_{\rm
vac}$. Imposing the bound on the size of the backreaction,
$\mathcal{B}_A<10^{-2}$, as well as on $\epsilon_A/\epsilon<0.9$,
we find that the maximum possible value of $(r-r_{\rm vac})/r_{\rm vac}$
can be at most $5$ for $r=3\times 10^{-3}$, and smaller for smaller
$r_{\rm vac}$ because of $r_{\rm source}/r_{\rm vac} \propto r_{\rm
vac}$ for a given $\epsilon_A/\epsilon$; see \eqref{r-rvac-ratio}.
The constraints weaken when we impose a weaker bound on the
backreaction, $\mathcal{B}_A<10^{-1}$. For
larger values of $r_{\rm vac}=3\times 10^{-2}$, the strongest constraint
comes from the BKP upper bound on $r<0.07$ rather than from
backreaction, yielding $(r-r_{\rm vac})/r_{\rm vac}<1.3$.

%the allowed value of $\xi_A$ is limited to $$\xi_A\simeq 2.7,$$ 
%while if we relax the backreaction level to $\mathcal{B}_A<10^{-1}$, this limit increases to $\xi_A\simeq 3.2$.
%Moreover,

The allowed parameter space we find in this section is much more
constrained than that found in the literature for the spectator
axion-SU(2) model \cite{Dimastrogiovanni:2016fuu, Fujita:2017jwq}. The
reason is two folds. First, the upper limit of $\epsilon_{B} \simeq
10^{-2}$ adopted by the previous study is too conservative to satisfy
the consistency of the slow-roll parameters given by
$\epsilon_B/\epsilon=16\epsilon_B/r_{\rm vac}<1$ for a given value of
$r_{\rm vac}<0.07$.  Second, the region of the strong backreaction was
defined as $\mathcal{J}_A/(\ga\lambda\psi^2\dot{\varphi}/f)<1$, which is
too conservative to satisfy the slow-roll dynamics of the gauge field.
The field equation of $\psi$ for the massless case is given in \eqref{eq--psi}
\be\label{eq--psiIII}
\frac{(a\psi\ddot{)}}{a}+\frac{H(a\psi\dot{)}}{a} + 2\ga^2\psi^3 -  \frac{\ga\lambda\dot\varphi}{f}\psi^2=\mathcal{J}_A.
\ee
Assuming slow-roll dynamics of the gauge field, i.e., $\frac{\ddot\psi}{H^2\psi}\ll \frac{\dot\psi}{H\psi}\ll 1$, we can write it as
\be\label{eq--psiIV}
3H \dot\psi + \dot{H}\psi + V_{{\rm{eff}},\psi}(\psi)\simeq 0,
\ee
where the field derivative of the effective potential of $\psi$ is 
\be\label{eq--psiV}
V_{{\rm{eff}},\psi}(\psi) \simeq 2 H^2\psi(1+\xp^2) -  \frac{\ga\lambda\dot\varphi}{f}\psi^2.
\ee
Slow-roll demands $V_{{\rm{eff}},\psi}(\psi) \ll 1$, while
each of the terms in the right hand side can be much larger,
e.g., $\frac{\ga\lambda\dot\varphi}{f}\psi^2/V_{{\rm{eff}},\psi}\gg1
$. On the other hand, $\mathcal{J}_A$ should be at most on the order of
the slow-roll suppressed terms, i.e., $\frac{\mathcal{J}_A}{H^2\psi} \equiv
\mathcal{B}_A \ll 1$, which is more restrictive.

%
% This is redundant.
%
%In figure \ref{BR-r}, we have used this fact to define
%the region of the strong backreaction, $\mathcal{B}_A < 10^{-2}$
%As we
%see, the smallness of the backreaction
%requires that the sourced part of the tensor perturbation be at most
%$r_{\rm source}\simeq 1.3 r_{\rm vac}$. Therefore, assuming slow-roll dynamics in the axion-SU(2) gauge field sector, the spin-2 field's backreaction provides a strong constrain for the parameter space of the massless models. This analytic study is based on assuming slow-roll dynamics for the VEV of the gauge field and the slow-roll inflation. In case of violation of these assumption, full numerical analysis is required. }}

Before we leave this section, let us comment on the Higgsed models,
e.g., Higgsed gauge-flation and Higgsed chromo-natural models \cite{Nieto:2016gnp,Adshead:2017hnc,Adshead:2016omu}. Both scalar and tensor perturbations are amplified by
the Higgs VEV in this set up, with the scalar ones being more strongly amplified; thus,
the Higgs VEV, quantified by $\xz$, reduces the tensor-to-scalar
ratio \cite{Adshead:2017hnc, Adshead:2016omu}. On the other hand, in
section \ref{backreact}, we find that the size of the backreaction is
much stronger in the Higgsed models with $\xz\neq 0$. Hence, requiring
slow-roll dynamics in the gauge field sector during (quasi)-de Sitter
expansion, the Higgsed models cannot evade $r/r_{\rm vac}<{\mathcal O}(1)$ bound
either.

\section{Conclusion}\label{conclusion}

A background of axion and $SU(2)$ gauge fields produces a copious amount of spin-2 particles during inflation. In this paper, we have calculated the number [Eq. \eqref{eq:nPairs-III}] and energy densities [Eq. \eqref{delta-rho-B}] of the spin-2 particles as well as their backreaction on the equations of motion of the axion and gauge field backgrounds [Eqs. \eqref{cJA} and \eqref{cJP}]. We provided analytical formulae which are valid for all the inflation models with $SU(2)$ gauge fields studied in the literature (see Eq.s \eqref{hs-eq} and \eqref{eq-h} for the definition of the model parameters and Table \ref{T} for their correspondence to the literature). The former results are new. The latter results were presented only numerically for a single model in this family but in \cite{Fujita:2017jwq}. 
With that exception, this is the first time that the backreaction constrains this family of models on the equations of motion. The analytical formulae derived in this paper allow us to easily estimate the importance of the backreaction for any parameters and constrain the parameter space that is consistent with perturbation theory.  Moreover, it enables us to relate the backreaction in the gauge field and axion background equations to the number density of the spin-2 field as $\mathcal{J}_A \sim \ga n_{\rm{pairs}}$ and $\mathcal{P}_{\varphi} \sim \frac{\lambda H}{f} n_{\rm{pairs}}$ respectively. 

These spin-2 particles mix with gravitational waves. We related the number density of the spin-2 particles to the power spectrum [Eq. \eqref{power-gw}] and energy density [Eq. \eqref{delta-rho-GW}] of primordial gravitational waves from inflation as well as to the size of the backreaction [Eq. \eqref{JA-npair}]. The relation to the energy density $\delta\rho_{\rm s}^{\rm GW}$ is intuitive: $\delta\rho_{\rm s}^{\rm GW}$ is given by the number density of spin-2 particles times particle's physical momentum at horizon crossing (i.e., $k/a\sim H$), times the coupling strength squared, i.e., $\delta\rho_{\rm s}^{\rm GW}\sim (\psi/\mpl)^2 H n_{\rm pairs}$. Moreover, the ratio of the power spectra of sourced and vacuum gravitational waves is also proportional to the number density of the spin-2 field and the VEV of the $SU(2)$ gauge field, i.e., $P^{\rm s}_{T}/P_T^{\rm vac} \sim (\psi/\mpl)^2 (n_{\rm pairs}/H^3)$. That gives us a physical insight into how the strength of gravitational waves from $SU(2)$ gauge fields is determined.

Finally, we constrained the parameter space of the massless
models in this class of inflationary scenarios. We find that the
backreaction and the consistency of the slow-roll condition exclude most
of the parameter space. In particular, the tensor-to-scalar ratio of the
gravitational waves sourced by the gauge field can at most be on the
order of that of the standard vacuum contribution. Going beyond the
massless models, we argue that the Higgsed models cannot evade this
conclusion either.

The analytical study presented in this paper is based on quasi-de Sitter
expansion and slow-roll dynamics of the background gauge field. For more general
situations, full numerical analysis is required. At second order in
perturbation, the spin-2 field couples to the scalar sector and
contributes to the scalar power spectrum and non-Gaussianity. This
non-linear effect may be important \cite{Papageorgiou:2018rfx}. We expect that the loop contribution to the
scalar power spectrum is related to the number density of the spin-2
field, $n_{\rm{pairs}}^2/H^6$, that we computed in this paper.

\acknowledgments

We are grateful to Kaloian D. Lozanov for his help with figures~\ref{BR-r} and \ref{BR-r-II} and useful discussion.

\appendix

\section{Symmetry of the VEV $SU(2)$ field}\label{BG-VEV}
In section \ref{review}, we present the metric and the VEV of the gauge field in a specific coordinate system in which the spatial metric is $a^2\delta_{ij}$. Here we present the general solution of the VEV of the gauge field which generates a homogeneous and isotropic energy-momentum tensor in general spatial coordinates.

Fixing the time-function, $t$, under a global rotation, we have 
\be\label{local-R}
x^{'\mu} \mapsto x^{\mu} = \Lambda^{\mu}_{~\nu} x^{'\nu},
\ee
where $\Lambda^{0}_{~\nu}=0$ and the tetrad fields ($\eta^{\alpha\beta}=\textbf{e}^{\alpha}_{\mu}\textbf{e}^{\beta\mu}$) transform as 
\be\label{tetrad}
\textbf{e}^{'\alpha}_{~\mu}(t) \mapsto \textbf{e}^{\alpha}_{~\mu}(t) = \Lambda^{\nu}_{~\mu} \textbf{e}^{'\alpha}_{~\nu}(t).
\ee
Since we choose to use the tetrad system with $\textbf{e}^{'a}_{0}(t)=0$ and $\textbf{e}^{'a}_{i}(t)=a \delta^a_i$, under the action of \eqref{local-R}, we have $\textbf{e}^{a}_{0}(t)=0$.
The general form of the gauge field's VEV in the temporal gauge that can generate a homogeneous and isotropic energy-momentum tensor (hence respect the symmetries of the FLRW background) is given by
\be\label{VEV-frw}
\bar{A}^a_{\mu}(t,\vec{x}) = \psi(t) \textbf{e}^a_{\mu}(t) .
\ee
More precisely, using \eqref{tetrad} and the above solution, the field strength tensor is 
\be
\bar{F}^a_{0i} = \frac{1}{a} \p_0 \big(a\psi\big) \textbf{e}^a_{i}(t) \an \bar{F}^a_{ij}= \ga \psi^2 \epsilon^a_{bc} \textbf{e}^b_{i}(t)\textbf{e}^c_{j}(t),
\ee 
which leads to a homogeneous and isotropic energy-momentum tensor for the gauge field sector. In addition to the above global rotational symmetry, there is a residual gauge symmetry, a continuous global $SU(2)$ symmetry, as well
\be\label{U}
U(\alpha) = \exp(i \alpha^a T_a), 
\ee
which respects the temporal gauge and the form of the VEV in \eqref{VEV-frw} 
\be\label{A-U}
\bar{A}_{\mu} \mapsto \frac{1}{-i\ga} U \bar{D}_{\mu}U^{-1} = \psi(t) \textbf{e}^a_{\mu}(t) U T_a U^{-1} = \psi(t) \textbf{e}^a_{\mu}(t) T'_a.
\ee
It is straightforward to see that the VEV gauge field in \eqref{VEV-frw} also satisfies
\be 
\nabla_{\mu} \bar{A}^{\mu}(t)=0.
\ee
Finally, action has a $Z_2$ symmetry, Parity, as
$$\psi \mapsto -\psi \quad (\textmd{and for} \quad \mathcal{L}_A=\mathcal{L}_{Cn} \quad  \varphi \mapsto - \varphi),$$
 which can be spontaneously broken by the VEV while it is still the symmetry of the \textit{background} energy-momentum tensor.

\section{Transverse-traceless field}\label{app-spin-2}
In this part, we expand the gauge field's action around the VEV as \footnote{Interestingly, the linear Einstein equations combine the $B_{\mu}$ with the GWs. However, it is independent of the other parts of the perturbed gauge field. In other words, the linearized Einstein equations do not combine $B_i$ and $A^{SV}_i$ which makes the decomposition \eqref{decom} physically meaningful and possible.} 
\be\label{decom}
A_{\mu}(x^{\nu}) = \bar{A}_{\mu}(t)  + A^{\rm SV}_{\mu}(x^{\nu}) + B_{\mu}(x^{\nu}),
\ee
where $A^{\rm SV}_{\mu}(x^{\nu})$ parameterizes the scalar and vector modes in $A^a_{\mu}$ (longitudinal modes) \footnote{More precisely, after fixing the gauge, the perturbed gauge fields have $3\times 4-3=9$ degrees of freedom which can be decomposed in terms of 3 scalars, 2 vectors and one tensor fluctuation. In particular, in the temporal gauge, we have
\be
\delta A^a_i = \delta^a_i Q + \delta^{ak} \p_{ik} \tilde{Z} + \ga \psi a \epsilon^{a~k}_{~i} \p_k (Z-\tilde{Z}) + \delta^j_a \p_i v_j + \epsilon^{a~j}_{~i} w_j + a\delta^{aj} \tilde{\gamma}_{ij},
\ee
where $Q$, $Z$, and $\tilde{z}$ are scales, $v_i$, $w_i$ and $\tilde{\gamma}_{ij}$ are transverse fields and $\tilde{\gamma}_{ij}$ is symmetric. However, $\tilde{\gamma}_{ij}$ is the only field that contributes to the transverse part of the $A^a_i$.}, while $B_{\mu}(x^{\nu})$ is the transverse part of the gauge field 
\be
\nabla_{\mu} B^{\mu}(x^{\nu})=0.
\ee
Notice that under the action of the continuous global symmetry \eqref{U}, $B_{\mu}$ transforms similar to \eqref{A-U}. 
 We have 
\be\label{BA}
{\rm{tr}}(B_{\mu} \bar{A}^{\mu})=0.
\ee
Perturbing action \eqref{Gf} and \eqref{Cn} and using \eqref{BA}, the quadratic action of $B_{\mu}(x^{\nu})$ is \footnote{For an Abelian field with action \eqref{Cn}, the quadratic action of $B_{\mu}$ is 
\be\label{ActionU1}
\da \mathcal{L}_{A}[\varphi, A] = -\frac12 \nabla_{\mu}B_{\nu}\nabla^{\mu}B^{\nu} + \frac{\lambda}{2f} \epsilon^{\mu\nu\lambda\sigma} \p_{\mu}\varphi B_{\nu}\p_{\lambda} B_{\sigma} - \frac{\alpha_H}{2} \ga^2Z_0^2 B_{\mu}B^{\mu}. 
\ee}
\be\label{Action-quad}
\da\mathcal{L}_{A}[\varphi, A] &=& -\frac12 [\tD_{\mu}B_{\nu}]^a[\tD^{\mu}B^{\nu}]^a +  \epsilon^{\mu\nu\lambda\sigma} \p_{\mu}\bar{\alpha}_A~ B^a_{\nu}\p_{\lambda} B^a_{\sigma} -\frac12 \alpha_{\rm{H}}~ \ga^2 Z^2_0 B_{\mu}^aB^{\mu}_a \nonumber\\
&-& \frac12 [M^2_{\mu\nu}]^a[B^{\mu}B^{\nu}]^a - \frac12 [M_{\mu}B_{\nu}]^a[M^{\nu}B^{\mu}]^a,
\ee
where $\da$ denotes terms quadratic order in $B_{\mu}$, $\bar{\alpha}_A$ is defined in \eqref{alpha-bar}, while the explicit form of the covariant derivative, $\tD_{\mu}$, and the mass terms are \footnote{Note that comparing to the background covariant derivative $ \bar{D}_{\mu} = \nabla_{\mu} - i\ga \bar{A}_{\mu}$, the gauge field has a factor of 2 in \eqref{Dmu}. That comes from the fact that $[X_{\mu}Y_{\nu}]^a=\frac{1}{2}[X_{\mu},Y_{\nu}]^a$ and therefore $([\bar{D}_{\mu},B_{\nu}])^a = (\tD_{\mu}B_{\nu})^a$.}
\be\label{Dmu}
\tD_{\mu} &\equiv&  \nabla_{\mu} -2i \ga \bar{A}_{\mu} ,\\
M_{\mu\nu}^{2} &\equiv&   4i \ga  \epsilon_{\mu\nu}^{~~\lambda\sigma} \p_{\lambda}\bar{\alpha}_A~ \bar{A}_{\sigma}^a T_a,\\
M_{\mu} &\equiv& 2\ga \bar{A}^a_{\mu}T_a.
 \ee
 Note that $(M_{\mu}B_{\nu})^2$ cancels the term proportional to $\bar{A}_{\mu}\bar{A}_{\nu}$ in the $(\tD_{\mu}B_{\nu})^2$. Therefore, the only actual masses are the one proportional to $\bar{\alpha}_A$ and $\alpha_{H}$.

One can determine $\dot{\bar{\alpha}}$ as a function of $\bar{A}_{\mu}$ by  the gauge field background equation in \eqref{field-eq-A}.
Notice that a given value of $\bar{A}_{\mu}$ and $Z_0$ gives the same value for $\bar{\alpha}_A$ regardless of whether $\mathcal{L}_A$  is $\mathcal{L}_{Gf}$ or $\mathcal{L}_{Cn}$. That indicates that their tensor sectors are the same at the linear order. 

We emphasize that unlike the $U(1)$ field in which its transverse part is a spin-1 field, the transverse part of $A_{\mu}^a$ (in the temporal gauge) can be written as
\be\label{Bai}
B^a_{i} = \delta^a_{j} B_{ij} = a \delta^a_{j} \tilde{\gamma}_{ij}.
\ee
In the following, we prove that $\tilde\gamma_{ij}$ is a (pseudo) spin-2 field.

\subsection{(pseudo) spin-2 in perturbed $SU(2)$ field?} \label{spin-2-SU(2)}

At first, it may come as a surprise that there is a (pseudo) spin-2 degree of freedom in a spin-1 $SU(2)$ gauge field as in \eqref{pert-A}. In this appendix, we show that once the gauge field is perturbed around its isotropic and homogeneous solution \eqref{BG-A}, there is a sector in the perturbed field that transforms as a spin-2 field under rotations and is an odd eigenstate of parity. \footnote{Another way to see the pseudo-tensor nature of the spin-2 degree of freedom is the fact that it always couples with the tensor metric perturbation with a factor of $\psi$ which is a pseudo-scalar.}

It is convenient to write the fields in the complex spherical coordinates $(r,z,\bar{z})$ which are related to the standard spherical coordinates $(r,\theta,\phi)$ as
\bea
z=e^{i\phi}\tan\frac{\theta}{2}  \quad \textmd{and} \quad \bz= e^{-i\phi} \tan\frac{\theta}{2}.
\eea
The FLRW background geometry in this coordinate system is
\bea
ds^2=-dt^2+a^2(t)\bigg(dr^2+2r^2\eta_{z\bz}dzd\bz\bigg),
\eea
where 
\be
\eta_{z\bz}=\frac{2}{(1+z\bz)^2}.
\ee
Moreover, consider a choice of the spatial triads ($\textbf{e}_i^\alpha\textbf{e}_{\alpha j}=g_{ij}$) as
\be
\textbf{e}^\alpha_r=a(t)\delta^{\alpha}_r \an  \textbf{e}^\alpha_z=a(t)\delta^\alpha_z,
\ee
where $\alpha$ is the index of the $SO(3)$ algebra and runs from 1 to 3. 
One can write $B^a_i$ in \eqref{Bai} as
\bea
B^a_i=\tilde{\gamma}_{ij}(t,\vec{x})\textbf{e}^{\alpha j}\delta^a_{\alpha}.
\eea
Then in the $(r,z,\bz)$ coordinates and for a wave propagating in the direction $\hat{k}=-\hat{r}$, we have $B^a_r=0$ while $B^a_z$ and $B^a_{\bz}$ are non-zero. Moreover, given the fact that ${\rm{tr}}(\tilde{\gamma}_{ij})=0$, the only non-zero components of $\tilde{\gamma}_{ij}$ are $\tilde{\gamma}_{zz}$ and $\tilde{\gamma}_{\bz\bz}$. Under the action of a rotation around the $x^3-$direction, we have
\bea
z\xmapsto{R} w=e^{i\delta}z \quad \textmd{and} \quad \bz\xmapsto{R} \bw=e^{-i\delta}\bz,
\eea
and 
\bea\label{tr-A}
  \textbf{e}^{\alpha z} & \xmapsto{R} & \textbf{e}^{\alpha w}=e^{i\delta}\textbf{e}^{\alpha z}\\\label{tr-e}
 \delta_{\!_{T}}A^a_z &\xmapsto{R} & \delta_{\!_{T}}A^a_w=e^{-i\delta}\delta_{\!_{T}}A^a_z .
\eea
Finally, from the combination of \eqref{tr-A} and \eqref{tr-e} we arrive at the desired result
\bea \label{R-tg}
 \tilde{\gamma}_{zz}\xmapsto{R} \tilde{\gamma}_{ww}=e^{-2i\delta}\tilde{\gamma}_{zz},
\eea
which shows that $\tilde{\gamma}_{ij}$ transforms as a spin-2 field under the action of rotations. Moreover, under the action of parity, we have
\be
 \delta A^a_i \xmapsto{P} - \delta A^a_i,
\ee
which leads to 
\bea \label{P-tg}
 \tilde{\gamma}_{ij}\xmapsto{P} -\tilde{\gamma}_{ij}.
\eea
Thus, $\tilde{\gamma}_{ij}$ is a pseudo-tensor. This completes the proof that $\tilde{\gamma}_{ij}$ is a (pseudo) spin-2 degree of freedom.

\section{Mathematical Supplement} \label{Math}

Here, we present some mathematical formulae and relations which we need throughout this work including some properties of Gamma and Whittaker functions as well as the asymptotic form of the Meijer G-functions in the large argument limit.

The Gamma function has simple poles for non-positive integers
\be\label{Gamma-r} {\rm{Res}}(\Gamma,-n)= \frac{(-1)^n}{n!} \quad (n\in \mathbb{N}).
\ee
Moreover, for any complex $z$, it satisfies 
\be\label{zGamma}
\Gamma(z+1)=z\Gamma(z),
\ee
which for non-integer values of $z$, gives
\be\label{Gamma-Gamma}
\Gamma(z)\Gamma(-z) = -\frac{\pi}{z \sin(\pi z)}    \quad (z\notin \mathbb{Z}).
\ee
The derivative of the $\Gamma$-function can be written as a polygamma function 
\be
\psi^{(d-1)}(z) = \frac{d^d}{dz^d} \ln\Gamma(z),
\ee
which has the following series representation
\be\label{psi-harmonic}
\psi^{(d-1)}(z) = (-1)^{d} (d-1)! \sum^{\infty}_{j=0} \frac{1}{(z+j)^{d}},
\ee
 which holds for $d > 1$ and any complex $z$ not equal to a negative integer.  
Therefore, the harmonic series can be written as
\be\label{har-series}
\sum_{q=1}^n \frac{1}{q}= -\psi^{(0)}(1) +\psi^{(0)}(n+1).
\ee
Furthermore, the asymptotic series \be
\lim_{\lvert z \rvert\rightarrow \infty}\Gamma(z)\simeq \sqrt{2\pi} z^{z-\frac12}\exp(-z) \bigg(1+\frac{1}{12z}+\frac{1}{288z^2}-\frac{139}{51840z^3} + \dots \bigg), 
\ee
which is valid in the sector $\lvert \textmd{arg}(z) \rvert<\pi$,
leads to the following asymptotic expansion of the digamma function
\be\label{exp-dgamma}
\psi^{(0)}(z) = \ln(z)-\frac{1}{2z}-\frac{1}{12z^2}+\frac{1}{120z^4} + \dots\,\,\,, \where \lvert z \rvert \rightarrow \infty\,. 
\ee
For complex values of $z$ in which $z=x+i y$ with finite real $x$ and $y\rightarrow \infty$, we have \cite{Nist}
\be\label{G-z-}
 \Gamma(x+i y)  \simeq \sqrt{2\pi} \lvert y \rvert^{x-\frac12} e^{-\pi y /2} e^{-i(y+\frac{\pi}{2}(\frac12-x))}.
\ee
Finally, in the limit that $z$ goes to zero, Gamma functions satisfy 
\be\label{Gamma-lim}
\lim_{z \rightarrow0} z \Gamma(-z-n) =- \frac{(-1)^n}{n!} \an \lim_{z \rightarrow0} \frac{d}{dz}(z \Gamma(-z-n))=  \frac{(-1)^n}{n!} \psi^{(0)}(n+1),
\ee
where $n\in \mathbb{N}$. For later convenience, we recall that in the complex analysis, if $f(z)$
has a pole of order $k$ at $z = z_0$ then the residues are given as
\be\label{pole-k}
{\rm{Res}}(f,z_0)= \frac{1}{(k-1)!} \frac{d^{k-1}}{dz^{k-1}}\bigg( (z-z_0)^k f(z)\bigg)\bigg\rvert_{z=z_0}.
\ee

The $W$ and $M$ Whittaker functions, which are the solutions of \eqref{Whittaker}, have the asymptotic expansions
\Beq\label{WM-asymp}
W_{\kappa,\mu} & \sim  e^{-z/2} z^{\kappa}(1+\mathcal{O}(\frac{1}{z})) \qquad\quad &\textmd{for} \quad \lvert z\rvert \rightarrow \infty, \\
M_{\kappa,\mu} & \sim  \quad z^{\mu+\frac12}(1+\mathcal{O}(z)) \qquad \quad &\textmd{for} \quad \lvert z\rvert \rightarrow 0,
\Eeq
implying that $W$/$M$ functions correspond to positive frequency modes in the asymptotic past/future limits of de Sitter, respectively. 

The $W$ function satisfies the following relation 
$$W_{\kappa,-\mu}(z)=W_{\kappa,\mu}(z).$$
In our setup in which $z$, $\mu$ and $\kappa$ are all imaginary quantities, we have $W^{*}_{\kappa,\mu}(z)=W_{-\kappa,\mu}(-z)$. The $W$ and $M$ functions are related as 
\be\label{MvsW}
M_{\kappa,\mu}(z)= \frac{\Gamma(2\mu+1)}{\Gamma(\frac12+\mu+\kappa)} e^{-i(\frac12+\mu-\kappa)\pi} W_{\kappa,\mu}(z) + \frac{\Gamma(2\mu+1)}{\Gamma(\frac12+\mu-\kappa)} e^{i\kappa \pi} W_{-\kappa,\mu}(e^{i\pi}z),
\ee
which holds when $2\mu$ is not an integer and $ -\frac{3\pi}{2}< \lvert \rm{arg}z\rvert <\frac{\pi}{2}$ \cite{Sp-book}.
The W-Whittaker functions have the Mellin-Barnes integral representation \cite{Nist}
\be\label{Mellin-Barnes}
W_{\kappa,\mu}(z) = \frac{e^{-\frac{z}{2}}}{2i\pi} \int^{i\infty}_{-i\infty} \frac{\Gamma(\frac12+\mu+s)\Gamma(\frac12-\mu+s)\Gamma(-\kappa-s)}{\Gamma(\frac12+\mu-\kappa)\Gamma(\frac12-\mu-\kappa)} z^{-s}ds  \where \lvert {\rm{arg}}(z)\rvert <\frac32\pi, \nonumber\\
\ee 
which holds when $\frac12\pm \mu -\kappa \neq 0, -1, -2 ,\dots$,
and the contour of the integration separates the poles of $\Gamma(\frac12+\mu+s)
\Gamma(\frac12-\mu+s)$ 
from poles of $\Gamma(-\kappa-s)$.

The Whittaker functions satisfy the following integral identities
\bea\label{Wh-intI}
\int x^n e^{-ix}W_{\kappa,\mu}(-2ix)dx&=&\frac{x^{n+1}\textmd{G}^{2,2}_{2,3}\biggl(-2ix\bigg|\begin{matrix}
  -n, & 1+\kappa & \\
  \frac12-\mu, & \mu+\frac12, & -n-1\\
  \end{matrix}\biggl)}{\Gamma(\frac12-\kappa-\mu)\Gamma(\frac12-\kappa+\mu)},\\\label{Wh-intII}
\int x^n e^{ix}W_{\kappa,\mu}(-2ix)dx&=&x^{n+1}\textmd{G}^{2,1}_{2,3}\biggl(-2ix\bigg|\begin{matrix}
  -n, & 1-\kappa & \\
  \frac12-\mu, & \mu+\frac12, & -n-1\\
  \end{matrix}\biggl).
\eea 

The Meijer G-functions with $Re(p)>0$, $Re(q)>0$ and $p-q\neq0$, has the following asymptotic form for $x\gg 1$
\bea\label{asym-I}
&&\frac{1}{x^{p-1}}\textmd{G}^{2,1}_{2,3}\bigg(-2ix\bigg|\begin{matrix}
 p, & q & \\
 \frac12-\mu, & \frac12+\mu, & p-1\\
 \end{matrix}\bigg)\simeq\frac{i(-2i)^p}{2}\Gamma(\frac32-p-\mu)\Gamma(\frac32-p+\mu)\Gamma(p-q),\nonumber\\
&+&\frac{i(-2i)^q}{2(q-p)}\Gamma(\frac32-q-\mu)\Gamma(\frac32-q+\mu)x^{q-p}.\\ \label{asym-II}
&&\frac{1}{x^{p-1}}\textmd{G}^{2,2}_{2,3}\bigg(-2ix\bigg|\begin{matrix}
  p, & q & \\
 \frac12-\mu, & \frac12+\mu, & p-1\\
  \end{matrix}\bigg)\simeq\frac{i(-2i)^p}{2}\frac{\Gamma(\frac32-p-\mu)\Gamma(\frac32-p+\mu)}{\Gamma(1-p+q)}.
\eea

\section{Computation of the Induced Currents} \label{Current-appendix}
In this appendix, we compute the momentum integral $\mathcal{K}[X]$ in \eqref{cJ-} which is necessary for the induced currents and energy density. \footnote{The scalar induced current in the same setup has been worked out in \cite{Lozanov:2018kpk}. The scalar induced current due to a $U(1)$ case has been worked out in \cite{Kobayashi:2014zza}.} First, we work out the total integral which is divergent. In appendix \ref{Adiabatic Subtraction}, we regularize this current by using adiabatic subtraction. Finally, in appendix \ref{energy-app}, we compute the energy density of the spin-2 fluctuations.

It is convenient to decompose $\mathcal{K}[X]$ in terms of polarizations as
$$\mathcal{K}[X] = \sum_{\sigma=\pm 2} \mathcal{K}_{\sigma}[X].$$
Using \eqref{eq:usExactSolution} and \eqref{cJ-}, we can write $\mathcal{K}_{\sigma}[X]$ as 
\be\label{cJ-A}
\mathcal{K}_{\sigma}[X]  = \lim_{\Lambda \rightarrow\infty}  e^{i\kappa_{\sigma}\pi} \int^{\Lambda}_{0} \frac{\x d\x}{2} \big(-\lambda_{\sigma} \x +  X \big) \lvert W_{\kappa_{\sigma},\mu}(-2i\x)\rvert^2,
\ee
where $\x$ is a rescaled physical momentum and $\Lambda$ is the physical UV cutoff which in the end will be sent to infinity
 $$\x \equiv \frac{k}{aH} \an  \Lambda= \frac{k_{\rm{UV}}}{aH}.$$  
Note that in our setup, both $\kappa$ and $\mu$ are pure imaginary. In the following, for notational convenience, we drop the argument $X$ of $\mathcal{K}_{\sigma}[X]$ and $\sigma$ subscript in $\kappa_{\sigma}$ and $\lambda_{\sigma}$, unless otherwise stated. Upon using the integral representation of Whittaker functions in \eqref{Mellin-Barnes}, we find
\be\label{mG-1}
&& \mathcal{K}_{\sigma} =  \lim_{\Lambda \rightarrow\infty}  \frac{1}{8(2\pi)^2} e^{i \kappa\pi} \bigg[\Gamma(\frac12+\mu-\kappa)\Gamma(\frac12-\mu-\kappa)\Gamma(\frac12+\mu^*+\kappa)\Gamma(\frac12-\mu^*+\kappa)
\bigg]^{-1}\nonumber\\
&&  \int^{i\infty}_{-i\infty} ds \int^{i\infty}_{-i\infty} ds' e^{i(s-s')\frac{\pi}{2}}(2\Lambda)^{2-s-s'}\bigg( -\frac{\lambda\Lambda}{3-s-s'} + \frac{X}{2-s-s'}\bigg) \Gamma(\frac12+\mu+s)\Gamma(\frac12-\mu+s) \nonumber\\
&& \Gamma(-\kappa-s) \Gamma(\frac12+\mu^{*}+s')\Gamma(\frac12-\mu^*+s')\Gamma(\kappa-s').
\ee
The integrand has singularities at $s'=-\frac12 \pm \mu  -n$, $\kappa+n$, $3-s$, and $2-s$. Moreover, it is proportional to $\Lambda^{2-s-s'}$ which vanishes for ${\rm{Re}}(s')>3-{\rm{Re}}(s)$ in the limit $\Lambda \rightarrow\infty$. Upon choosing the contour of $s$ such that ${\rm{Re}}(s)>-1$ and closing the $s'$-contour in the right-half plane without passing through the poles, \footnote{Note that the integral of \eqref{mG-1} over a finite path along the real axis vanishes at $\lim \rm{Im}(s') \rightarrow \pm \infty$.} we are left with the following six poles
\be
s'_1=\kappa, ~ s'_2=\kappa+1, ~s'_3=\kappa+2, ~s'_3=\kappa+3, ~s'_5=2-s, \an s'_6=3-s.
\ee
Doing the $s'$-integral, we obtain 
\be\label{tG-tot}
&& \mathcal{K}_{\sigma}  =  \frac{e^{i\kappa\pi}}{24(2\pi)^2} \bigg[\Gamma(\frac12+\mu-\kappa)\Gamma(\frac12-\mu-\kappa)\Gamma(\frac12+\mu^*+\kappa)\Gamma(\frac12-\mu^*+\kappa)
\bigg]^{-1}
\nonumber\\
&& \times \lim_{\Lambda\rightarrow \infty}\int^{i\infty}_{-i\infty} ~ds ~ \Gamma(\frac12+\mu+s)\Gamma(\frac12-\mu+s)  \Gamma(-\kappa-s)  K_{\sigma}(s,\Lambda).
\ee
where $K_{\sigma}(s,\Lambda)$ is 
\be\label{G}
&&  K_{\sigma}(s,\Lambda)=  (2i\pi) e^{-i\kappa \pi} e^{i(s+\kappa)\frac{\pi}{2}}\bigg[ 3X  \Gamma(\frac52+\mu^{*}-s)\Gamma(\frac52-\mu^*-s)\Gamma(\kappa-2+s)\nonumber\\
&& + \frac32 i \lambda  \Gamma(\frac72+\mu^{*}-s)\Gamma(\frac72-\mu^*-s)\Gamma(\kappa-3+s) \nonumber\\
&& + ~~~(2\Lambda)^{2-s-\kappa}\bigg(\frac{3\lambda\Lambda}{s+\kappa-3}- \frac{3X}{s+\kappa-2}\bigg) 
\Gamma(\frac12+\mu^{*}+\kappa)\Gamma(\frac12-\mu^*+\kappa)  \nonumber \\
&& +~~~i (2\Lambda)^{1-s-\kappa}\bigg( \frac{3\lambda\Lambda}{s+\kappa-2} - \frac{3X}{s+\kappa-1}\bigg) 
\Gamma(\frac32+\mu^{*}+\kappa)\Gamma(\frac32-\mu^*+\kappa)   \nonumber \\
&& -~~~ \frac{1}{2!} (2\Lambda)^{-s-\kappa}\bigg( \frac{3\lambda\Lambda}{s+\kappa-1} - \frac{3X}{s+\kappa}\bigg) 
\Gamma(\frac52+\mu^{*}+\kappa)\Gamma(\frac52-\mu^*+\kappa) \nonumber \\
&& -~~~\frac{i}{4} (2\Lambda)^{-s-\kappa} \frac{\lambda}{s+\kappa} 
\Gamma(\frac72+\mu^{*}+\kappa)\Gamma(\frac72-\mu^*+\kappa)  \bigg].
\ee
This integral is divergent and includes terms proportional to $\Lambda^3$, $\Lambda^2$ and $\Lambda$.

Let us first compute the (first two) finite terms in \eqref{G}. Using \eqref{Gamma-Gamma}, we write them as
\be
\mathcal{K}_{\sigma,1} = \frac{1}{24}  \bigg[\Gamma(\frac12+\mu-\kappa)\Gamma(\frac12-\mu-\kappa)\Gamma(\frac12+\mu^*+\kappa)\Gamma(\frac12-\mu^*+\kappa)
\bigg]^{-1}  \mathcal{I}_{\sigma},
\ee
where 
\be
\mathcal{I}_{\sigma} &=&  \frac{ 3\pi }{(2i\pi)} \int_{-i\infty}^{+i\infty} ds ~  \frac{e^{i(s+\kappa)\pi}}{\sin(\pi(s+\kappa))}  \Gamma(\frac12+\mu^{*}-s)\Gamma(\frac12-\mu^*-s)\Gamma(\frac12+\mu+s)\Gamma(\frac12-\mu+s)\nonumber\\
&&  \frac{\big((\frac32-s)^2-\mu^2\big)\big((\frac12-s)^2-\mu^2\big)}{(\kappa+s)(\kappa+s-1)(\kappa+s-2)}  \bigg[X + i\lambda \frac{(\frac52-s)^2-\mu^2}{2(\kappa+s-3)} \bigg].
\ee 
It can be further simplified as 
\be\label{G5-G6}
\mathcal{I}_{\sigma}  & = &  \frac{\pi}{(2i\pi)}  \int_{-i\infty}^{+i\infty} ds ~  \Gamma(\frac12+\mu^{*}-s)\Gamma(\frac12-\mu^*-s)\Gamma(\frac12+\mu+s)\Gamma(\frac12-\mu+s) \frac{ e^{i(s+\kappa)\pi}}{\sin(\pi(\kappa+s))}  \nonumber\\
&& \times  \bigg[ \mC(s)-\mC(s-1) + \frac{\C(X)}{(\kappa+s)} \bigg],
\ee
in which $\C(X)$ and $\mC(s)$ are
\be
&& \C(X) \equiv  3X(\frac12+6\kappa^2-2\mu^2)-\frac32 (7+20\kappa^2-12\mu^2) \vert \kappa \vert , \\ 
&& \mC(s) \equiv  \frac{c_1(s+3)-3c_1(s+2)+3c_1(s+1)-c_2(s+2)+2c_2(s+1)}{\kappa+s} \nonumber\\
&&+  \frac{c_1(s+2)-3 c_1(s+1)-c_2(s+1)}{\kappa+s-1} + \frac{c_1(s+1)}{\kappa+s-2} +9X s(s+1-2\kappa) \nonumber\\
&&  +\frac32 i\lambda(7+20\kappa^2-12\mu^2)s + 5i \lambda s(s+1)(1+2s-3\kappa).
\ee
Here, $c_1(s)$ and $c_2(s)$ are the following functions of $s$
\be
c_1(s)&=& -\frac{i\lambda }{4} \bigg[(\frac52-s)^2-\mu^2\bigg] \bigg[(\frac32-s)^2-\mu^2\bigg]\bigg[(\frac12-s)^2-\mu^2\bigg],\\
c_2(s)&=&\frac{3X}{2} \bigg[(\frac32-s)^2-\mu^2\bigg]\bigg[(\frac12-s)^2-\mu^2\bigg]\,.
\ee
Recalling that $\mu$ is pure imaginary, the first two terms in integral \eqref{G5-G6} can be written as
\be\label{I1-0}
&& \mathcal{I}_{\sigma,1}=  \frac{\pi}{(2i\pi)}\bigg( \int_{-i\infty}^{+i\infty} ds - \int_{-i\infty-1}^{+i\infty-1} ds \bigg) ~\Gamma(\frac12+\mu^{*}-s)\Gamma(\frac12-\mu^*-s)\Gamma(\frac12+\mu+s)\Gamma(\frac12-\mu+s) \nonumber\\
&& \frac{ e^{i(s+\kappa)\pi}}{\sin(\pi(\kappa+s))} \mC(s).
\ee
which has poles at 
\be
s_{n,\pm}&=&-\frac12\pm \mu -n ,\quad s_{n,0}=-\kappa+n,\\
\tilde{s}_{n,\pm}&=& \frac12\pm \mu+n , \qquad \tilde{s}_{n,0}= -\kappa-n,~~~ \quad (n\in \mathbb{N}).
\ee
Notice that $s_{n,0}$ poles with $n=0,1,2$ are 2nd rank while the rest are simple poles. Therefore, it is more convenient to close the contour path of $s$ on the left half-plane which only includes simple poles below 
$$s_1=-1-\kappa , \quad  s_{0,\pm}=-\frac12 \pm \mu.$$
Doing the complex integral, we have 
\be\label{I1}
 \mathcal{I}_{\sigma,1} &=&   \Gamma(\frac12+\mu^{*}+\kappa)\Gamma(\frac12-\mu^*+\kappa)\Gamma(\frac12+\mu-\kappa)\Gamma(\frac12-\mu-\kappa) \bigg\{  \mC(-1-\kappa) \nonumber\\
& + & \frac{i}{2}  \bigg[ \frac{ e^{2i\pi\kappa} + e^{2i\pi\mu}}{\sin(2\pi\mu)} \mC(-\frac12+\mu) - \frac{e^{2i\pi\kappa} + e^{-2i\pi\mu}}{\sin(2\pi\mu)} \mC(-\frac12-\mu) \bigg] \bigg\},
\ee
in which the explicit forms of $\mC(-\frac12+\mu)$ and $\mC(-\frac12-\mu)$ are
\be
\frac12\big[ \mC(-\frac12+\mu) + \mC(-\frac12-\mu) \big] &=& \frac94 X(-1+4\kappa+4\mu^2)-\frac{3i\lambda}{4}(7+20\kappa^2-12\mu^2+5\kappa(-1+4\mu^2)),\nonumber\\
\frac12\big[ \mC(-\frac12+\mu) - \mC(-\frac12-\mu) \big]&=& 2\lambda \lvert \mu \rvert \big[ - 9X \lvert \kappa \rvert -4 - 15\kappa^2 +4\mu^2 \big].
\ee
Now we turn to the last term in \eqref{G5-G6} which is
\be\label{I2}
\mathcal{I}_{\sigma,2}  & = &  \frac{\C(X)}{(2i\pi)} \pi \int_{-i\infty}^{+i\infty} ds ~  \Gamma(\frac12+\mu^{*}-s)\Gamma(\frac12-\mu^*-s)\Gamma(\frac12+\mu+s)\Gamma(\frac12-\mu+s) \nonumber \\
&& \times \frac{ e^{i(s+\kappa)\pi}}{(\kappa+s)\sin(\pi(\kappa+s))}.
\ee
Closing the contour on the left half-plane with an infinite radius semicircle, \footnote{More precisely, we use $(\kappa+s)^{d}~ (d>1)$ in the denominator of \eqref{I2} and then compute the $d\rightarrow1$ limit solution. As a result, the added infinite radius semicircle integral vanishes.} we have the poles $s_{n,\pm}$ and $\tilde{s}_{n+1,0}$, which are an infinite number of simple poles. Summing up the contribution of $s_{n}$ poles and using \eqref{psi-harmonic}, we have 
\be
\label{I2-t1}
\mathcal{I}_{\sigma,2}\big\rvert_{\tilde{s}_{n+1,0}-poles} =  \C(X)~\psi(1)~ \Gamma(\frac12+\mu^{*}+\kappa)\Gamma(\frac12-\mu^*+\kappa)\Gamma(\frac12+\mu-\kappa)\Gamma(\frac12-\mu-\kappa). ~~~~
\ee
Moreover, using \eqref{psi-harmonic}, we find the contribution of $s_{n,\pm}$ poles as
\be
\label{I2-t2}
&& \mathcal{I}_{\sigma,2}\big\rvert_{s_{n,\pm}-poles} =    \C(X)~   \Gamma(\frac12+\mu^{*}+\kappa)\Gamma(\frac12-\mu^*+\kappa)\Gamma(\frac12+\mu-\kappa)\Gamma(\frac12-\mu-\kappa) \nonumber\\
&& \frac{i}{2\sin(2\mu\pi)}  \bigg( (e^{2i\kappa\pi}+e^{2i\mu\pi}) \psi^{(0)}(\frac12-\kappa-\mu) - (e^{2i\kappa\pi}+e^{-2i\mu\pi}) \psi^{(0)}(\frac12-\kappa+\mu)\bigg),~~~
\ee
where $\psi^{(0)}(z)=\frac{d}{dz}\ln\Gamma(z)$ is the digamma function.
Finally, adding \eqref{I1}, \eqref{I2-t1} and \eqref{I2-t2}, we obtain
\be\label{K1}
&& \mathcal{K}_{\sigma,1}   = \frac{1}{24}   \bigg[   \mC(-1-\kappa) +   2i \lambda \lvert \mu \rvert (4\mu^2-15\kappa^2-4 - 9 \lvert \kappa \rvert X) \big[ \frac{\cos(2\pi\mu)+e^{2i\pi\kappa}}{\sin(2\pi\mu)}\big] -\frac94(4\mu^2-1)X    \nonumber\\
&& + \C(X) \bigg( \psi^{(0)}(1) + \frac{i}{2}  \frac{(e^{2i\kappa\pi}+e^{2i\mu\pi})}{\sin(2\mu\pi)} \psi^{(0)}(\frac12-\kappa-\mu) - \frac{i}{2}  \frac{(e^{2i\kappa\pi}+e^{-2i\mu\pi})}{\sin(2\mu\pi)} \psi^{(0)}(\frac12-\kappa+\mu) \bigg) 
\nonumber \\
&& +\frac{3i\lambda}{4}(7+20\kappa^2-12\mu^2+12X \lvert \kappa \rvert)  -\frac{15}{4} \lvert \kappa\rvert (1-4\mu^2) \bigg]. 
\ee

%-----
Now we turn to the remaining 4 ($\Lambda$-dependent) lines in \eqref{G}. Here, we close the contour in the right-half plane which encloses the following poles at the $\Lambda\rightarrow\infty$ limit
$$s_1=-\kappa, ~s_2=1-\kappa, ~ s_3=2-\kappa, \an s_4=3-\kappa.$$
Some of the above poles are second rank and therefore the integral includes a derivative of the Gamma function, i.e. $\psi^{(0)}(z)$. Doing the second complex integral in \eqref{tG-tot} and using \eqref{Gamma-lim} and \eqref{pole-k}, we find 
\be\label{K2}
\mathcal{K}_{\sigma,2} &=& \frac{1}{24} \lim_{\Lambda\rightarrow \infty} \bigg[ -4\lambda  \Lambda^{3} + 6 (X- \vert \kappa \vert ) \Lambda^{2}  +  3\lambda \Lambda \bigg(4\vert \kappa \vert X + \frac12(1+12\kappa^2-4\mu^2)\bigg)   - \C(X) \ln(2\Lambda)    \nonumber\\
&+& \C(X) \bigg( \frac{i \pi}{2} -\psi^{(0)}(1) + \psi^{(0)}(\frac12+\mu-\kappa) + \psi^{(0)}(\frac12-\mu-\kappa) \bigg) - \mC(-1-\kappa) \nonumber\\
&-& \frac{3i\lambda}{4} \big( 7 + 20 \kappa^2 -12\mu^2 + 12  X\lvert \kappa \rvert\big) - \lvert \kappa \rvert (20+37\kappa^2) + X\big(\frac32+21\kappa^2+6\mu^2\big) \bigg].
\ee
Adding \eqref{K1} and \eqref{K2}, we finally find the desired $\mathcal{K}_{\sigma}$ as 
\be\label{G-tot-lambda}
&& \mathcal{K}_{\sigma} = \frac{1}{24} \lim_{\Lambda\rightarrow \infty} \bigg[ -4\lambda_{\sigma}  \Lambda^{3} + 6 (X- \vert \kappa \vert ) \Lambda^{2}  +  3\lambda_{\sigma} \Lambda \bigg(4\vert \kappa \vert X + \frac12(1+12\kappa^2-4\mu^2)\bigg)    \nonumber\\
&-&  \C(X) \ln(2\Lambda)  - \lvert \kappa \rvert (\frac{95}{4}+37\kappa^2-15\mu^2)  +   2i\lambda_{\sigma} \lvert \mu \rvert (4\mu^2-15\kappa^2-4 - 9 \lvert \kappa \rvert X) \big[ \frac{\cos(2\pi\mu)+e^{2i\pi\kappa_{\sigma}}}{\sin(2\pi\mu)}\big]   \nonumber\\ 
&+& \frac{i}{2} \C(X) \bigg( \pi + \frac{(e^{2i\kappa_{\sigma}\pi}+e^{-2i\mu\pi})}{\sin(2\mu\pi)} \psi^{(0)}(\frac12-\kappa_{\sigma}-\mu) -   \frac{(e^{2i\kappa_{\sigma}\pi}+e^{2i\mu\pi})}{\sin(2\mu\pi)} \psi^{(0)}(\frac12-\kappa_{\sigma}+\mu) \bigg)  \nonumber\\
&+& 3 X\big(\frac54+7\kappa^2-\mu^2\big) \bigg].
\ee
It has divergent terms of the order $3$, $2$, $1$ and log of $\Lambda$. Summing over the polarization states, we find 
\be\label{K-tot}
&& \mathcal{K} = \frac{1}{12} \lim_{\Lambda\rightarrow \infty} \bigg\{ 6 (X- \vert \kappa \vert ) \Lambda^{2}   - \C(X) \ln(2\Lambda) - \lvert \kappa \rvert (\frac{95}{4}-37\lvert \kappa \rvert^2 + 15 \lvert\mu \rvert^2) +3 X\big(\frac54- 7 \lvert\kappa\rvert^2 + \lvert\mu\rvert^2\big)  \nonumber\\
&+&  2 \lvert \mu \rvert (-4 \lvert\mu\rvert^2 + 15\lvert\kappa\rvert^2-4-9 \lvert \kappa \rvert X)  \frac{\sinh(2\pi \lvert \kappa \rvert)}{\sinh(2\pi\lvert \mu \rvert)}  +  \frac{\C(X)}{4} {\rm{Re}}\bigg[ \frac{(e^{2 \lvert \kappa \rvert \pi}+e^{2 \lvert \mu \rvert \pi})}{\sinh(2\lvert \mu \rvert\pi)} \psi^{(0)}(\frac12+i \lvert \kappa \rvert - i \lvert\mu\rvert) \nonumber\\ 
&-& \frac{(e^{2\lvert \kappa\rvert \pi}+e^{-2 \lvert\mu\rvert \pi})}{\sinh(2\lvert \mu \rvert\pi)} \psi^{(0)}(\frac12+ i \lvert\kappa \rvert + i \lvert\mu\rvert) + \frac{(e^{-2 \lvert \kappa \rvert \pi}+e^{2 \lvert \mu \rvert \pi})}{\sinh(2\lvert \mu \rvert\pi)} \psi^{(0)}(\frac12-i \lvert \kappa \rvert - i \lvert\mu\rvert) \nonumber\\
&-& \frac{(e^{-2 \lvert \kappa \rvert \pi}+e^{-2 \lvert \mu \rvert \pi})}{\sinh(2\lvert \mu \rvert\pi)} \psi^{(0)}(\frac12-i \lvert \kappa \rvert + i \lvert\mu\rvert)\bigg]   \bigg\}.
\ee
We find that $\mathcal{K}$ is real and has divergent terms of the order $2$ and log of $\Lambda$.

Before renormalizing $\mathcal{K}$ and removing $\Lambda$ terms, let us take a closer look at the finite terms in \eqref{K-tot}. Recalling \eqref{xp-condition} and \eqref{kappa-mu--} ($\lvert \kappa \rvert > 3.5$), we realize that the dominant finite terms are proportional to $e^{2(\lvert \kappa \rvert - \lvert \mu \rvert)\pi} > e^{\pi\xp} \gg 1$. Thus, we can approximate $\mathcal{K}[X]$ as 
\be\label{K-tot-}
&& \mathcal{K}[X] = \frac{1}{6} e^{2(\lvert \kappa \rvert - \lvert \mu \rvert)\pi} \bigg\{ \lvert \mu \rvert (-4\lvert\mu\rvert^2 + 15\lvert\kappa\rvert^2-4 - 9 \lvert \kappa \rvert X)  +  \frac{\C(X)}{4} {\rm{Re}}\bigg[ \psi^{(0)}(\frac12 + i \lvert \kappa \rvert - i \lvert \mu \rvert ) \nonumber\\ 
&& - \psi^{(0)}(\frac12 + i \lvert \kappa \rvert + i \lvert \mu \rvert ) \bigg] \bigg\} + \lim_{\Lambda\rightarrow \infty} \bigg[ \frac{1}{2} (X- \vert \kappa \vert ) \Lambda^{2}   - \frac{\C(X)}{12} \ln(2\Lambda) \bigg] + \mathcal{O}(\xp^{3-m}\xz^m),
\ee
where $\mathcal{O}(\xp^{3-m}\xz^m)$ with $m=0,1,2,3$ are the next leading terms and hence negligible. The quantity $(\lvert \kappa \rvert - \lvert \mu \rvert)$ is presented in figure \ref{kappa-mu}. In the next section, we renormalize the above and find the physical quantity, $\mathcal{K}_{reg}[X]$, as
\be
\mathcal{K}_{reg}[X] \equiv \mathcal{K}[X] - \mathcal{K}^{c.t.}[X],
\ee
where $\mathcal{K}^{c.t.}[X]$ is the counter-term.

\subsection{Regularized current} \label{Adiabatic Subtraction}
In this section, we use the adiabatic subtraction technique in curved QFT to remove the divergent terms in the current. 
The mode function $ِB_{\sigma,\bk}(\tau)$ has the following WKB form
\be\label{WKBform}
ِB^{WKB}_{\sigma,\bk}(\tau)=\frac{1}{(2\pi)^{\frac32}\sqrt{2W_{\sigma,\bk}}(\tau)} e^{- i \int^{\tau}_{-\infty} d\tilde{\tau} W_{\sigma,\bk}(\tilde\tau)},
\ee
where $W_{\sigma,\bk}^2 $ can be written in terms of the instantaneous frequency, $\omega^2_{\sigma,\bk}$, defined by
\be
\omega^2_{\sigma,\bk}  = k^2-\lambda_{\sigma} \delta_{\rm{c}} k \mH + \frac{\tilde{m}^2}{H^2} \mH^2,
\ee
as
\be\label{W}
W_{\sigma,\bk}^2 = \omega^2_{\sigma,\bk} -\frac{a''}{a} +\frac34 \bigg(\frac{W'_{\sigma,\bk}}{W_{\sigma,\bk}}\bigg)^2 -\frac12 \frac{W''_{\sigma,\bk}}{W_{\sigma,\bk}}.
\ee

If $W_{\sigma,\bk}$ is real and positive, then $B^{WKB}_{\sigma,\bk}(\tau)$ corresponds to canonically normalized positive frequency modes in the asymptotic past.
For regularization in 4 dimension, we need to expand $W_{\sigma,\bk}$ up to the second order of time derivatives with respect to $a$ as
\be\label{W-2}
W^2_{\sigma,\bk}  = \omega_{\sigma,\bk}^2 -\frac{a''}{a} +\frac34  \bigg(\frac{\omega'_{\sigma,\bk}}{\omega_{\sigma,\bk}}\bigg)^2 -\frac12 \frac{\omega''_{\sigma,\bk}}{\omega_{\sigma,\bk}}.
\ee
Using \eqref{W-2} in \eqref{cJ-}, we obtain $\mathcal{K}^{c.t.}_{\sigma}[X]$ as \footnote{Note that the instantaneous frequency squared is negative in the interval $\x\in (\x_2,\x_1)$ in \eqref{q12}. Therefore, technically, we have to consider an IR cut-off for the momentum integral, $\Lambda_{IR}\gtrsim \x_1$. In principle it can be a problem since $\Lambda_{IR}$ explicitly appears in the finite terms. However, this effect and the other finite terms in $\mathcal{K}^{c.t.}$ are at most of the order $\mathcal{O}(\xp^{3-m}\xz^m)$ with $m=0,1,2,3$ and are subleading comparing with the dominant finite terms of $\mathcal{K}$ in \eqref{K-tot-}. As a result, a more careful regularization process would not improve our results.}
\be 
\mathcal{K}_{\sigma}^{c.t.} &=& \lim_{\Lambda \rightarrow\infty} \int^{\Lambda} \x'^2 d\x'\frac{\mH}{2\omega_{\sigma,\bk}} \big( -\lambda_{\sigma} \x' + X \big) \bigg( 1 + \frac{\mH^2}{ \omega^2_{\sigma,\bk}}  + \frac{\mH^4}{4\omega^4_{\sigma,\bk}}\big( - 2 \lambda_{\sigma} \lvert \kappa \rvert \x' + \frac{27}{4} - 3\mu^2 \big) \nonumber\\
&-&\frac58 \frac{\mH^6}{\omega^6_{\sigma,\bk}}( - \lambda_{\sigma} \lvert \kappa \rvert \x' + \frac94 -\mu^2 )^2 \bigg).
\ee

Doing the integral, and summing over the polarization states, we obtain 
\be\label{cG-WKB}
&& \mathcal{K}_{\sigma}^{c.t.} = \frac{1}{12} \lim_{\Lambda\rightarrow \infty} \bigg[ 6 (X- \vert \kappa \vert ) \Lambda^{2}  - \
\C(X) \ln(2\Lambda) \bigg] + \mathcal{O}(\xp^{3-m}\xz^m).
\ee
The above counter term has divergences of the order $2$ and log of $\Lambda$ which cancel with the divergences of the total $\mathcal{K}$ in \eqref{K-tot-}. Moreover, it has finite terms of order $\mathcal{O}(\xp^{3-m}\xz^m)$ which are subleading compared to the dominate terms in $\mathcal{K}$ and we neglect them here. Finally, subtracting \eqref{cG-WKB} from \eqref{K-tot-}, we have the desired regularized $\mathcal{K}[X]$ as
\be\label{K-reg}
\mathcal{K}_{reg}[X] &=&
\frac{1}{6} e^{2(\lvert \kappa \rvert - \lvert \mu \rvert)\pi} \bigg\{ \lvert \mu \rvert (-4\lvert\mu\rvert^2 + 15\lvert\kappa\rvert^2-4 - 9 \lvert \kappa \rvert X)  +  \frac{\C(X)}{4} {\rm{Re}}\bigg[ \psi^{(0)}(\frac12 + i \lvert \kappa \rvert - i \lvert \mu \rvert ) \nonumber\\ 
&& - \psi^{(0)}(\frac12 + i \lvert \kappa \rvert + i \lvert \mu \rvert ) \bigg] \bigg\} + \mathcal{O}(\xp^{3-m}\xz^m), 
\ee
where $\mathcal{O}(\xp^{3-m}\xz^m)$ with $m=0,1,2,3$ are the next leading terms and hence negligible.

\subsection{Energy density} \label{energy-app}
Here we compute the energy density in the spin-2 fluctuations of the gauge field
which is presented in \eqref{drho-2}. Going to Fourier space, we can write $\langle \da\rho \rangle$ as 
\be
\langle \da\rho \rangle = \frac{1}{2a^4} \sum_{\sigma} \int^{\Lambda}_0 dk^3 \bigg[ \lvert B'_{\sigma} \rvert^2 + \bigg(k^2 - 2\lambda_{\sigma} \xp k\mH  +\alpha_{H} \xz^2 \mH^2 \bigg) \lvert B_{\sigma} \rvert^2 \bigg],
\ee
where in the end we send $\Lambda$ to infinity.
From \eqref{eq:usExactSolution}, we can write $\Bs$ as
$$\Bs(\tau,\vec{k}) = \frac{1}{\sqrt{k}}\bar{\Bs}(\x).$$
We then write the energy density as 
\be
\langle \da\rho \rangle = (2\pi)H^4 \sum_{\sigma} \int^{\Lambda}_{0} \x d\x \bigg[ \x^2 \big(\lvert \p_{\x}\bar{\Bs} \rvert^2 + \lvert \bar{\Bs} \rvert^2  \big) + \big( - 2\lambda_{\sigma} \xp \x  + \alpha_{H} \xz^2 \big) \lvert \bar{\Bs} \rvert^2 \bigg].~~
\ee
The first term in the integral can be written as
\be\label{eq-dot}
\int^{\Lambda}_{0} \x^3 d\x  \lvert \p_{\x}\bar{\Bs} \rvert^2 = \int^{\Lambda}_{0}  d\x \bigg[  \p_{\x}\bigg( \x^3 \bar{\Bs} \p_{\x} \bar{\Bs}  - \frac32  \x^2 \bar{\Bs}^2\bigg) + 3\x \bar{\Bs}^2 -\x^3 \bar{\Bs} \p_{\x}^2 \bar{\Bs} \bigg].
\ee
Using the field equation of $\tilde{\Bs}$ in \eqref{eq-h}, we arrive at
\be
\int^{\Lambda}_{0} \x^3 d\x  \lvert \p_{\x}\bar{\Bs} \rvert^2 = \int^{\Lambda}_{0} \x d\x \bigg[  1 +  \x^2 -\lambda_{\sigma} \delta_c \x  + \frac{\tilde{m}^2}{H^2} \bigg] \lvert \bar{\Bs} \rvert^2  \bigg] + \tilde{\mathcal{C}}_{c.t.},
\ee
where $\tilde{\mathcal{C}}_{c.t.}$ comes from integrating the total derivative term
\be
\tilde{\mathcal{C}}_{c.t.} = \lim_{\Lambda \rightarrow \infty}\bigg( \x^3 \bar{\Bs} \p_{\x} \bar{\Bs}  - \frac32  \x^2 \bar{\Bs}^2 \bigg)\bigg\rvert_{\x=\Lambda}.
\ee

\begin{figure}[h!]
\begin{center}
\includegraphics[width=0.6\textwidth]{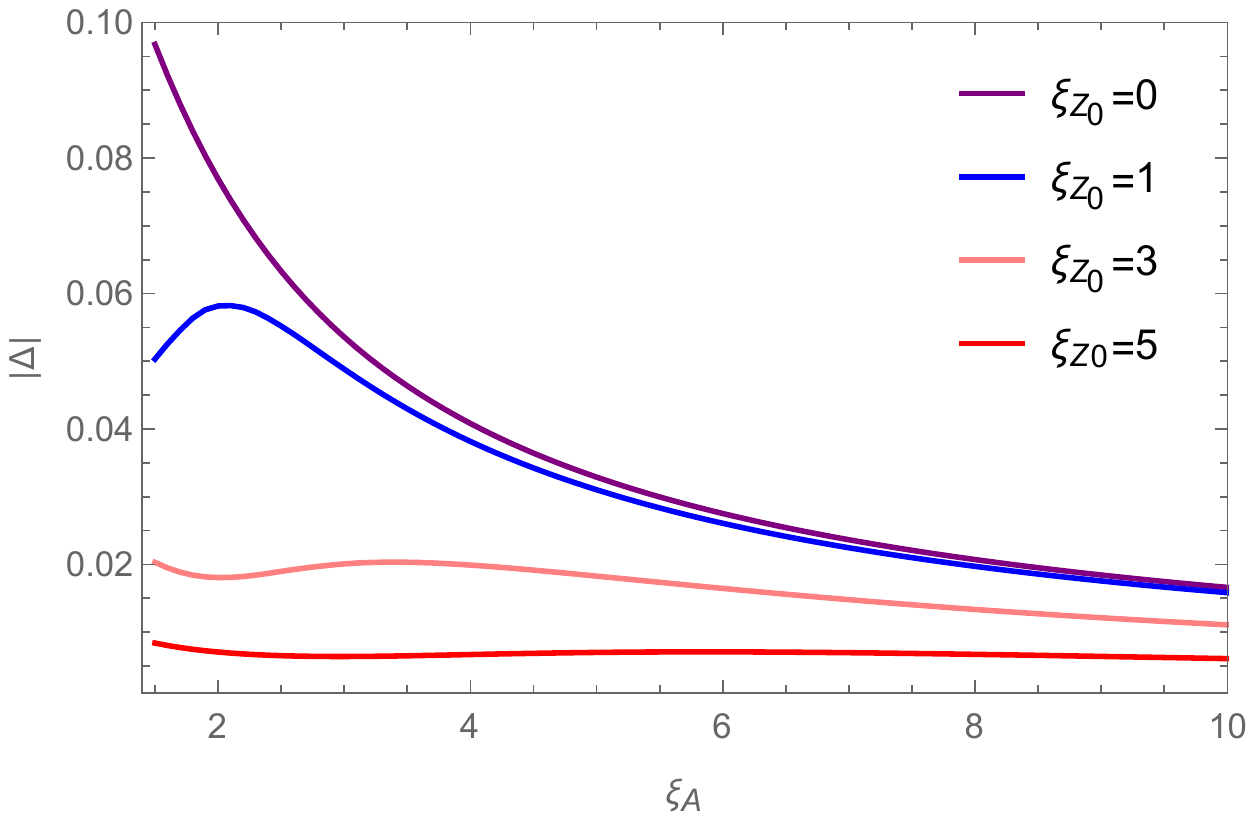} 
\caption{ The ratio $\Delta \equiv  \bigg( \frac{\int^{\x_{\rm{IR}}}_{0} d\x \x^3 \lvert \bar{B}_{+} \rvert^2 }{ (\delta_c+2\xp)  \int^{\x_{\rm{IR}}}_{0} d\x \x^2 \lvert \bar{B}_{+} \rvert^2}\bigg)$ as a function of $\xp$ and $\xz$. Here $\x_{\rm{IR}}$ is the physical momentum in which the deviation from adiabaticity is $\Omega_{+}(\x_{\rm{IR}})=0.1$. The ratio $\Delta$ for the minus polarization is smaller with the same order of magnitude.}\label{Delta}
\end{center}
\end{figure}

Using \eqref{eq-dot}, we can write the bare energy density as
\be\label{delta-rh}
\langle \da\rho \rangle = (2\pi)H^4 \bigg\{ \sum_{\sigma}   \int^{\Lambda}_{0} \x d\x \bigg[ (\delta_c + 2\xp)\bigg( -\lambda_{\sigma} \x  + \frac{1 + \frac{\tilde{m}^2}{H^2}  + \alpha_{H} \xz^2}{\delta_c + 2\xp} \bigg) +  2\x^2 \bigg] \lvert \bar{\Bs} \rvert^2 + \mathcal{C}_{c.t.}\bigg\}. \nonumber\\
\ee
From \eqref{cJ-}, the first two terms can be written in terms of $\mathcal{K}\big[\frac{1 + \frac{\tilde{m}^2}{H^2}  + \alpha_{H} \xz^2}{\delta_c + 2\xp}\big]$ and we have its renormalized form in \eqref{K-reg}. The \textit{regularized} part of the last term inside the brackets satisfies the following inequality (see figure \ref{Delta})
\be\label{eq-x3int}
 \sum_{\sigma} \int d\x \x^3 \lvert \bar{\Bs} \rvert^2  \lesssim 0.1 \times (\delta_c+2\xp)  \sum_{\sigma}   \int \lvert \bar{\Bs} \rvert^2  \x^2 d\x,
\ee
which implies that it is negligible comparing to the first term in \eqref{delta-rh}. Thus, the \textit{regularized} energy density can be well approximated as 
\be\label{rho-reg}
\langle \da\rho \rangle_{reg} \approx \frac{H^4}{(2\pi)^2}  (\delta_c + 2\xp) \mathcal{K}_{reg}\big[\frac{ 1 + \frac{\tilde{m}^2}{H^2}  + \alpha_{H} \xz^2}{\delta_c + 2\xp} \big],
\ee
which, as we see, is given in terms of the regularized $\mathcal{K}_{reg}[X]$ in \eqref{K-reg} with $X=\frac{ 1 + \frac{\tilde{m}^2}{H^2}  + \alpha_{H} \xz^2}{\delta_c + 2\xp}$.

\section{Sourced graviational waves}\label{GW-app}

In this appendix, we work out the analytical form of the gravitational waves sourced by the gauge field for the general action \eqref{g-action}. The derivation given here follows closely \cite{Maleknejad:2016qjz}.

\begin{figure}[h!]
\begin{center}
\includegraphics[width=0.45\textwidth]{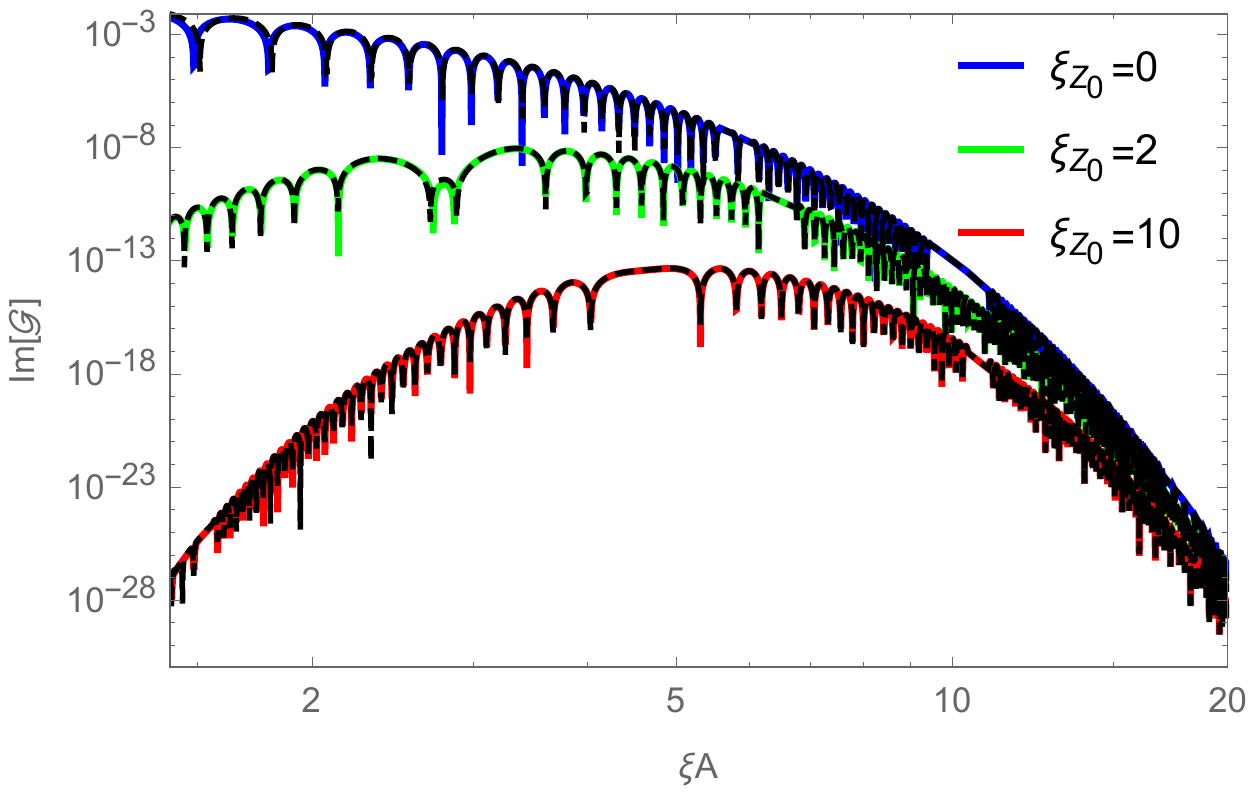} 
\includegraphics[width=0.45\textwidth]{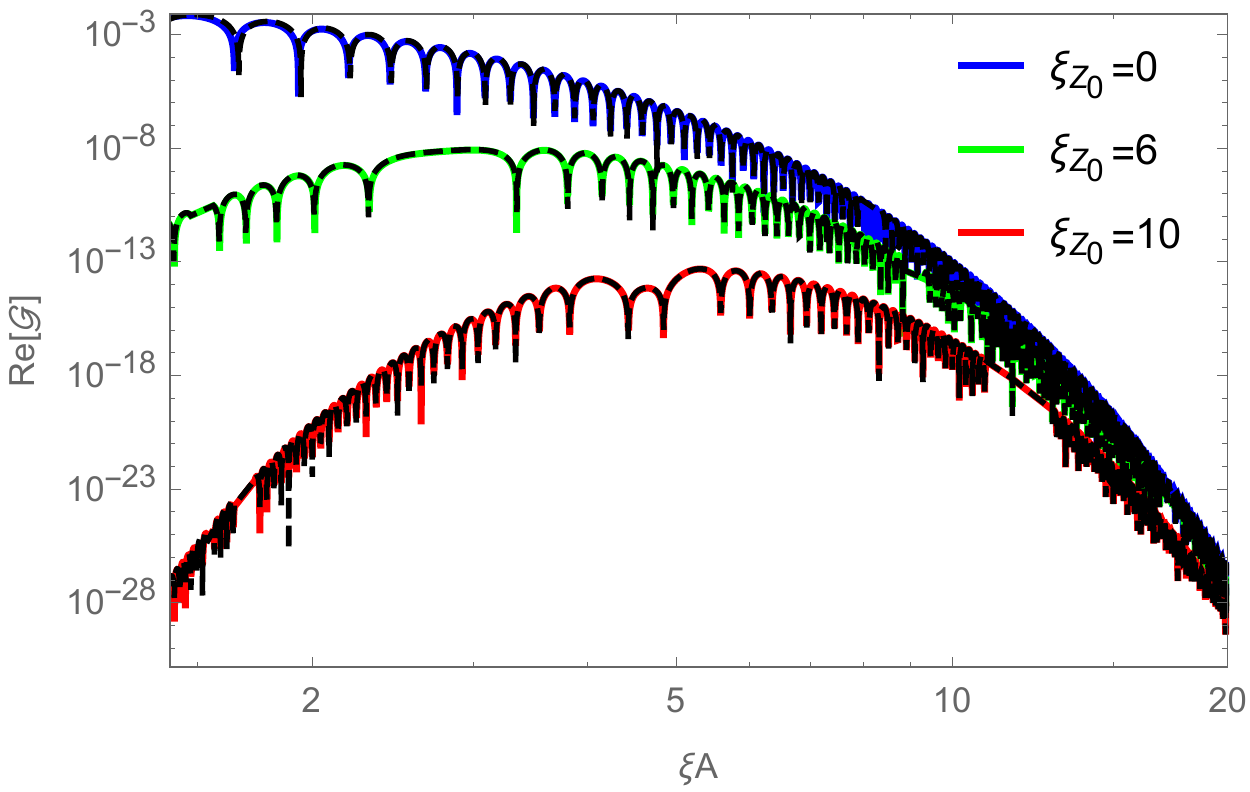} 
\caption{The exact form of $\mathcal{G}_{\sigma}(\xp,\xi_{Z_0})$ in \eqref{Int-IV} compared to the approximated from in \eqref{Int-V} with respect to $\xi_A$ for different values of $\xi_{Z_0}$. The solid (black) lines are the exact forms while the dashed line on the top of each curve is its approximated form.}\label{Curly-G-E6-E7}
\end{center}
\end{figure}

The inhomogeneous solution of \eqref{hs-eq} sourced by $B_{\sigma}$ is given as
\be\label{hs-Green}
 \hs^s(\tau,\vk) = \int^{\Lambda}_{\x} G(\x,\x') S^T_{\sigma}(\x') d\x',
\ee
where the source term, $S^T_{\sigma}(\x')$, and the retarded Green's function, $G(\x,\x')$, are given by
\be
S^T_{\sigma}(\x') &=& \frac{2}{\x'} \bigg(\frac{\psi}{\mpl}\bigg) \bigg[ (-\lambda_{\sigma}\beta_c + \frac{\theta_c}{\x'})B_{\sigma}(\x',\vk) + \p_{\x'} B_{\sigma}(\x',\vk)\bigg],\\
G(\x,\x') &=& \bigg( \frac{\x'-\x}{\x'\x} \cos(\x'-\x) - (1+\frac{1}{\x\x'})\sin(\x'-\x)\bigg) \Theta(\x'-\x),
\ee
respectively. Here $\Theta(\x'-\x)$ is the Heaviside step function. 
Using the integral relations \eqref{Wh-intI} and \eqref{Wh-intII} and doing the integral \eqref{hs-Green} for $-k\tau =\x\ll 1$, we obtain
\bea\label{Int-I}
&&  \hs^s(\tau,\vk) \simeq \frac{e^{i\kappa_{\sigma}\!\pi/2}}{(2\pi)^{\frac32}}\bigg[\! -(i+\lambda_{\sigma}\beta_c)\big\{\frac{\textmd{G}^{2,1}_{2,3}\biggl(-2i\xL\bigg|\begin{matrix}
     1, & 1+\kappa_{\sigma} & \\
      \frac12-\mu, & \frac12+\mu, & 0\\
   \end{matrix}\bigg)}{\Gamma(\frac12-\kappa_{\sigma}-\mu)\Gamma(\frac12-\kappa_{\sigma} +\mu)}
  +\textmd{G}^{2,2}_{2,3}\biggl(-2i\xL\bigg|\begin{matrix}
  1, & 1-\kappa_{\sigma} & \\
 \frac12-\mu, & \frac12+\mu, & 0\\
  \end{matrix}\bigg)\big\}\nonumber\\
  &-& \frac{1}{\xL} \big\{ \textmd{G}^{2,2}_{2,3}\biggl(-2i\xL\bigg|\begin{matrix}
  2, & -\kappa_{\sigma} & \\
 \frac12-\mu, & \frac12+\mu, & 1\\
   \end{matrix}\biggl) - (1-\kappa_{\sigma} -i\lambda_{\sigma} \beta_c+\theta_c)\textmd{G}^{2,2}_{2,3}\biggl(-2i\xL\bigg|\begin{matrix}
               2, & 1-\kappa_{\sigma} & \\
               \frac12-\mu, & \frac12+\mu, & 1\\
               \end{matrix}\biggl)\nonumber\\
&+& \frac{\textmd{G}^{2,1}_{2,3}\biggl(-2i\xL\bigg|\begin{matrix}
       2, & 2+\kappa_{\sigma} & \\
      \frac12-\mu, & \frac12+\mu, & 1\\
        \end{matrix}\bigg)}{\Gamma(-\frac12-\kappa_{\sigma}-\mu)\Gamma(-\frac12-\kappa_{\sigma}+\mu)}
          + (1+\kappa_{\sigma}-i\lambda_{\sigma} \beta_c-\theta_c)\frac{\textmd{G}^{2,1}_{2,3}\biggl(-2i\xL\bigg|\begin{matrix}
           2, & 1+\kappa_{\sigma} & \\
          \frac12-\mu, & \frac12+\mu, & 1\\
         \end{matrix}\bigg)}{\Gamma(\frac12-\kappa_{\sigma}-\mu)\Gamma(\frac12-\kappa_{\sigma}+\mu)} \big\}\nonumber\\
&+&\frac{1}{\xL^2} \big\{ i(\theta_c-\kappa_{\sigma})\big[ \textmd{G}^{2,2}_{2,3}\biggl(-2i\xL\bigg|\begin{matrix}
 3, & 1-\kappa_{\sigma} & \\
  \frac12-\mu, & \frac12+\mu, & 2\\
  \end{matrix}\bigg)-\frac{\textmd{G}^{2,1}_{2,3}\biggl(-2i\xL\bigg|\begin{matrix}
               3, & 1+\kappa_{\sigma} & \\
              \frac12-\mu, & \frac12+\mu, & 2\\
             \end{matrix}\bigg)}{\Gamma(\frac12-\kappa_{\sigma}-\mu)\Gamma(\frac12-\kappa_{\sigma}+\mu)}\big]\nonumber\\
 &-& i \textmd{G}^{2,2}_{2,3}\biggl(-2i\xL\bigg|\begin{matrix}
  3, & -\kappa_{\sigma} & \\
 \frac12-\mu, & \frac12+\mu, & 2\\
  \end{matrix}\bigg)
 + i \frac{\textmd{G}^{2,1}_{2,3}\biggl(-2i\xL\bigg|\begin{matrix}
     3, & 2+\kappa_{\sigma} & \\
    \frac12-\mu, & \frac12+\mu, & 2\\
   \end{matrix}\bigg)}{\Gamma(-\frac12-\kappa_{\sigma}-\mu)\Gamma(-\frac12-\kappa_{\sigma}+\mu)} \big\} \bigg] \times \bigg(\frac{\psi}{\mpl}\bigg) \bigg( \frac{aH}{\sqrt{2}k^{\frac32}}\bigg).\nonumber\\
\eea
which implies that the sourced part of the gravitational wave can be written as
\be
\hs^s(\tau,\vk) = \frac{e^{i\kappa_{\sigma}\!\pi/2}}{(2\pi)^{\frac32}} \bigg(\frac{\psi}{\mpl}\bigg) \bigg( \frac{aH}{\sqrt{2}k^{\frac32}}\bigg)
\mathcal{G}_{\sigma}(\xp,\xi_{Z_0}).
\ee
Using the asymptotic form of Meijer-G functions at $\x_{\Lambda}\gg 1$ in  \eqref{asym-I}-\eqref{asym-II}, we can simplify $\mathcal{G}_{\sigma}$ as 
\bea\label{Int-IV}
&&  \mathcal{G}_{\sigma}(\xp,\xi_{Z_0}) = \frac{\Gamma(\frac12-\mu)\Gamma(\frac12+\mu)}{\Gamma(-\kappa_{\sigma})} \bigg\{\! \bigg[\frac{(i+\lambda_{\sigma}\beta_c)}{\kappa_{\sigma}} - \frac{2(\lambda_{\sigma} \beta_c+i(2+\theta_c))}{(\frac14-\mu^2)} +  \frac{4 i(2+\theta_c)(1+\kappa_{\sigma})}{(\frac94-\mu^2)(\frac14-\mu^2)} \bigg]  \nonumber\\
&& +  \bigg[  i - \lambda_{\sigma}\beta_c + \frac{2i \kappa_{\sigma}(i \lambda_{\sigma}\beta_c+ \theta_c)}{(\frac14-\mu^2)}+ \frac{4 i \kappa_{\sigma}\big(  \frac14-\mu^2 +2\kappa_{\sigma} +\theta(\kappa_{\sigma}-1)\big)}{(\frac94-\mu^2)(\frac14-\mu^2)} \bigg] \frac{\Gamma^2(-\kappa_{\sigma})}{\Gamma(\frac12-\kappa_{\sigma}-\mu)\Gamma(\frac12-\kappa_{\sigma}+\mu) }  \bigg\}.\nonumber\\
\eea
We show $\lvert \mathcal{G}_{+}(\xp,\xz) \rvert^2$ in figure \ref{A22}. As we see, this function decreases with the increase of $\xp$ and $\xz$ and due to the Gamma function has infinite number of roots on the $\xp$ axis.
We can well approximate the above as (see figure \ref{Curly-G-E6-E7})
\bea\label{Int-V}
&&  \mathcal{G}_{\sigma}(\xp,\xi_{Z_0}) \simeq \frac{\pi}{\cos(\pi\mu)\Gamma(-\kappa_{\sigma})} \bigg[\! \frac{(i+\lambda_{\sigma}\beta_c)}{\kappa_{\sigma}} +  \frac{( i - \lambda_{\sigma}\beta_c )\Gamma^2(-\kappa_{\sigma})}{\Gamma(\frac12-\kappa_{\sigma}-\mu)\Gamma(\frac12-\kappa_{\sigma}+\mu) }  \bigg].~~
\eea

Finally, after using \eqref{G-z-} in the limit that $\lvert \mu \rvert \gg1$, and up to a phase factor, we have 
\bea\label{Int-VI}
&&  \mathcal{G}_{\sigma}(\xp,\xi_{Z_0}) \simeq  \mathcal{A}_{\sigma} \sqrt{2\pi\lvert \kappa \rvert}  e^{\pi (\frac12\lambda_{\sigma}\lvert \kappa \rvert - \lvert \mu \rvert)},
\eea
where $\mathcal{A}_{\sigma}$ is the following quantity
\be\label{cal-A}
\mathcal{A}_{\sigma} \equiv  \bigg(\frac{(i+\lambda_{\sigma}\beta_c)}{\kappa_{\sigma}} +  \frac{( i - \lambda_{\sigma}\beta_c )\Gamma^2(-\kappa_{\sigma})}{\Gamma(\frac12-\kappa_{\sigma}-\mu)\Gamma(\frac12-\kappa_{\sigma}+\mu) }\bigg).
\ee 
In figure \ref{A22}, we present $\lvert \mathcal{A}_{+} \rvert^2$ as a function of $\xp$ and $\xz$.

\bibliography{references}

\end{document}